\documentclass[a4paper,11pt]{article}
\pdfoutput=1

\usepackage{jheppub}

\usepackage{subfig}
\usepackage{xspace}
\usepackage[countmax]{subfloat}
\usepackage{slashed}

\usepackage{bbm}

\usepackage{color}
\definecolor{darkblue}{rgb}{0,0,0.5}
\definecolor{darkred}{rgb}{0.5,0,0}

\newcommand{\ecf}[2]{e_{#1}^{(#2)}} 
\newcommand{\ecfnobeta}[1]{e_{#1}} 

\newcommand{\Dobs}[2]{D_{#1}^{(#2)}} 
\newcommand{\Dobsnobeta}[1]{D_{#1}} 

\newcommand{\Xobs}[2]{X_{#1}^{(#2)}} 

\newcommand{\Cobs}[2]{C_{#1}^{(#2)}} 
\newcommand{\Cobsnobeta}[1]{C_{#1}}

\DeclareRobustCommand{\Sec}[1]{Sec.~\ref{#1}}

\DeclareRobustCommand{\Tab}[1]{Table~\ref{#1}}

\DeclareRobustCommand{\Fig}[1]{Fig.~\ref{#1}}
\DeclareRobustCommand{\Figs}[2]{Figs.~\ref{#1} and \ref{#2}}
\DeclareRobustCommand{\Eq}[1]{Eq.~(\ref{#1})}
\DeclareRobustCommand{\Eqs}[2]{Eqs.~(\ref{#1}) and (\ref{#2})}
\DeclareRobustCommand{\Ref}[1]{Ref.~\cite{#1}}

\newcommand{\Nsub}[2]{\tau_{#1}^{(#2)}}

\newcommand{\pythia}[1]{\textsc{Pythia\xspace #1}}
\newcommand{\madgraph}[1]{\textsc{MadGraph5\xspace #1}}
\newcommand{\fastjet}[1]{\textsc{FastJet\xspace #1}}

\newcommand{\herwigpp}[1]{\textsc{Herwig++\xspace #1}}

\preprint{MIT--CTP 4588}

\title{Power Counting to Better Jet Observables}

\author{Andrew J. Larkoski,}
\author{Ian Moult,}
\author{and Duff Neill}

\affiliation{Center for Theoretical Physics, Massachusetts Institute of Technology, Cambridge, MA 02139, USA}

\emailAdd{larkoski@mit.edu}
\emailAdd{ianmoult@mit.edu}
\emailAdd{dneill@mit.edu}

\abstract{
Optimized jet substructure observables for identifying boosted topologies will play an essential role in maximizing the physics reach of the Large Hadron Collider. Ideally, the design of discriminating variables would be informed by analytic calculations in perturbative QCD. Unfortunately, explicit calculations are often not feasible due to the complexity of the observables used for discrimination, and so many validation studies rely heavily, and solely, on Monte Carlo. In this paper we show how methods based on the parametric power counting of the dynamics of QCD, familiar from effective theory analyses, can be used to design, understand, and make robust predictions for the behavior of jet substructure variables.  As a concrete example, we apply power counting for discriminating boosted $Z$ bosons from massive QCD jets using observables formed from the $n$-point energy correlation functions. We show that power counting alone gives a definite prediction for the observable that optimally separates the background-rich from the signal-rich regions of phase space.  Power counting can also be used to understand effects of phase space cuts and the effect of contamination from pile-up, which we discuss.  As these arguments rely only on the parametric scaling of QCD, the predictions from power counting must be reproduced by any Monte Carlo, which we verify using \pythia{8} and \herwigpp.  We also use the example of quark versus gluon discrimination to demonstrate the limits of the power counting technique.

}

\begin{document} 
\maketitle

\section{Introduction}
\label{sec:intro}

Over the past several years there has been an explosion in the number of jet observables and techniques developed for discrimination and grooming \cite{Abdesselam:2010pt,Altheimer:2012mn,Altheimer:2013yza}.  Several of these are used by the ATLAS and CMS experiments, and their performance has been validated directly on data \cite{CMS:2011xsa,Miller:2011qg,ATLAS-CONF-2012-066,Chatrchyan:2012mec,ATLAS:2012jla,Aad:2012meb,ATLAS:2012kla,ATLAS:2012am,Aad:2013gja,Aad:2013fba,TheATLAScollaboration:2013tia,TheATLAScollaboration:2013sia,TheATLAScollaboration:2013ria,TheATLAScollaboration:2013pia,CMS:2013uea,CMS:2013kfa,CMS:2013wea,CMS-PAS-JME-10-013,CMS-PAS-QCD-10-041,Aad:2014gea,LOCH:2014lla,CMS:2014fya} and employed in new physics searches in highly boosted regimes \cite{CMS:2011bqa,Fleischmann:2013woa,Pilot:2013bla,TheATLAScollaboration:2013qia,Chatrchyan:2012ku,Chatrchyan:2012sn,CMS:2013cda,CMS:2014afa,CMS:2014aka}.  Analyses using jets will become increasingly important at the higher energies and luminosities of Run 2 of the LHC.

While the proliferation of jet observables is exciting for the field, the vast majority of proposed observables and procedures have been analyzed exclusively in Monte Carlo simulation.  Monte Carlos are vital for making predictions at the LHC, but should not be a substitute for an analytical understanding, where possible.  Because Monte Carlos rely on tuning the description of non-perturbative physics to data, this can obscure what the robust perturbative QCD predictions are and hide direct insight into the dependence of the distributions on the parameters of the observable.  This is especially confusing when different Monte Carlo programs produce different results.

Perturbative predictions of distributions have traditionally been constrained to only the simplest observables, such as the jet mass \cite{Catani:1991bd,Chien:2010kc,Chien:2012ur,Dasgupta:2012hg,Jouttenus:2013hs}, but to high accuracy. Such high-order calculations are important for reducing the systematic theoretical uncertainties. More recently, resummation has been applied to some simple jet substructure variables \cite{Feige:2012vc,Dasgupta:2013ihk,Dasgupta:2013via,Larkoski:2014pca}, and an understanding of some of the subtleties of resummation for ratio observables, as often used in jet substructure, has been developed \cite{Larkoski:2013paa,Larkoski:2014tva,Larkoski:2014bia}. Even the simplest calculations have suggested new, improved techniques, like the modified Mass Drop Tagger \cite{Dasgupta:2013ihk,Dasgupta:2013via}, or uncovered unexpected structures in perturbative QCD, like Sudakov Safety \cite{Larkoski:2013paa,Larkoski:2014wba,Larkoski:2014bia}.  For more complex observables, however, an analytic calculation may be essentially impossible, and we must rely on Monte Carlo simulations. Because of the wide variety of jet observables, some of which can be calculated analytically and some that cannot, it is necessary to find an organizing principle that can be used to identify the robust predictions of QCD, without requiring a complete calculation to a given perturbative accuracy.

In this paper we show how power counting methods can be used to design and understand the behavior of jet substructure variables. With minimal computational effort, power counting accurately captures the parametric predictions of perturbative QCD. The dynamics of a QCD jet are dominated by soft and collinear emissions and so by identifying the parametric scaling of soft and collinear contributions to a jet observable, we are able to make concrete and justified statements about the performance of jet substructure variables.  Formal parametric scaling, or power counting, is widely used in the formalism of soft-collinear effective theory (SCET) \cite{Bauer:2000ew,Bauer:2000yr,Bauer:2001ct,Bauer:2001yt}, an effective field theory of QCD in the soft and collinear limits. However, in this paper, we will not rely on any results from SCET so as to make the discussion widely accessible.  Similar techniques were employed in \Ref{Walsh:2011fz}, but with the goal of determining which jet observables are calculable.

As a concrete application of the soft and collinear power counting method, we will focus on observables formed from the generalized $n$-point energy correlation functions $\ecf{n}{\beta}$ \cite{Larkoski:2013eya}, relevant for discriminating massive QCD jets from boosted, heavy objects.  Measuring multiple energy correlation functions on a jet defines a multi-dimensional phase space populated by signal and background jets.  By appropriately power counting the dominant regions of phase space, we are able to identify the signal- and background-rich regions and determine powerful observables for discrimination.  In addition, from power counting arguments alone, we are able to predict the effect of pile-up contamination on the different regions of phase space.  We apply power counting to the following:
\begin{itemize}

\item {\bf Boosted Z Bosons vs.~QCD} The two- and three-point energy correlation functions, $\ecf{2}{\beta}$ and $\ecf{3}{\beta}$, have been shown to be among the most powerful observables for identifying the hadronic decays of boosted $Z$ bosons \cite{Larkoski:2013eya}.  We  discuss the phase space defined by $\ecf{2}{\beta}$ and $\ecf{3}{\beta}$, and determine which regions are populated by signal and background jets. Using this understanding of the phase space, we propose a powerful discriminating variable to identify boosted two prong jets, given by
\begin{equation}
\Dobs{2}{\beta}= \frac{\ecf{3}{\beta}}{(\ecf{2}{\beta})^3}\,.
\end{equation}
This should be contrasted with the variable $\Cobs{2}{\beta}=\ecf{3}{\beta}/(\ecf{2}{\beta})^2$ originally proposed in \Ref{Larkoski:2013eya}. We also show that power counting can be used to understand the impact of pile-up radiation on the different regions of phase space, and in turn to understand the susceptibility of signal and background distributions to pile-up.

\item {\bf Quarks vs.~Gluons} Quark versus gluon jet discrimination is somewhat of a non-example for the application of power counting because there is nothing parametrically distinct between quark and gluon jets.  However, this will illustrate why quark versus gluon discrimination is such a hard problem, and why different Monte Carlos can have wildly different predictions \cite{Larkoski:2014pca}.

\end{itemize}

The outline of this paper is as follows.  In \Sec{sec:modes} we will precisely define what we mean by ``collinear'' and ``soft'' modes of QCD and introduce the observables used throughout this paper.  While we will mostly focus on the energy correlation functions, we will also discuss the $N$-subjettiness observables \cite{Thaler:2010tr,Thaler:2011gf} as a point of reference.  In \Sec{sec:boostz}, we apply power counting to the study of $Z$ versus QCD discrimination using the two- and three-point energy correlation functions $\ecf{2}{\beta}$ and $\ecf{3}{\beta}$.  We argue that the single most powerful observable for discrimination is $\ecf{3}{\beta}/(\ecf{2}{\beta})^3$. Power counting is used to understand how the addition of pile-up radiation effects the distributions of this variable, and show that they are more robust to pile-up than for previously proposed variables formed from the energy correlation functions.\footnote{The CMS study of \Ref{CMS:2013uea} found that the observable $C_2^{(\beta)}\equiv\ecf{3}{\beta}/(\ecf{2}{\beta})^2$ suggested in \Ref{Larkoski:2013eya} for boosted $Z$ identification is very sensitive to pile-up contamination.} We verify that these predictions are borne out in Monte Carlo.  In \Sec{sec:qvg}, we attempt to apply power counting to quark versus gluon jet discrimination.  Na\"ively, this should be the simplest case, however, power counting arguments are not applicable because all qualities of quarks and gluons only differ by order-1 numbers.  Finally, we conclude in \Sec{sec:conc} by re-emphasizing that power counting is a useful predictive tool for jet observables that are too complicated for direct analytic calculations, and suggest some problems to which it may prove fruitful.

\section{Observable Basis and Dominant Physics of QCD}
\label{sec:modes}

\subsection{Observables}

Throughout this paper, our analyses will be focused around the (normalized) $n$-point energy correlation functions $\ecf{n}{\beta}. $\footnote{The notation $\ecf{n}{\beta}$ differs from the original notation $\text{ECF}(n,\beta)$ presented in \Ref{Larkoski:2013eya} where the energy correlation functions were defined, but we hope that this notation used here is more compact.  Specifically, the relationship is
\begin{equation}
\ecf{n}{\beta} = \frac{\text{ECF}(n,\beta)}{\left(
\text{ECF}(1,\beta)
\right)^n} \ .
\end{equation}
} The two-, and three-point energy correlation functions are defined as
\begin{align}
\ecf{2}{\beta} &= \frac{1}{p_{TJ}^2}\sum_{1\leq i<j\leq n_J} p_{Ti}p_{Tj} R_{ij}^\beta \ ,\nonumber \\
\ecf{3}{\beta} &= \frac{1}{p_{TJ}^3}\sum_{1\leq i<j<k\leq n_J} p_{Ti}p_{Tj}p_{Tk} R_{ij}^\beta R_{ik}^\beta R_{jk}^\beta \ , \nonumber \\
\end{align}
where $p_{TJ}$ is the transverse momentum of the jet with respect to the beam, $p_{Ti}$ is the transverse momentum of particle $i$, and $n_J$ is the number of particles in the jet.  The boost-invariant angle $R_{ij}^2 = (\phi_i-\phi_j)^2+(y_i-y_j)^2$ is the Euclidean distance in the azimuth-rapidity plane and for infrared and collinear (IRC) safety, the angular exponent $\beta>0$.  In this paper we will only study up through $\ecf{3}{\beta}$, but higher-point energy correlation functions are defined as the natural generalization. We will often omit the explicit dependence on $\beta$, denoting the $n$-point energy correlation function simply as $\ecfnobeta{n}$.

The energy correlation functions have many nice properties that make them ideal candidates for defining a basis of jet observables.  First, the energy correlation functions are defined such that $\ecf{n}{\beta}\to0$ in any of the soft or collinear limits of a configuration of $n$ particles.  Second, because all angles in the energy correlation functions are measured between pairs of particles, $\ecf{n}{\beta}$ is insensitive to recoil or referred to as  ``recoil-free'' \cite{Catani:1992jc,Dokshitzer:1998kz,Banfi:2004yd,Larkoski:2013eya,Larkoski:2014uqa}. This means that it is not sensitive to the angular displacement of the hardest particle (or jet core) from the jet momentum axis due to soft, wide angle radiation in the jet.  The effects of recoil decrease the sensitivity of an observable to the structure of radiation about the hard core of the jet, making it less efficient for discrimination purposes.

Depending on the application, different energy correlation functions are useful as discriminating observables.  As discussed in \Ref{Larkoski:2013eya}, the two-point energy correlation function is sensitive to radiation about a single hard core, and so is useful for quark versus gluon discrimination. Similarly, the three- and four-point energy correlation functions are useful for 2- or 3-prong jet identification, respectively, corresponding to boosted electroweak bosons ($W/Z/H$) or hadronically decaying top quarks.  By measuring appropriate energy correlation functions we define a phase space, populated by signal and background jets.

As a point of reference, we will also study the $N$-subjettiness observables and compare the structure of their phase space with that of the energy correlation functions.  The (normalized) $N$-subjettiness observable $\Nsub{N}{\beta}$ is defined as
\begin{equation}\label{eq:nsubdef}
\Nsub{N}{\beta} = \frac{1}{p_{TJ}}\sum_{1\leq i \leq n_J} p_{Ti}\min\left\{
R_{i1}^\beta,\dotsc,R_{iN}^\beta
\right\} \ .
\end{equation}
The angle $R_{iK}$ is measured between particle $i$ and subjet axis $K$ in the jet.  Thus, $N$-subjettiness partitions a jet into $N$ subjet regions and measures the $p_T$-weighted angular distribution with respect to the subjet axis of each particle.  There are several different choices for how to define the subjet axes; here, we will define the subjet axes by the exclusive $k_T$ jet algorithm \cite{Cacciari:2008gp} with the winner-take-all (WTA) recombination scheme \cite{Bertolini:2013iqa,Larkoski:2014uqa,Salambroadening}.  In contrast to the traditional $E$-scheme recombination \cite{Blazey:2000qt}, which defines the (sub)jet axis to coincide with the net momentum direction, the WTA recombination scheme produces (sub)jet axes that are recoil-free and nearly identical to the $\beta = 1$ minimized axes.\footnote{The $\beta = 1$ minimized axes are also referred to as ``broadening axes'' \cite{Thaler:2011gf,Larkoski:2014uqa} as they correspond to axes that minimize the value of broadening \cite{Rakow:1981qn,Ellis:1986ig,Catani:1992jc}.}  With this definition, the observables $\ecf{2}{\beta}$ and $\Nsub{1}{\beta}$ are identical through NLL accuracy for all $\beta > 0$ \cite{Larkoski:2014uqa}.

Since $N$-subjettiness directly identifies $N$ subjet directions in a jet, it is a powerful variable for $N$-prong jet discrimination.  In particular, the $N$-subjettiness ratios
$$
\Nsub{2,1}{\beta}\equiv \frac{\Nsub{2}{\beta}}{\Nsub{1}{\beta}} \quad\text{and}\quad \Nsub{3,2}{\beta}\equiv \frac{\Nsub{3}{\beta}}{\Nsub{2}{\beta}} \ ,
$$
relevant for boosted $W/Z/H$ and top quark identification, respectively, are widely-used in jet studies at the ATLAS and CMS experiments.  Numerical implementations of the energy correlation functions and $N$-subjettiness are available in the \texttt{EnergyCorrelator} and \texttt{Nsubjettiness} \fastjet{contrib}s \cite{Cacciari:2011ma,fjcontrib}.

\subsection{Soft and Collinear Modes of QCD}
\label{sec:softcollqcd}

At high energies, QCD is approximately a weakly-coupled conformal gauge theory and so jets are dominated by soft and collinear radiation.  Because it is approximately conformal, there is no intrinsic energy or angular scale associated with this radiation.  To introduce a scale, and so to determine the dominant soft and collinear emissions, we must break the conformal invariance by making a measurement on the jet.  The scale of the soft and collinear emissions is set by the measured value of the observable.\footnote{It is important to note that since QCD is not a conformal field theory, we can only use the power counting presented here to study the phase space defined by a set of IRC safe observables.  If we considered IRC unsafe observables, then generically, we would need to power count contributions from non-perturbative physics such as hadronization.
}

This observation can be exploited to make precise statements about the energy and angular structure of a jet, depending on the value of observables measured on that jet. This reasoning is often implicitly understood in the jet community and literature, and is formalized in SCET.  Nevertheless, these precise power-counting arguments are not widely used outside of SCET, and so we hope that the applications in this paper illustrate their effectiveness and relative simplicity.

We begin by defining a soft emission, $s$, as one for which
\begin{equation}
z_s \equiv \frac{p_{Ts}}{p_{TJ}}\ll1 \ , \qquad R_{sj}\sim 1 \ ,
\end{equation}
where $j$ is any other particle in the jet and $R_{sj}\sim 1$ means that $R_{sj}$ is not associated with any parametric scaling.  Similarly, a collinear emission, $c$, is defined as having a $p_T$ fraction
\begin{equation}
\frac{p_{Tc}}{p_{TJ}}\sim1 \ ,
\end{equation}
but with an angle to other particles which depends on whether they are also collinear or soft:
\begin{equation}
R_{cc}\ll 1\ , \qquad R_{cs}\sim 1 \ .
\end{equation}
Here, $R_{cc}$ is the angle between two collinear particles, while $R_{cs}$  is the angle between a soft particle and a collinear particle.  The precise scalings of $R_{cc}$ and $z_s$ will depend on the observable in question, as will be explained shortly.  Soft emissions also implicitly include radiation that is simultaneously both soft and collinear.

To introduce these ideas concretely, we use the example of the two-point energy correlation function:
\begin{equation}\label{eq:2pt_ex}
\ecf{2}{\beta} =  \frac{1}{p_{TJ}^2}\sum_{1\leq i<j\leq n_J} p_{Ti}p_{Tj} R_{ij}^\beta \ .
\end{equation}
Consider performing a measurement of $\ecf{2}{\beta}$ on a jet and further requiring $\ecf{2}{\beta}\ll1$. Because the energy flow in jets is in general collimated, this defines a non-trivial region of phase space, with a large fraction of jets satisfying this requirement.  A large value of $\ecf{2}{\beta}$ would mean that there is a hard, perturbative splitting in the jet which is suppressed by the small value of $\alpha_s$. From the definition of $\ecf{2}{\beta}$ in \Eq{eq:2pt_ex}, we see that a measurement of $\ecf{2}{\beta}\ll1$ forces all particles in the jet to either have small $p_{Ti}$ or small $R_{ij}$. In other words, the observable is dominated by soft and collinear emissions. The precise scaling of $p_{Ti}$ and $R_{ij}$ is then determined  by the measured value of $\ecf{2}{\beta}$.\footnote{In SCET, $R_{cc}$ and $z_s$ are often immediately assigned a related scaling. While this is true for this example, it is not in general true in the case of multiple measurements, and we wish to emphasize in this section how the measurement sets both scalings.}

There are three possible configurations that contribute to $\ecf{2}{\beta}$: soft-soft correlations, soft-collinear correlations, and collinear-collinear correlations.  Therefore, $\ecf{2}{\beta}$ can be expressed as
\begin{equation}
\ecf{2}{\beta} \sim \frac{1}{p_{TJ}^2}\sum_{s} p_{Ts}p_{Ts} R_{ss}^\beta + \frac{1}{p_{TJ}^2}\sum_{s,c} p_{Ts}p_{Tc} R_{cs}^\beta + \frac{1}{p_{TJ}^2}\sum_{c} p_{Tc}p_{Tc} R_{cc}^\beta  \ ,
\end{equation}
where we have separated the contributions to $\ecf{2}{\beta}$ into the three different correlations.  To determine the dominant contributions to $\ecf{2}{\beta}$, we will throw away those contributions that are parametrically smaller, according to our definitions of soft and collinear above.  First, $p_{Ts} \ll p_{Tc}$, and so we can ignore the first term to leading power.  Because $R_{cs}\sim 1$, we set $R_{cs}= 1$ in the second term.  Also, note that $p_{Tc}\sim p_{TJ}$ and so we can replace the instances of $p_{Tc}$ with $p_{TJ}$ in the second and third terms.  Making these replacements, we find
\begin{equation}\label{eq:fact_e2}
\ecf{2}{\beta} \sim \sum_{s}  z_s +\sum_{c}  R_{cc}^\beta  \ ,
\end{equation}
where we have ignored any corrections arising at higher power in the soft and collinear emissions' energies and angles. We wish to emphasize with the explicit summation symbols that we have not restricted to a single soft or collinear emission, but consider an arbitrary number of emissions. Furthermore, we do not assume a strongly ordered limit, but instead explore the complete phase space arising from soft and collinear emissions, including regions where such ordering is explicitly broken.

\Eq{eq:fact_e2} demonstrates the dominant structure of a jet on which we have measured $\ecf{2}{\beta}\ll 1$.  The contribution to $\ecf{2}{\beta}$ from soft and collinear emissions do not mix to this accuracy; that is, they factorize from one another.  Also, because there is no measurement to distinguish the soft and collinear contributions to $\ecf{2}{\beta}$, we then have that 
\begin{equation}
\ecf{2}{\beta} \sim  z_s \sim R_{cc}^\beta  \,.
\end{equation}
That is, the measured value of $\ecf{2}{\beta}$ sets the $p_T$ of the soft particles and the splitting angle of the collinear particles, and therefore defines the structure of the jet. 

In \Eq{eq:fact_e2}, we have explicitly written a summation over the particles with soft and collinear scalings. To determine the scalings of the different contributions, it is clearly sufficient to consider the scaling of an individual term in each sum. In the remainder of this paper we will drop the explicit summation for notational simplicity.

Scaling arguments similar to the power counting approach discussed here are often used in other approaches to QCD resummation to identify the relevant soft and collinear scales, and could also be used to analyze observables. For example, in the method of regions \cite{Beneke:1997zp,Smirnov:2001in}, the regions of integration over QCD matrix elements which contribute dominantly to a given observable are determined, and an expansion about each of these regions is performed. These regions of integration, and the scaling of the momenta in these regions, correspond to the modes of the effective theory determined through the power counting approach.\footnote{More precisely, only on-shell modes appear as degrees of freedom in the effective theory.}

Similarly, in the CAESAR approach to resummation \cite{Banfi:2004yd}, implemented in an automated computer program, the first step of the program is the identification of the relevant soft and collinear scales. This is performed by expanding a given observable in the soft and collinear limits, and considering the region of integration for a single emission. This procedure is similar to that used in the case of $\ecf{2}{\beta}$ just discussed, and would identify the same dominant contributions and scalings. Using the knowledge of the behavior of the QCD splitting functions, CAESAR then performs a resummed calculation of the observable. However, the CAESAR computer program is currently restricted to observables for which the relevant scales, and hence the logarithmic structure, is determined by a single emission, and further, to single differential distributions. 

When considering observables relevant for jet substructure, one is interested in variables such as $\ecf{n}{\beta}$, $n>2$, whose behavior is not determined by the single emission phase space. For such observables, the 
 single-emission analysis is not sufficient
 and the explicit analysis of QCD matrix elements to determine the dominant regions of integration which contribute becomes quite complicated. For these cases, we find the power counting approach of the effective field theory paradigm to be a particularly convenient organizing principle. Using the knowledge that on-shell soft and collinear modes dominate, a consistent power counting can be used to determine the relevant scalings of these modes in terms of the measured observables, which is reduced to a simple algebraic exercise. Although the evaluation of QCD matrix elements in these scaling limits is of course required for a complete calculation, it is not required to determine the power counting, and we will see that power counting alone will often be sufficient for constructing discriminating observables for jet substructure studies.

Throughout the rest of this paper, we will employ these power-counting arguments to determine the dominant structure of jets on which multiple measurements have been made, for example $\ecf{2}{\beta}$ and $\ecf{3}{\beta}$.  In this case, the phase space that results is much more complicated than the example of $\ecf{2}{\beta}$ discussed above, but importantly, appropriate power counting of the contributions from soft and collinear emissions will organize the phase space into well-defined regions automatically.

\section{Power Counting Boosted Z Boson vs. QCD Discrimination}
\label{sec:boostz}

As a detailed example of the usefulness of power counting, we consider the problem of discriminating hadronically-decaying, boosted $Z$ bosons from massive QCD jets.  Because $Z$ boson decays have a 2-prong structure, we will measure the two- and three-point energy correlation functions, $\ecfnobeta{2}$ and $\ecfnobeta{3}$, on the jets, defining a two-dimensional phase space.  We will find that there are two distinct regions of this phase space corresponding to jets with one or two hard prongs.  QCD jets exist dominantly in the former region while boosted $Z$ bosons exist dominantly in the latter.  Power counting these phase space regions will allow us to determine the boundaries of the regions and to define observables that separate the signal and background regions most efficiently.

Both because it is a non-trivial application, as well as still being tractable, we will present a detailed analysis of the phase space regions for boosted $Z$ identification.  This will require several pieces.  First, we will study the full phase space of perturbative jets defined by $\ecfnobeta{2}$ and $\ecfnobeta{3}$ and identify signal and background regions via power counting. This will lead us to define a discriminating variable, $\Dobs{2}{\beta}$. Second, any realistic application of a boosted $Z$ tagger includes a cut on the jet mass in the window around $m_Z$, and the effect of the mass cut on the discrimination power can also be understood by a power counting analysis of the phase space.  Third, at the high luminosities of the LHC, contamination from pile-up is important and can substantially modify distributions for jet substructure variables.  By appropriate power counting of the pile-up radiation, we can understand the effect of pile-up on the perturbative phase space and determine how susceptible the distributions of different discrimination variables are to pile-up contamination.  As a reference, throughout this section we will contrast the energy correlation functions to the $N$-subjettiness observables $\Nsub{1}{\beta}$ and ${\Nsub{2}{\beta}}$ \cite{Thaler:2010tr,Thaler:2011gf}.  A full effective theory analysis and analytic calculation of $\Dobs{2}{\beta}$ will be presented in  \Ref{usD2}.

\subsection{Perturbative Radiation Phase Space}\label{sec:pert}

\begin{figure}
\begin{center}
\subfloat[]{\label{fig:unresolved}
\includegraphics[width=6cm]{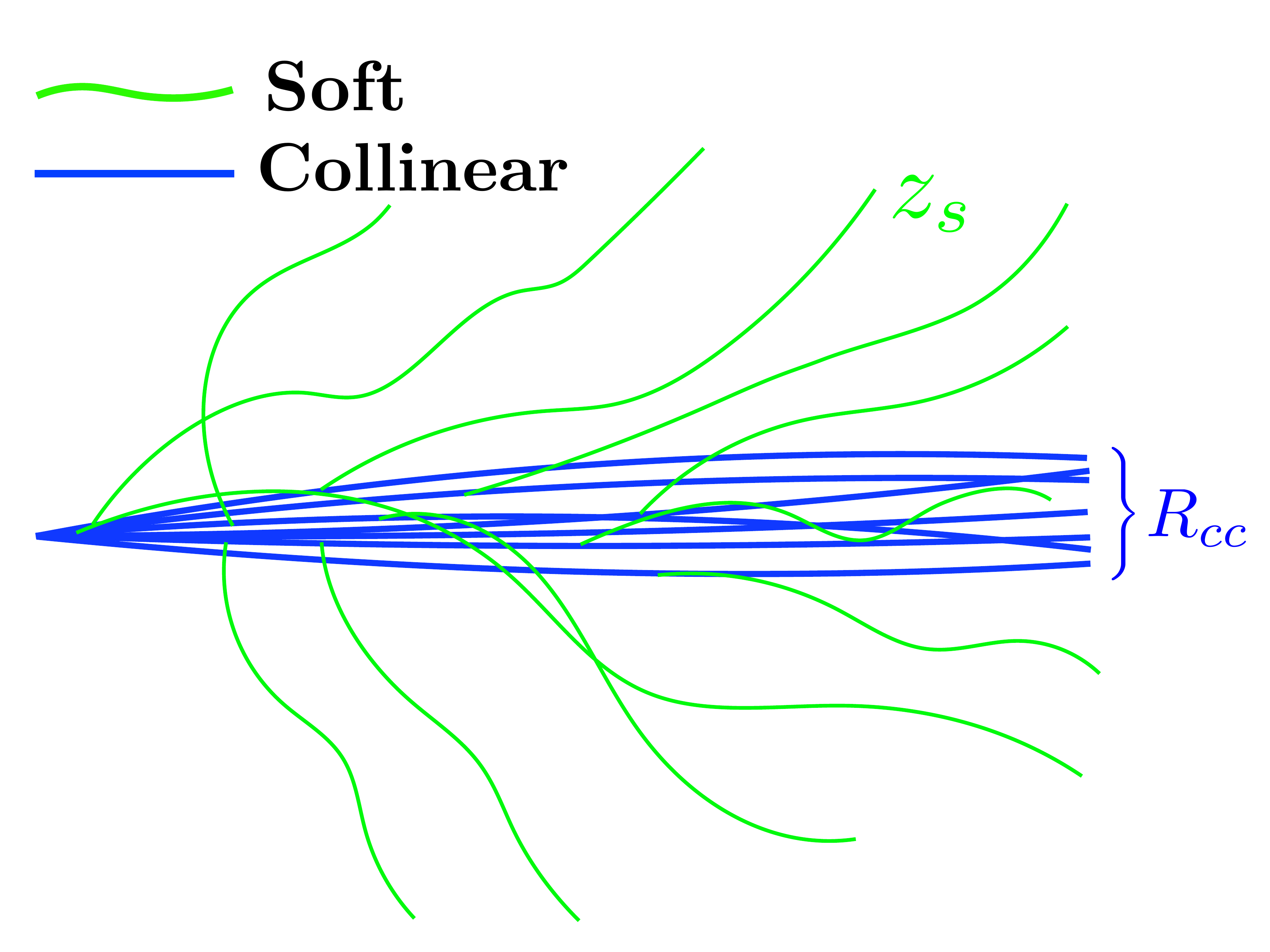}    
}\qquad
\subfloat[]{\label{fig:NINJA}
\includegraphics[width=6cm]{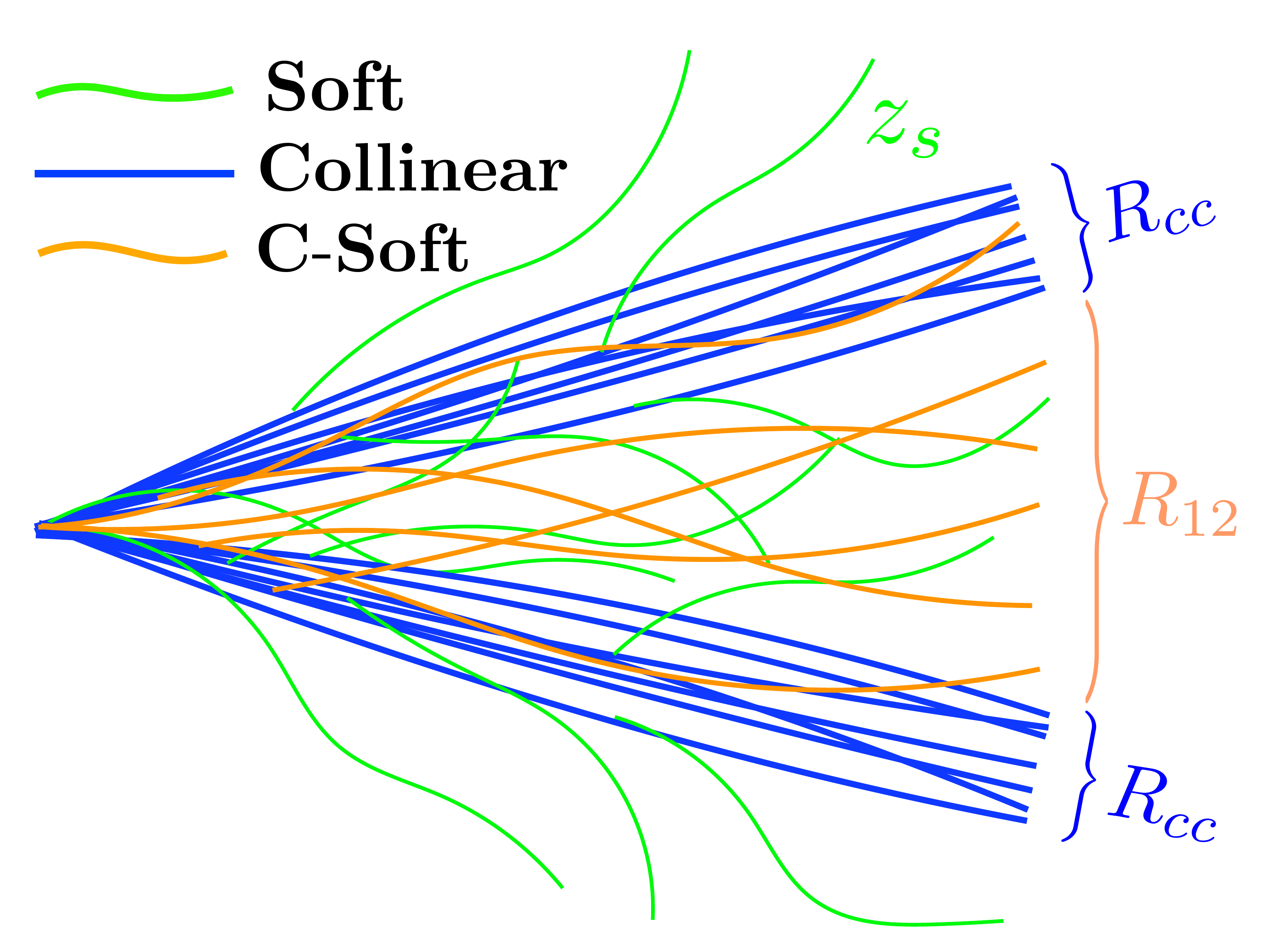}
}
\end{center}
\caption{a) 1-prong jet, dominated by collinear (blue) and soft (green) radiation. The angular size of the collinear radiation is $R_{cc}$ and the $p_T$ fraction of the soft radiation is $z_s$.  b) 2-prong jet resolved into two subjets, dominated by collinear (blue), soft (green), and collinear-soft (orange) radiation emitted from the dipole formed by the two subjets.  The subjets are separated by an angle $R_{12}$  and the $p_T$ fraction of the collinear-soft radiation is $z_{cs}$.
}
\label{fig:pics_jets}
\end{figure}

We begin by studying the $(\ecfnobeta{2}, \ecfnobeta{3})$ phase space arising from perturbative radiation from the jet. The measurement of $\ecfnobeta{2}$ and $\ecfnobeta{3}$ on a jet can resolve at most two hard subjets. The phase space for the variables $\ecfnobeta{2}$ and $\ecfnobeta{3}$ is therefore composed of jets which are unresolved by the measurement, dominantly from the QCD background, and shown schematically in \Fig{fig:unresolved}, and jets with a resolved 2-prong structure, as from boosted $Z$ decays, shown schematically in \Fig{fig:NINJA}. We will find that the resolved and unresolved jets live in parametrically different regions of the phase space, and the boundary between the two regions can be understood from a power counting analysis.

First, consider the case of the measurement of $\ecf{2}{\beta}$ and $\ecf{3}{\beta}$ on a jet with a single hard core of radiation, as in \Fig{fig:unresolved}, which is dominated by soft radiation with characteristic $p_T$ fraction $z_s \ll 1$, and collinear radiation with a characteristic angular size $R_{cc}\ll1$.  All other scales are order-1 numbers that we will assume are equal to 1 without further discussion. With these assumptions, we are able to determine the scaling of the contributions to $\ecf{2}{\beta}$ and $\ecf{3}{\beta}$ from collections of soft and collinear particles.  The scalings are given in \Tab{tab:pc} for contributions from three collinear particles ($CCC$), two collinear and one soft particle ($CCS$), one collinear and two soft particles ($CSS$), and three soft particles ($SSS$).\footnote{The contributions in \Tab{tab:pc} are from \emph{any} subset of three particles in the jet. We do not single out an initial parton from which the others arise  as in a showering picture. }

\begin{table}[t]
\begin{center}
\begin{tabular}{c|c|c}
modes&$\ecf{2}{\beta}$ &$\ecf{3}{\beta}$ \\ 
\hline
$CCC$ & $R_{cc}^{\beta}$ & $R_{cc}^{3\beta}$ \\
$CCS$ & $R_{cc}^{\beta}+z_s$ &$z_s R_{cc}^{\beta}$ \\
$CSS$ & $z_s+z_s^2$ & $z_s^2$ \\
$SSS$ & $z_s^2$  & $z_s^3$
\end{tabular}
\end{center}
\caption{Scaling of the contributions of 1-prong jets to $\ecf{2}{\beta}$ and $\ecf{3}{\beta}$ from the different possible configurations of soft ($S$) and collinear ($C$) radiation.
}
\label{tab:pc}
\end{table}

Dropping those contributions that are manifestly power-suppressed, the two- and three-point energy correlation functions measured on 1-prong jets therefore scale like
\begin{align}
\ecf{2}{\beta} & \sim  R_{cc}^\beta + z_s \,, \\
\ecf{3}{\beta} & \sim R_{cc}^{3\beta}+z_s^2 + R_{cc}^\beta z_s\,.
\end{align}
To go further, we must determine the relative size of $z_s$ and $R_{cc}^\beta$.  There are two possibilities, depending on the region of phase space identified by the measurement: either $z_s$ makes a dominant contribution to $\ecfnobeta{2}$, or its contribution is power suppressed with respect to $R_{cc}^\beta$.  In the case that $z_s$ contributes to $\ecfnobeta{2}$, this immediately implies that $\ecfnobeta{3}\sim (\ecfnobeta{2})^2$, regardless of the precise scaling of $R_{cc}^\beta$.\footnote{Note that on the true upper boundary of the phase space, the assumption of strong ordering of emissions is broken.}  If instead $z_s$ gives a  subleading contribution compared to $R_{cc}^\beta$ in $\ecfnobeta{2}$, then $\ecfnobeta{3}\sim (\ecfnobeta{2})^3$.\footnote{The existence of a consistent power counting does not guarantee a factorization theorem. Indeed, while a factorization theorem exists on the quadratic boundary, it does not exist at leading power on the cubic boundary.} Therefore, from this simple analysis, we have shown that 1-prong jets populate the region of phase space defined by $(\ecfnobeta{2})^3 \lesssim \ecfnobeta{3} \lesssim (\ecfnobeta{2})^2$. Fascinatingly, this also implies that the relative values of $\ecfnobeta{2}$ and $\ecfnobeta{3}$ provide a direct probe of the ordering of emissions inside the jet, so that assumptions about the measured values of $\ecfnobeta{2}$ and $\ecfnobeta{3}$ are observable proxies for the ordering of emissions. The scaling of $R_{cc}$ and $z_s$ on each boundary of the phase space can then easily be determined, but will not be important for our discussion.

\begin{figure}
\begin{center}
\includegraphics[width=6.5cm]{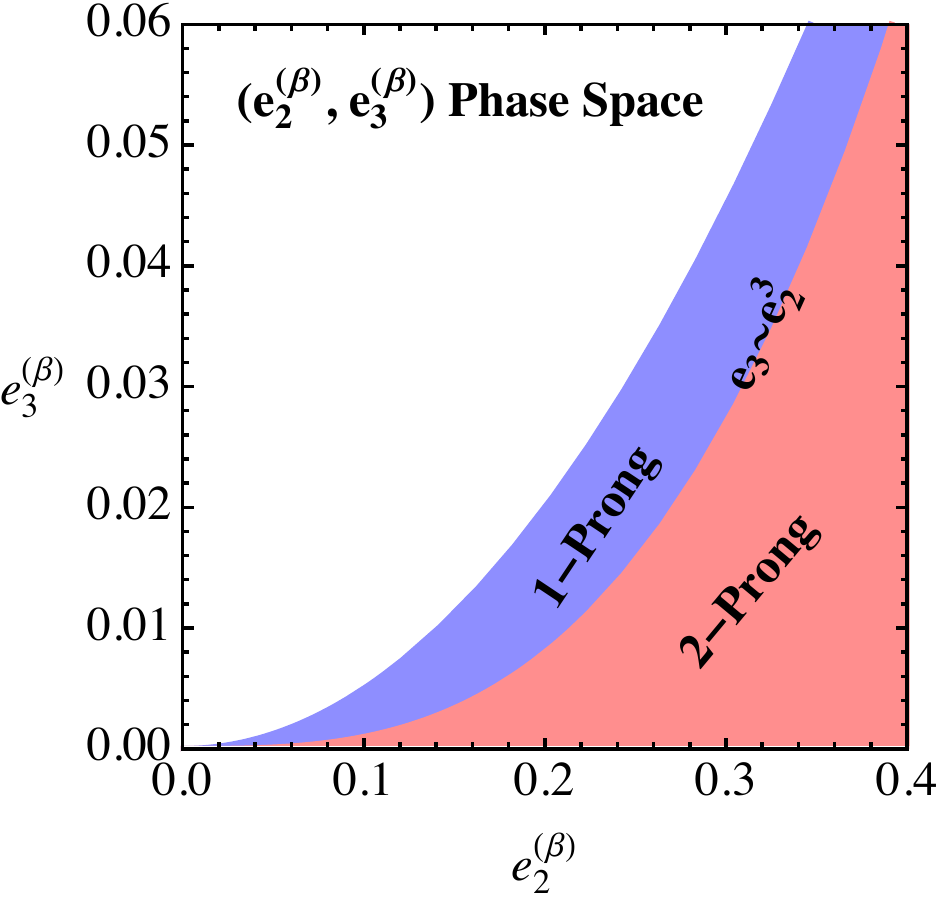}
\end{center}
\caption{ 
Phase space defined by the measurement of the energy correlation functions $ \ecfnobeta{2}$ and $\ecfnobeta{3}$. The phase space is divided into 1- and 2-prong regions with a boundary corresponding to the curve $ \ecfnobeta{3}\sim (\ecfnobeta{2})^3$.
}
\label{fig:ps_nolines}
\end{figure}

This analysis shows that 1-prong jets fill out a non-trivial region in the  $(\ecfnobeta{2}, \ecfnobeta{3})$ phase space, and of particular interest for the design of discriminating observables is the fact that this region of phase space has a lower boundary. This region is shown in blue in \Fig{fig:ps_nolines}. To understand the region of phase space for $\ecfnobeta{3} \ll  (\ecfnobeta{2}) ^3$ we must consider the case in which the measurement of $\ecfnobeta{2}$ and $\ecfnobeta{3}$ resolves two subjets within the jet.

The setup for the power counting of 2-prong jets is illustrated in \Fig{fig:NINJA}.  We consider a jet with two subjets, each of which carry ${\cal O}(1)$ of the jet $p_T$ and are separated by an angle $R_{12}\ll1$.  Each of the subjets has collinear emissions at a characteristic angle $R_{cc}\ll R_{12}$.  Because $R_{12}\ll 1$, there is in general global soft radiation at large angles with respect to the subjets with characteristic $p_T$ fraction $z_s\ll 1$.  For color-singlet jets, like boosted $Z$ bosons, this global soft radiation contribution comes purely from initial state radiation (ISR).\footnote{While this background is to a certain extent irreducible, given the important feature of ISR is its color-uncorrelated and soft nature, many of our observations about the effects of pile-up in \Sec{sec:pu} will be applicable to ISR.}  Finally, there is radiation from the dipole formed from the two subjets (called ``collinear-soft'' radiation), with characteristic  angle $R_{12}$ from the subjets, and with $p_T$ fraction $z_{cs}$.  The effective theory of this phase space region for the observable $N$-jettiness \cite{Stewart:2010tn} was studied in \Ref{Bauer:2011uc}.

We now consider the power counting of $\ecf{2}{\beta}$ and $\ecf{3}{\beta}$ for 2-prong jets. By the definition of this region of phase space, the hard splitting sets the value of $\ecf{2}{\beta}$. That is, we have $\ecf{2}{\beta}\sim R_{12}^\beta$, with all other contributions suppressed. For $\ecf{3}{\beta}$, it is clear that the leading contributions must arise from correlations between the two hard subjets with either the global soft, collinear or collinear-soft modes.  The scaling of these different contributions to $\ecf{3}{\beta}$ is given in \Tab{tab:pc_ninja}, from which we find that the scaling of the two- and three-point energy correlation functions for 2-pronged jets is
\begin{align}
\ecf{2}{\beta} & \sim  R_{12}^\beta \,, \\
\ecf{3}{\beta} & \sim R_{12}^{\beta}z_s+R_{12}^{2\beta} R_{cc}^{\beta}+R_{12}^{3\beta}z_{cs}\,.
\end{align}
There is no measurement performed to distinguish the three contributions to $\ecf{3}{\beta}$ and so we must assume that they all scale equally.

\begin{table}[t]
\begin{center}
\begin{tabular}{c|c}
modes &$\ecf{3}{\beta}$ \\ 
\hline
$C_1C_2\,S$ & $R_{12}^{\beta}z_s$  \\
$C_1C_2\, C$ & $R_{12}^{2\beta} R_{cc}^{\beta}$  \\
$C_1C_2\,C_s$ & $R_{12}^{3\beta}z_{cs}$  
\end{tabular}
\end{center}
\caption{Scaling of the contributions from global soft ($S$), collinear ($C$), and collinear-soft ($C_s$) radiation correlated with the two hard subjets (denoted by $C_1$ and $C_2$) in 2-prong jets to  $\ecf{3}{\beta}$ from the different possible configurations.
}
\label{tab:pc_ninja}
\end{table}

This result is sufficient to set the relative scaling of $\ecf{2}{\beta}$ and $\ecf{3}{\beta}$.  As we assume that the jet only has two hard subjets, we have that $z_{cs}\ll1$ and so
\begin{equation}
(\ecfnobeta{2})^3 \sim R_{12}^{3\beta} \gg R_{12}^{3\beta}z_{cs} \sim \ecfnobeta{3} \ ,
\end{equation}
which defines the 2-prong jet region of phase space as that for which $\ecfnobeta{3} \ll (\ecfnobeta{2})^3$.  With this identification, note the scaling of the various modes:
\begin{align}
R_{12}^\beta \sim \ecfnobeta{2}\, , \quad z_s \sim \frac{\ecfnobeta{3}}{\ecfnobeta{2}} \, , \quad R_{cc}^\beta \sim \frac{\ecfnobeta{3}}{(\ecfnobeta{2})^2} \, , \quad z_{cs} \sim \frac{\ecfnobeta{3}}{(\ecfnobeta{2})^3} \, .
\end{align}
While not important for our goals here, the fact that the energy correlation functions parametrically separate the scaling of the modes that contribute to the observables is vital for an effective theory analysis and calculability \cite{usD2}.  Note that because $\ecfnobeta{2}$ is first non-zero at a lower order in perturbation theory than $\ecfnobeta{3}$, $\ecfnobeta{3}$ can be zero while $\ecfnobeta{2}$ is non-zero.  Therefore, this 2-prong region of phase space extends down to the kinematic limit of $\ecfnobeta{3}  = 0$, as shown in red in \Fig{fig:ps_nolines}.

This power counting analysis, although very simple in nature, provides a powerful picture of the phase space defined by the measurement of $\ecfnobeta{2}$ and $\ecfnobeta{3}$, which is shown in \Fig{fig:ps_nolines}.  The 1- and 2-prong jets are defined to populate the phase space regions where
\begin{align*}
\text{1-prong jet: }& (\ecfnobeta{2})^3\lesssim \ecfnobeta{3} \lesssim (\ecfnobeta{2})^2\, ,\\
\text{2-prong jet: }& 0< \ecfnobeta{3} \ll (\ecfnobeta{2})^3\, .
\end{align*}
Background QCD jets dominantly populate the 1-prong region of phase space, while signal boosted $Z$ decays dominantly populate the 2-prong region of phase space.  This has important consequences for the optimal discrimination observable.  

An interesting observation about the boundary between 1- and 2-prong jets, defined by $ \ecfnobeta{3}\sim (\ecfnobeta{2})^3$, is that it is approximately invariant to boosts along the jet direction.  For a narrow jet, a boost along the jet direction by an amount $\gamma$ scales $p_T$s and angles as
\begin{equation}
p_T\to \gamma p_T \ ,\qquad R \to \gamma^{-1} R \ .
\end{equation}
Therefore, under a boost, $\ecf{2}{\beta}$ and $\ecf{3}{\beta}$ scale as
\begin{equation}
\ecf{2}{\beta} \to \gamma^{-\beta}\ecf{2}{\beta} \ , \qquad \ecf{3}{\beta} \to \gamma^{-3\beta} \ecf{3}{\beta} \ .
\end{equation}
Thus, the boundary between 1- and 2-prong jets, where $ \ecfnobeta{3}\sim (\ecfnobeta{2})^3$, is invariant to  boosts along the jet direction.  That is, under boosts, a jet will move along a contour of constant $ \ecfnobeta{3}/ (\ecfnobeta{2})^3$ in the ($\ecfnobeta{2},\ecfnobeta{3}$) plane.  

The analysis presented in this section is also the initial step in establishing rigorous factorization theorems in the different regions of phase space, allowing for analytic resummation of the double differential cross section of $\ecfnobeta{2}$ and $\ecfnobeta{3}$ \cite{Larkoski:2014tva,usD2}.

\subsubsection{Optimal Discrimination Observables}
\label{sec:opte2e3}

\begin{figure}
\begin{center}
\subfloat[]{\label{fig:e2e3_ps}
\includegraphics[width=6.5cm]{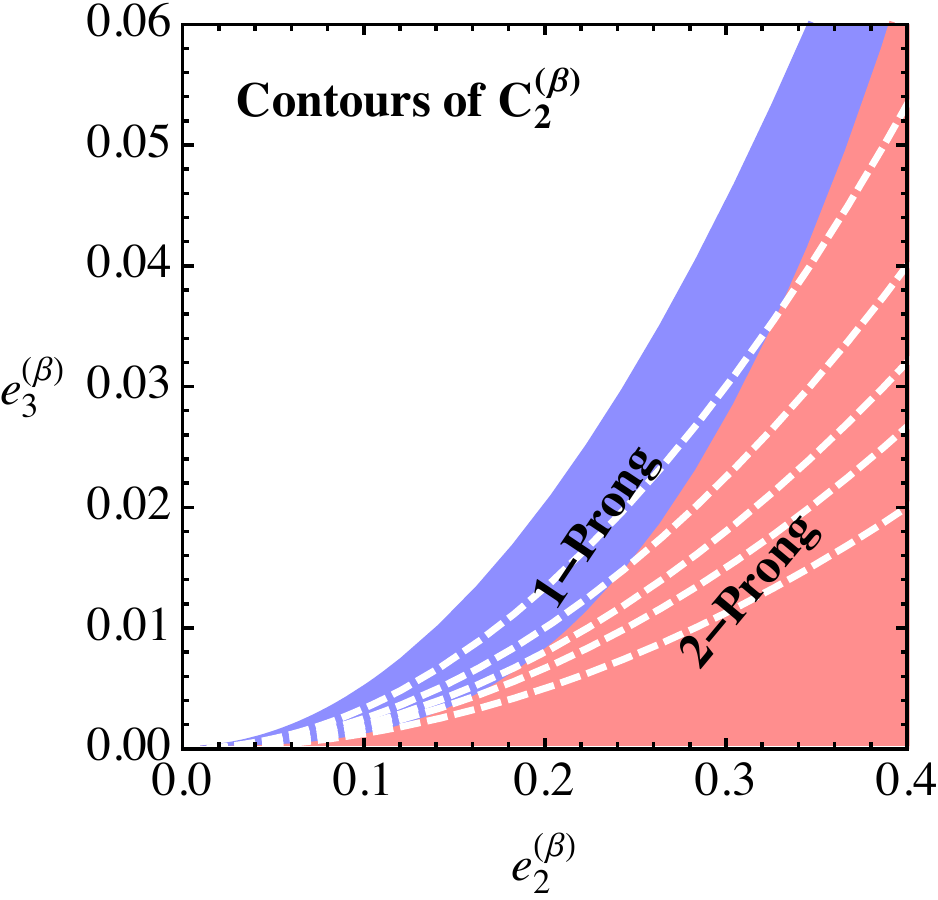}
}\qquad
\subfloat[]{\label{fig:contours_ps}
\includegraphics[width=6.5cm]{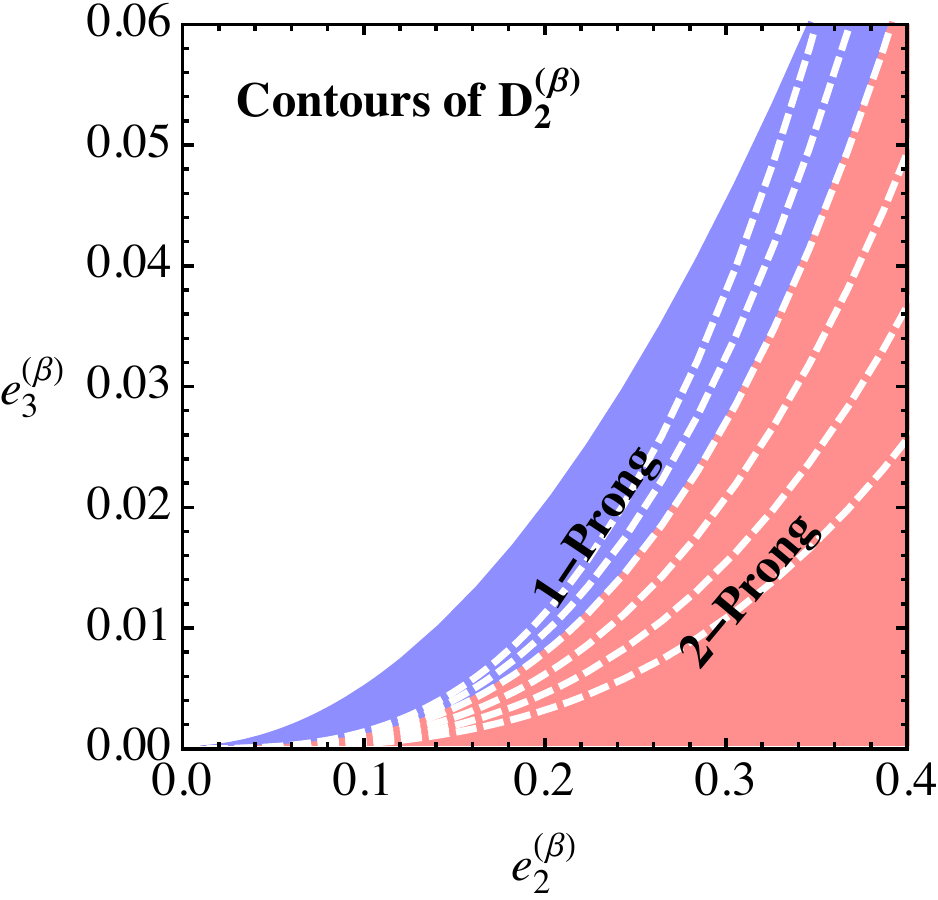}
}
\end{center}
\caption{ 
Contours of constant $\Cobs{2}{\beta}$ (left) and $\Dobs{2}{\beta}$ (right) in the phase space defined by $\ecf{2}{\beta},\ecf{3}{\beta}$.  The 1- and 2-prong regions of phase space are labeled, with their boundary corresponding to the curve $ \ecfnobeta{3}\sim (\ecfnobeta{2})^3$.
}
\label{fig:ps}
\end{figure}

The fact that the signal and background regions of phase space are parametrically separated implies that from power counting alone, we can determine the optimal observable for separating signal from background.  Because the boundary between the background-rich and signal-rich regions is $ \ecfnobeta{3}\sim (\ecfnobeta{2})^3$, this suggests that the optimal observable for discriminating boosted $Z$ bosons from QCD jets is\footnote{We thank Jesse Thaler for suggesting the notation ``$D$'' for these observables. Unlike $\Cobs{2}{\beta}$, whose name was motivated by its relation to the classic $e^+e^-$ event shape parameter C, $\Dobs{2}{\beta}$ is not related to the D parameter. }
\begin{equation}
\Dobs{2}{\beta} \equiv \frac{\ecf{3}{\beta}}{\left(
\ecf{2}{\beta}
\right)^3} \, .
\end{equation}
Signal jets will be characterized by a small value of $\Dobs{2}{\beta}$, while background jets will predominantly have large $\Dobs{2}{\beta}$.  With this observable, parametrically there is no mixing of the signal-rich and background-rich regions. Contours of constant $D_2^{(\beta)}$ lie entirely in the signal or background region, as is shown schematically in \Fig{fig:ps}.  Determining the precise discrimination power of $\Dobs{2}{\beta}$ requires an understanding of the ${\cal O}(1)$ details of the distributions of signal and background, beyond any purely power counting analysis.

The observation that the scaling relation $ \ecfnobeta{3}\sim (\ecfnobeta{2})^3$ is boost invariant provides further motivation for the variable $\Dobs{2}{\beta}$. Under boosts along the jet axis, jets can move along curves of constant $\Dobs{2}{\beta}$, but cannot cross the boundary between the 2-prong and 1-prong regions of phase space. This can be used to give a boost invariant definition of a 2-prong jet, as a jet with a small value of $\Dobs{2}{\beta}$, and a 1-prong jet, as a jet with large $\Dobs{2}{\beta}$.

\Ref{Larkoski:2013eya} used the two- and three-point energy correlation functions in the combination
\begin{equation}
C_2^{(\beta)} \equiv  \frac{\ecf{3}{\beta}}{\left({\ecf{2}{\beta}}\right)^2}
\end{equation}
for boosted $Z$ boson discrimination.  From the power counting analysis in this section, this variable is not a natural choice.  In particular, contours of constant $C_2^{(\beta)}$ pass through both the 1-prong and 2-prong regions of phase space, mixing the signal and background for any value of $C_2^{(\beta)}$, as shown in \Fig{fig:ps}. Therefore, from the power counting perspective, we would expect that $C_2^{(\beta)}$ is a poor boosted $Z$ boson discriminating observable.  Nevertheless, \Ref{Larkoski:2013eya} found that with a tight jet mass cut, and in the absence of pile-up, $C_2^{(\beta)}$ is a powerful boosted $Z$ discriminant.  A mass cut constrains the phase space significantly, which we will discuss in detail in \Sec{sec:masscute2e3}, allowing us to understand the result of \Ref{Larkoski:2013eya}. Pile-up will be addressed in \Sec{sec:pu}. 

It is important to recall that while $\ecf{2}{\beta}$ and $\ecf{3}{\beta}$ are IRC safe observables, so that their phase space can be analyzed with power counting techniques, ratios of IRC safe observables are not in general IRC safe \cite{Soyez:2012hv,Larkoski:2013paa,Larkoski:2014bia}. The observables $\Cobs{2}{\beta}$ and $\Dobs{2}{\beta}$ are however Sudakov safe \cite{Larkoski:2013paa,Larkoski:2014bia}, and therefore can be reliably studied with Monte Carlo simulation without applying any form of additional cut, such as a jet mass cut, on the phase space.

\subsubsection{Contrasting with $N$-subjettiness}
\label{sec:nsubcomp}

At this point, it is interesting to apply the power counting analysis to other observables for boosted $Z$ discrimination and see what conclusions can be made.  For concreteness, we will contrast the energy correlation functions with the $N$-subjettiness observables $\Nsub{1}{\beta}$ and $\Nsub{2}{\beta}$, defined as
\begin{equation}
\Nsub{N}{\beta} = \frac{1}{p_{TJ}}\sum_{1\leq i \leq n_J} p_{Ti}\min\left\{
R_{i1}^\beta,\dotsc,R_{iN}^\beta
\right\} \ .
\end{equation}
As with the energy correlation functions, we will consider $\Nsub{1}{\beta}$ and $\Nsub{2}{\beta}$ as measured on 1-prong and 2-prong jets and determine the regions of phase space where background and signal jets populate.  This can then be used to determine the optimal observable for boosted $Z$ discrimination from the $N$-subjettiness observables. We use the same notation for the scalings of the modes as in \Sec{sec:pert}.

\begin{table}[t]
\begin{center}
\begin{tabular}{c|c|c}
modes&$\Nsub{1}{\beta}$ &$\Nsub{2}{\beta}$ \\ 
\hline
$CCC$ & $R_{cc}^{\beta}$ & $R_{cc}^{\beta}$ \\
$CCS$ & $R_{cc}^{\beta}+z_s$ &$R_{cc}^\beta+z_s$ \\
$CSS$ & $z_s$ & $z_s$ \\
$SSS$ & $z_s$  & $z_s$
\end{tabular}
\end{center}
\caption{Scaling of the contributions of 1-prong jets to $\Nsub{1}{\beta}$ and $\Nsub{2}{\beta}$ from the different possible configurations of soft ($S$) and collinear ($C$) radiation.
}
\label{tab:pc_nsub}
\end{table}

Starting with 1-prong jets, and repeating the analysis of \Sec{sec:pert}, we find the dominant contributions to $\Nsub{1}{\beta}$ and $\Nsub{2}{\beta}$ as given in \Tab{tab:pc_nsub}.  For the configuration of two collinear particles and a soft particle ($CCS$), $\Nsub{2}{\beta}$ is either dominated by $z_s$ or by $R_{cc}^\beta$.  In this configuration, the two subjet axes can either lie on the two collinear particles or one axis can be on a collinear particle and the other on a soft particle.  Importantly, the measurement of $\Nsub{1}{\beta}$ and $\Nsub{2}{\beta}$ cannot distinguish these two possibilities and therefore cannot determine if the second axis in the 1-prong jet is at a small or large angle with respect to the first.\footnote{For this reason, soft and collinear contributions to $\Nsub{2}{\beta}$ on 1-prong jets do not factorize and therefore cannot be computed in SCET.}  With either configuration, $\Nsub{1}{\beta}$ and $\Nsub{2}{\beta}$ scale as
\begin{align}
\Nsub{1}{\beta} & \sim  R_{cc}^\beta + z_s \,, \\
\Nsub{2}{\beta} & \sim R_{cc}^\beta + z_s\,.
\end{align}
That is, for 1-prong jets, $\Nsub{1}{\beta}\sim\Nsub{2}{\beta}$.

\begin{table}[t]
\begin{center}
\begin{tabular}{c|c}
modes &$\Nsub{2}{\beta}$ \\ 
\hline
$C_1C_2\,S$ & $z_s$  \\
$C_1C_2\, C$ & $R_{cc}^{\beta}$  \\
$C_1C_2\,C_s$ & $R_{12}^{\beta}z_{cs}$  
\end{tabular}
\end{center}
\caption{Scaling of the contributions from global soft ($S$), collinear ($C$), and collinear-soft ($C_s$) radiation correlated with the two hard subjets (denoted by $C_1$ and $C_2$) in 2-prong jets to  $\Nsub{2}{\beta}$ from the different possible configurations.
}
\label{tab:pc_ninja_nsub}
\end{table}

For 2-prong jets, $\Nsub{1}{\beta}$ is dominated by the hard splitting, as was the case with the two-point energy correlation function, hence $\Nsub{1}{\beta}\sim R_{12}^\beta$.  For $\Nsub{2}{\beta}$, the two axes lie along the two hard prongs, so, just like with the three-point energy correlation function, $\Nsub{2}{\beta}$ is set by the radiation about those two hard prongs: global soft, collinear, or collinear soft.  \Tab{tab:pc_ninja_nsub} lists the contributions to $\Nsub{2}{\beta}$ from each of these modes, leading to the scaling
\begin{align}
\Nsub{1}{\beta} &\sim R_{12}^\beta \,, \\
\Nsub{2}{\beta} &\sim z_s +R_{cc}^\beta + R_{12}^\beta z_{cs}\, .
\end{align}
Demanding that the jet only has two hard prongs implies that $\Nsub{2}{\beta}\sim z_s\sim R_{cc}^\beta\sim R_{12}^\beta z_{cs} \ll R_{12}^\beta\sim \Nsub{1}{\beta}$, but no other conclusions can be made from power counting alone.  Unlike the well-defined division of phase space by the energy correlation functions, $N$-subjettiness has a much weaker division of 
\begin{align*}
\text{1-prong jet: }& \Nsub{2}{\beta}\sim \Nsub{1}{\beta}\, ,\\
\text{2-prong jet: }& \Nsub{2}{\beta}\ll \Nsub{1}{\beta}\, .
\end{align*}
This does suggest, however, that the optimal discrimination variable using $N$-subjettiness is $\Nsub{2,1}{\beta}\equiv\Nsub{2}{\beta}/ \Nsub{1}{\beta}$, which is what is widely used experimentally.  Nevertheless, the weaker phase space separation of $N$-subjettiness compared with that for the energy correlation functions would na\"ively imply that $\ecfnobeta{2}$ and $\ecfnobeta{3}$ provides better discrimination than $\Nsub{1}{\beta}$ and $\Nsub{2}{\beta}$; however, this statement requires an understanding of $\mathcal{O}(1)$ numbers, which is beyond the scope of a power counting analysis.

\subsubsection{Effect of a Mass Cut}
\label{sec:masscute2e3}

\begin{figure}
\begin{center}
\subfloat[]{\label{fig:beq2_C2}
\includegraphics[width=6.5cm]{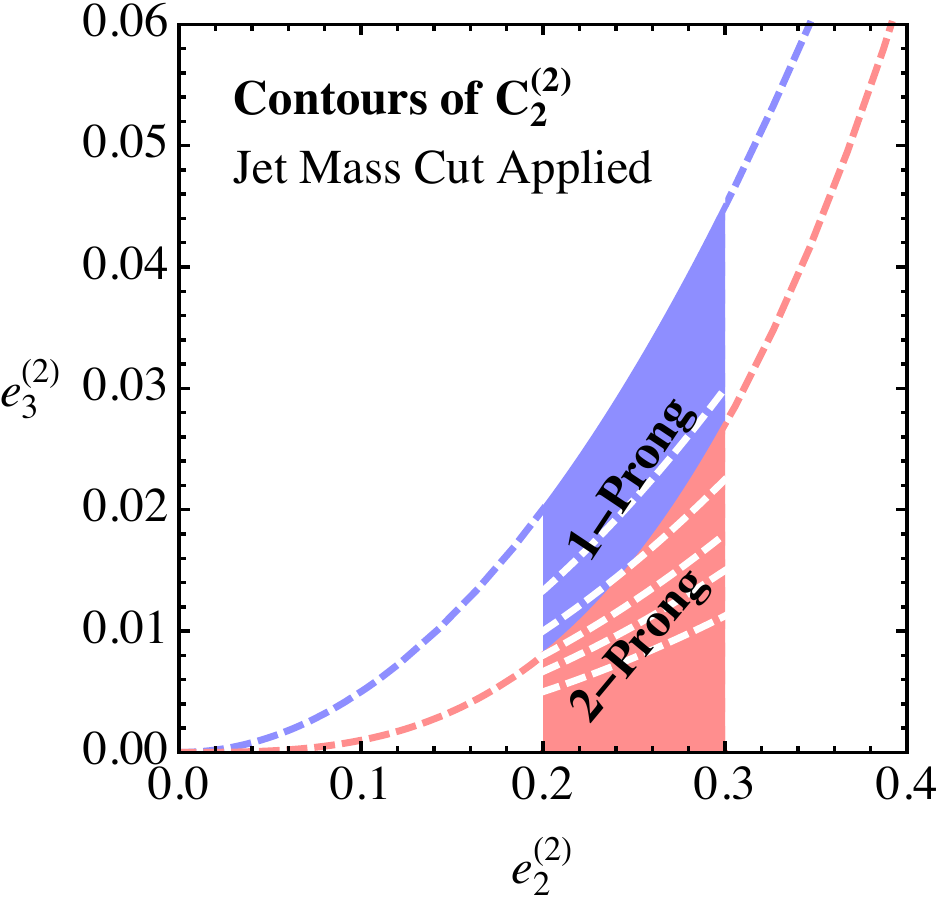}
}\qquad
\subfloat[]{\label{fig:beq2_D2}
\includegraphics[width=6.5cm]{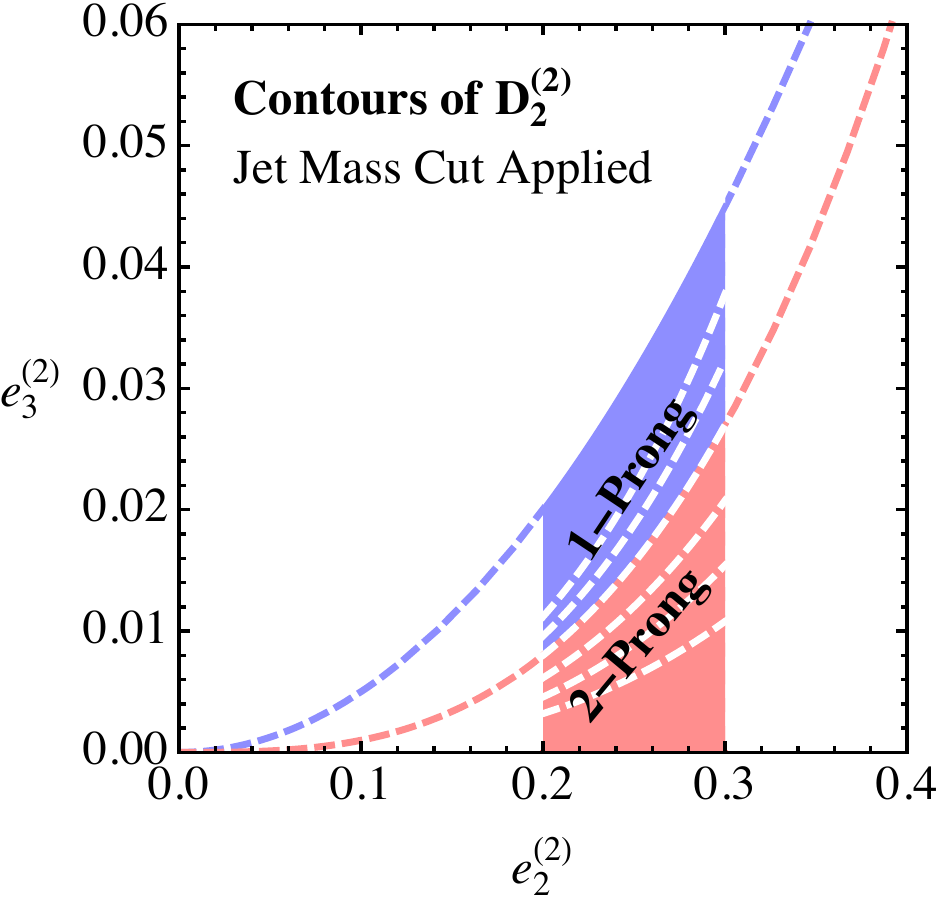}
}
\end{center}
\caption{Phase space defined by the energy correlation functions $\ecf{2}{2},\ecf{3}{2}$ in the presence of a mass cut. Contours of constant $\Cobs{2}{2}$ (left) and $\Dobs{2}{2}$ (right) are shown for reference. 
}
\label{fig:mcut}
\end{figure}

In an experimental application of $\Dobs{2}{\beta}$ to boosted Z discrimination, a mass cut is performed on the jet around the mass of the $Z$ boson.  In addition to removing a large fraction of the background, this cut also guarantees that the identified jets are actually generated from boosted $Z$ decays.  To fully understand the effect of the mass cut on the phase space requires analyzing the three-dimensional phase space of the mass, $\ecfnobeta{2}$, and $\ecfnobeta{3}$.  While complete, this full analysis would be distracting to the physics points that we wish to make in this section, and the impact of the mass cut can be understood without performing this analysis.  For $\beta = 2$, the two-point energy correlation function is simply related to the jet mass $m$ at fixed jet $p_T$:
\begin{equation}
\ecf{2}{2} \simeq \frac{m^2}{p_T^2} \ ,
\end{equation}
for central jets assuming that $m \ll p_T$ and up to overall factors of order 1.  Therefore, a cut on the jet mass is a cut on $\ecf{2}{2}$.  In this section, we will begin by discussing the simpler case of $\beta=2$, and then proceed to comment on the effect of a mass cut for general $\beta$.

The phase space in the $\ecf{2}{2},\ecf{3}{2}$ plane with the jet mass constrained to a window, and for some finite range of jet $p_T$ is shown schematically in \Fig{fig:mcut}.  Jets of a given mass can have that mass generated either by substantial soft radiation (for 1-prong jets) or by a hard splitting in the jet (a 2-prong jet), and so we want a discrimination observable that separates these two regions cleanly.  The boundary between the 1-prong and 2-prong jet regions is still defined by $\ecf{3}{2}\sim (\ecf{2}{2})^3$, and so we expect $\Dobs{2}{2}$ to be the most powerful discriminant.  However, by making a mass cut, the region of phase space at small masses, dominated by 1-prong jets, is removed.  Therefore, the fact that contours of the observable $\Cobs{2}{2}$ mix both 1- and 2-prong jets is much less of an issue.  Except at very high signal efficiencies, when one is sensitive to the functional form of the boundary between the signal and background regions, the discrimination performance of $\Cobs{2}{2}$ should be similar to that of $\Dobs{2}{2}$ when a tight mass cut is imposed. Indeed, in a sufficiently narrow window, any variable of the form $\ecf{3}{2}/(\ecf{2}{2})^n$, would provide reasonable discrimination, with all the discrimination power coming from $\ecf{3}{2}$ alone.  However, $\Dobs{2}{2}$ has the advantage that its discrimination power does not suffer from significant dependence on the value of the lower mass cut.

\begin{figure}
\begin{center}
\subfloat[]{\label{fig:bneq2_C2}
\includegraphics[width=6.5cm]{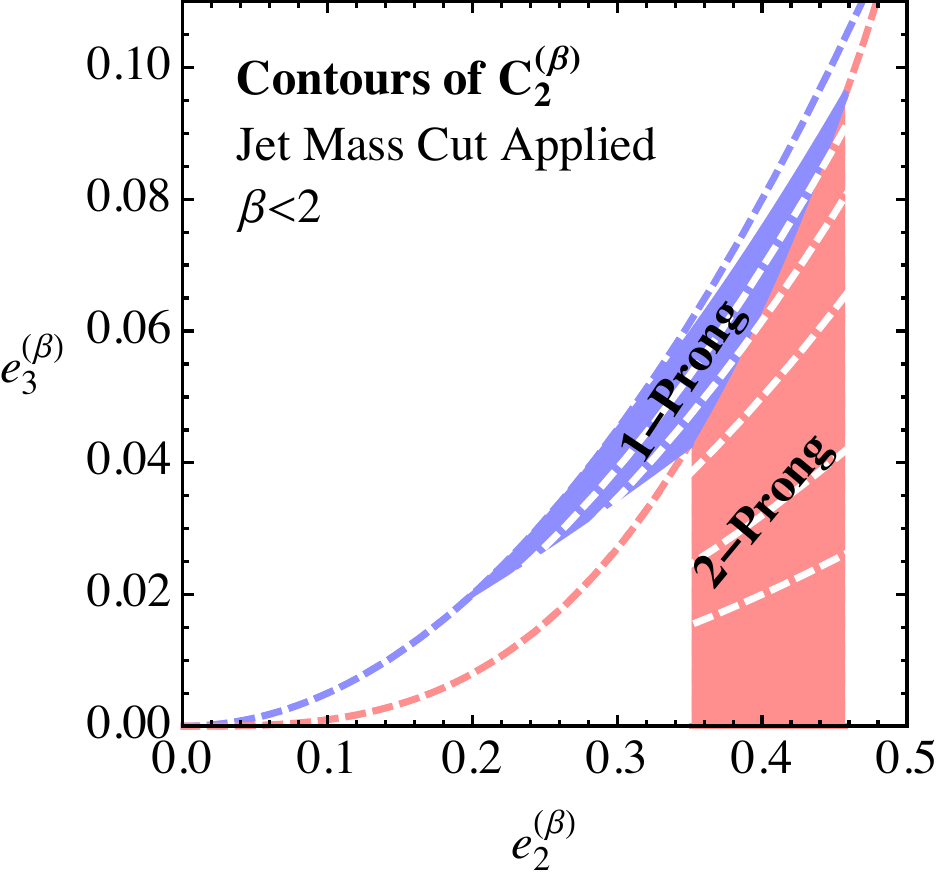}
}\qquad
\subfloat[]{\label{fig:bneq2_D2}
\includegraphics[width=6.5cm]{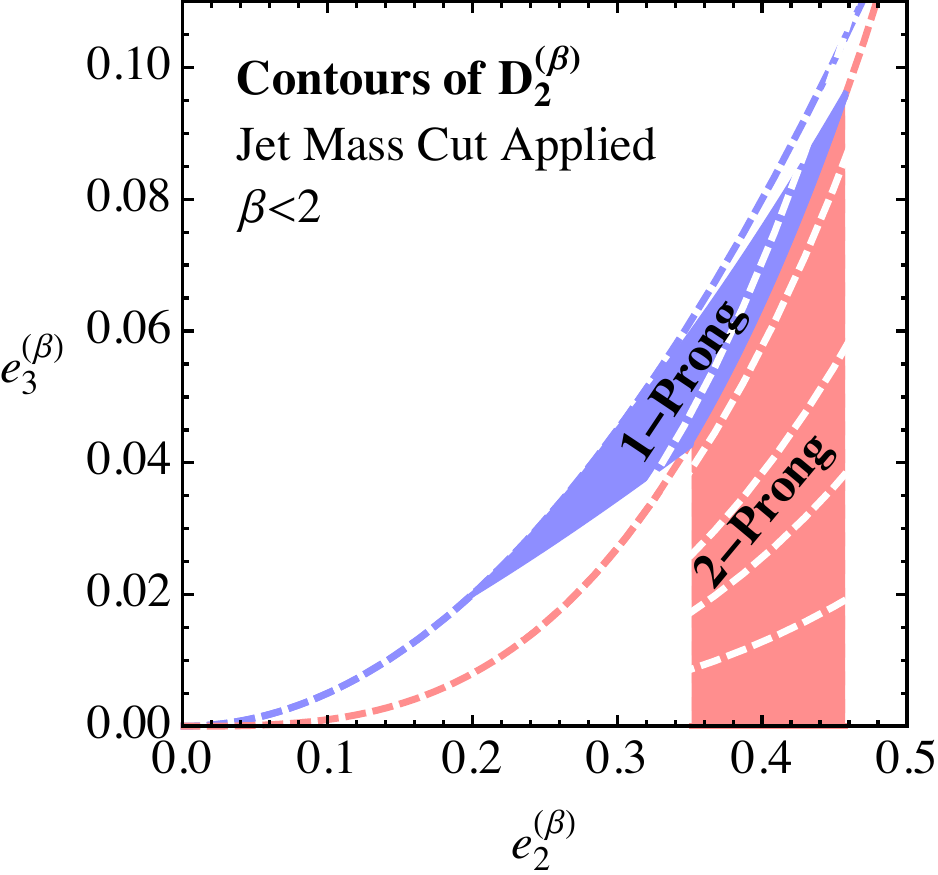}
}
\end{center}
\caption{Phase space defined by the energy correlation functions $\ecf{2}{\beta},\ecf{3}{\beta}$, for $\beta<2$, in the presence of a mass cut. Contours of constant $\Cobs{2}{2}$ (left) and $\Dobs{2}{2}$ (right) are shown for reference.
}
\label{fig:mcut_bneq2}
\end{figure}

While a lower mass cut is important for removing 1-prong background jets, an upper mass cut is also necessary for powerful discrimination.  The mass distribution of QCD jets has a long tail extending to masses of order the $p_T$ of the jet.  For these jets, the mass is generated by an honest hard splitting, and so these background jets look exactly like the signal from their substructure.  While the cross section for these high mass QCD jets is suppressed by $\alpha_s$, they can still be a significant background and therefore should be removed.  

Let's now consider the general $\beta$ case.  We will first consider the effect of a mass cut in the 2-prong region of the $\ecf{2}{\beta},\ecf{3}{\beta}$ plane.  Recall that in this region of phase space
\begin{equation}
\ecf{2}{\beta} \sim R_{12}^\beta \ ,
\end{equation}
where $R_{12}$ is the angle between the hard subjets.  Therefore, in this region of phase space $\ecf{2}{\beta}$ is simply related to the mass:
\begin{equation}
\ecf{2}{\beta} \sim \left(
\frac{m^2}{p_T^2}
\right)^{\beta/2} \ .
\end{equation}
A cut on the jet mass is therefore equivalent to an appropriate cut on $\ecf{2}{\beta}$ for 2-prong jets.

A mass cut in the 1-prong region of phase space is more subtle, as the dominant contributing mode to $\ecf{2}{\beta}$ changes throughout the phase space.  Recall that in this region, $\ecf{2}{\beta}$ has the scalings
\begin{align}
\ecf{2}{\beta}&\sim R_{cc}^\beta +z_s\,.
\end{align}
while
\begin{align}
 \ecf{2}{2}&\sim\frac{m^2}{p_T^2}\sim R_{cc}^2 +z_s\,.
\end{align}
While the soft contributions have the same scaling for both variables, the collinear contributions do not. There are two possibilities as for the relative scalings of $\ecf{2}{\beta}$ and the mass: if soft emissions do not contribute, then
\begin{equation}\label{eq:nosoft1prong}
\ecf{2}{\beta} \sim \left(
\frac{m^2}{p_T^2}
\right)^{\beta/2} \ ,
\end{equation}
which matches onto the relative scaling in the 2-prong region of phase space.  If instead soft emissions do contribute, then
\begin{equation}\label{eq:yessoft1prong}
\ecf{2}{\beta} \sim 
\frac{m^2}{p_T^2}\ ,
\end{equation}
which defines the upper boundary of the 1-prong phase space.  These phase space boundaries for jets on which two two-point energy correlation functions with different angular exponents (or recoil-free angularities \cite{Larkoski:2014uqa}) are measured is discussed in detail in \Ref{Larkoski:2014tva}.

Depending on whether $\beta$ is less than or greater than 2, the mass cut manifests itself differently.  For $\beta < 2$, note that from \Eqs{eq:nosoft1prong}{eq:yessoft1prong}, $\ecf{2}{\beta} > \ecf{2}{2}$ in the two-prong region, and so smaller values of $\ecf{2}{\beta}$ can correspond to the same mass.  Conversely, for $\beta > 2$, $\ecf{2}{\beta} < \ecf{2}{2}$ and so larger values of $\ecf{2}{\beta}$ can correspond to the same mass.  The effect of a mass cut on the allowed phase space for $\beta < 2$ is illustrated schematically in \Fig{fig:mcut}.  Because in this case small values of $\ecf{2}{\beta}$ can satisfy the mass cut, contours of $\Cobs{2}{\beta}$ can pass through the background region of phase space and significantly reduce the discrimination power.  Again, because it respects the parametric scaling of the phase space boundaries, we expect the discrimination power of $\Dobs{2}{\beta}$ to be more robust as $\beta$ decreases from 2.  However, the precise discrimination power depends on understanding the ${\cal O}(1)$ region around the 1-prong and 2-prong jet boundary as $\beta$ moves away from 2.  This observation also explains why \Ref{Larkoski:2013eya} found that the optimal choice for boosted $Z$ boson discrimination using $\Cobs{2}{\beta}$ with a tight mass cut was $\beta \simeq 2$.

\subsubsection{Summary of Power Counting Predictions}
\label{sec:sumpredict}

Here, we summarize the main predictions from our power counting analysis of boosted $Z$ discrimination, before a Monte Carlo study in \Sec{sec:mc_e2e3}.  We have:
\begin{itemize}

\item The parametric scaling of the boundary between 1-prong and 2-prong jets in the $(\ecfnobeta{2},\ecfnobeta{3})$ phase space is $\ecfnobeta{3}\sim (\ecfnobeta{2})^3$.  Therefore, $\Dobs{2}{\beta}$ should be a more powerful discrimination observable than $\Cobs{2}{\beta}$ because contours of constant $\Dobs{2}{\beta}$ do not mix signal and background regions, while contours of $\Cobs{2}{\beta}$ do mix signal and background regions.

\item When a mass cut is imposed on the jet, $\Cobs{2}{\beta}$ should have similar discrimination power to $\Dobs{2}{\beta}$, for $\beta \simeq 2$, except at high signal efficiency when the observable is sensitive to the boundary between the signal and background regions.  At high signal efficiency, $\Dobs{2}{\beta}$ should be a slightly better discriminant than $\Cobs{2}{\beta}$ for $\beta \simeq 2$.

\item The discrimination power of $\Cobs{2}{\beta}$ should decrease substantially as $\beta$ decreases from 2 when there is a mass cut on the jets.  By contrast, the discrimination power of $\Dobs{2}{\beta}$ should be more robust as $\beta$ decreases from 2.

\item The power counting predictions stated above should be robust to Monte Carlo tuning and reproduced by any Monte Carlo simulation, e.g. \herwigpp\ or \pythia{8}, since they are determined by parametric scaling of QCD dynamics.
\end{itemize}

\subsubsection{Monte Carlo Analysis}
\label{sec:mc_e2e3}

\begin{figure}
\begin{center}
\subfloat[]{
\includegraphics[width=6.5cm]{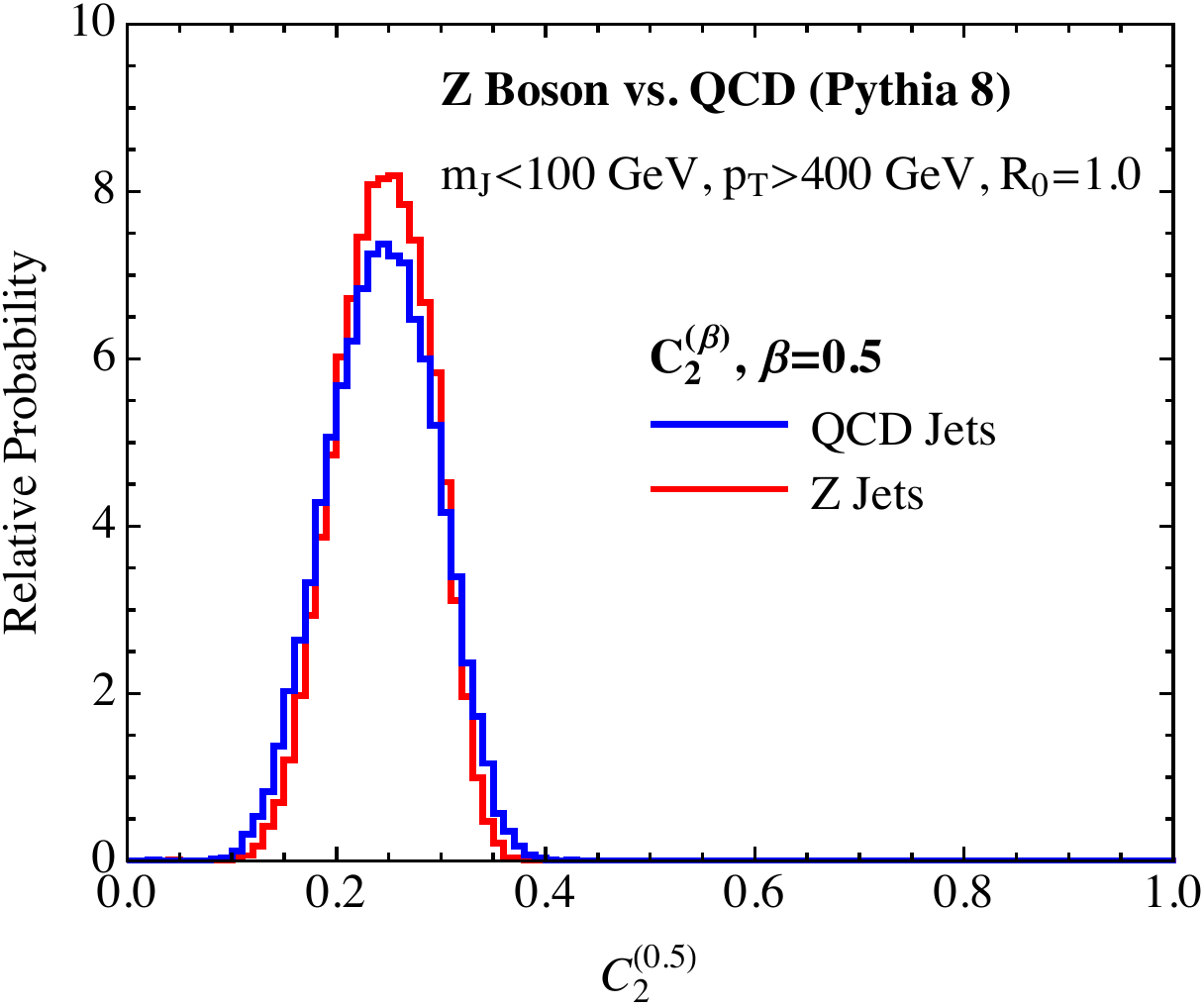}
}\qquad
\subfloat[]{
\includegraphics[width=6.5cm]{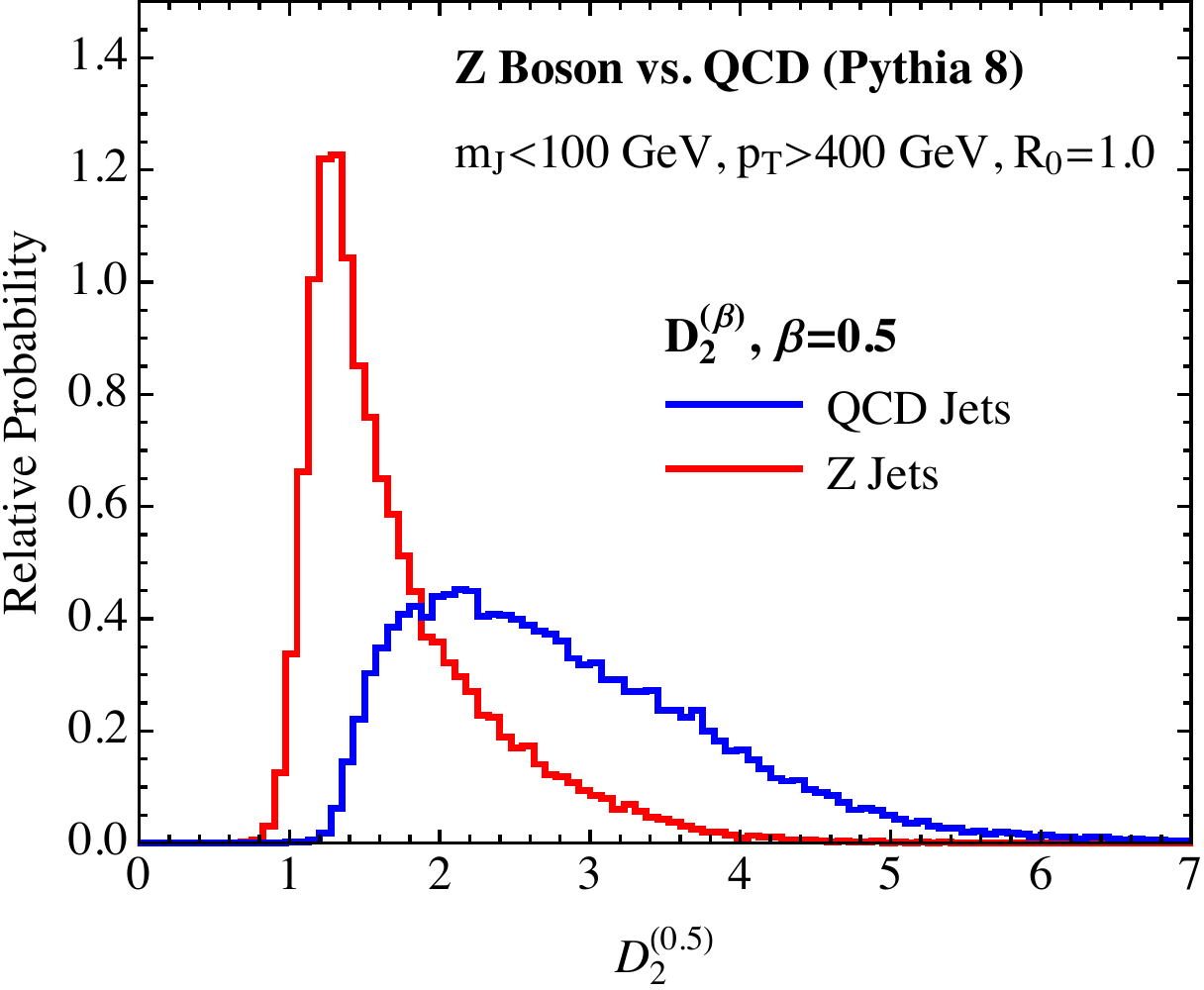}
}\qquad
\subfloat[]{
\includegraphics[width=6.5cm]{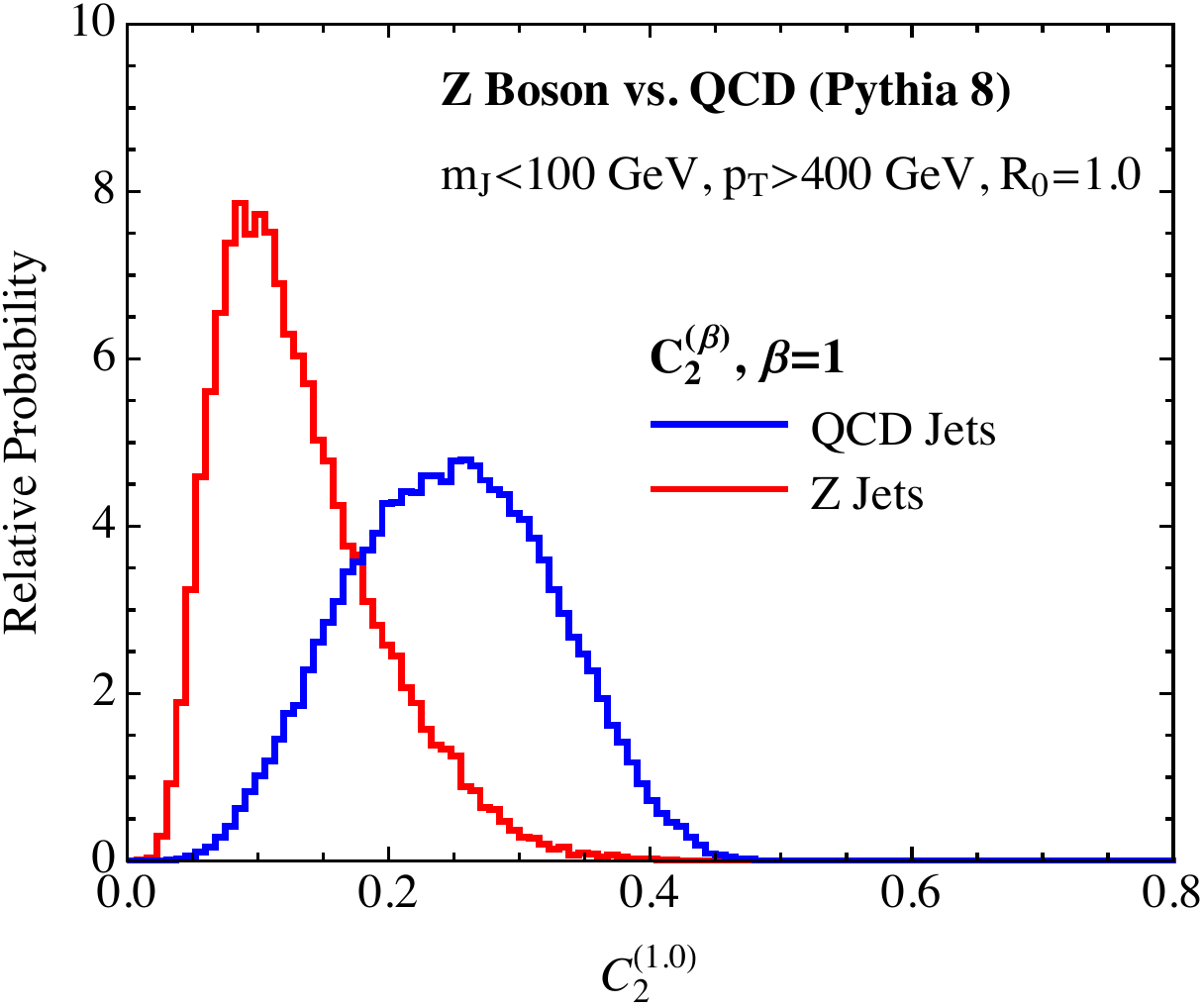}
}\qquad
\subfloat[]{
\includegraphics[width=6.5cm]{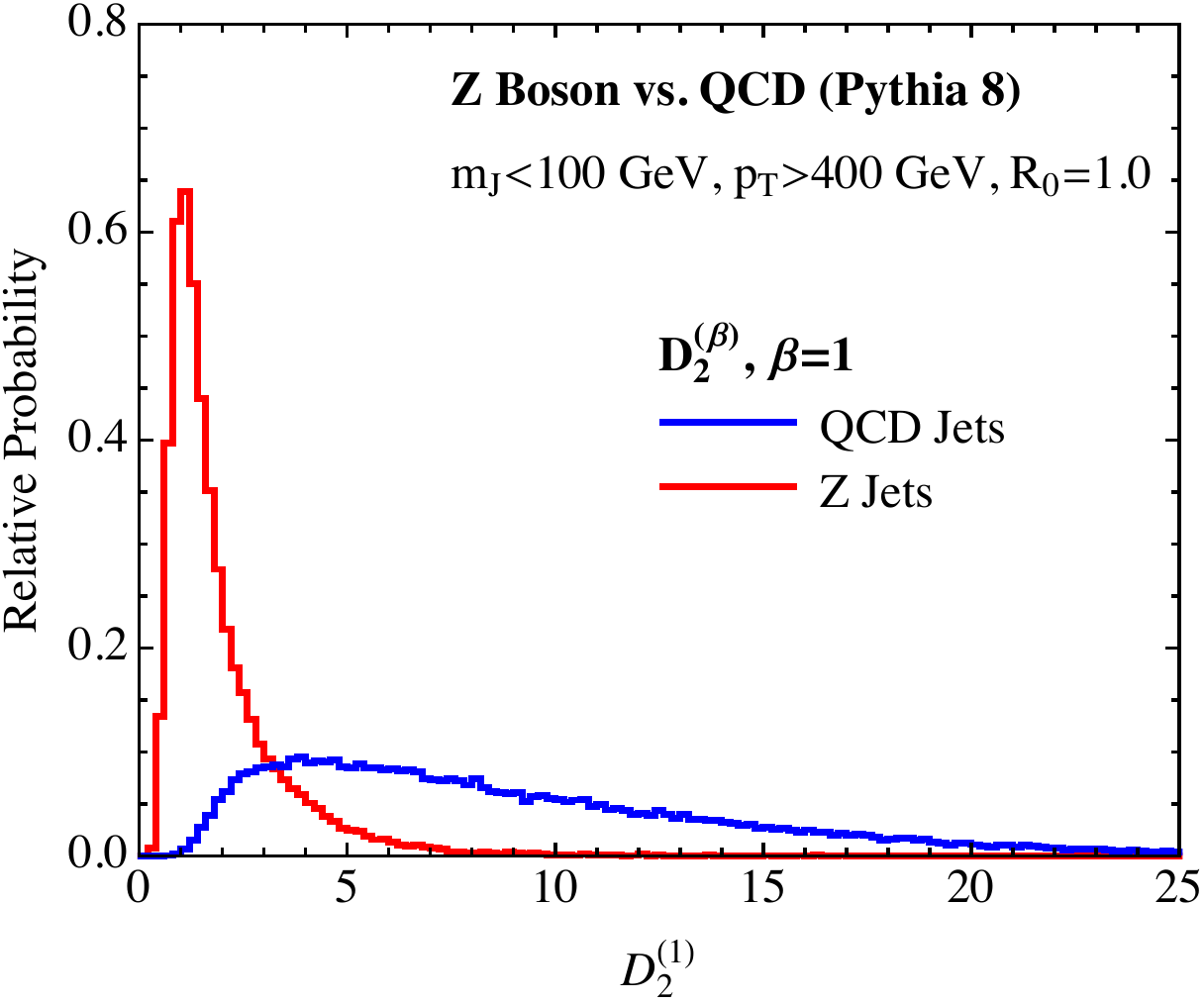}
}\qquad
\subfloat[]{
\includegraphics[width=6.5cm]{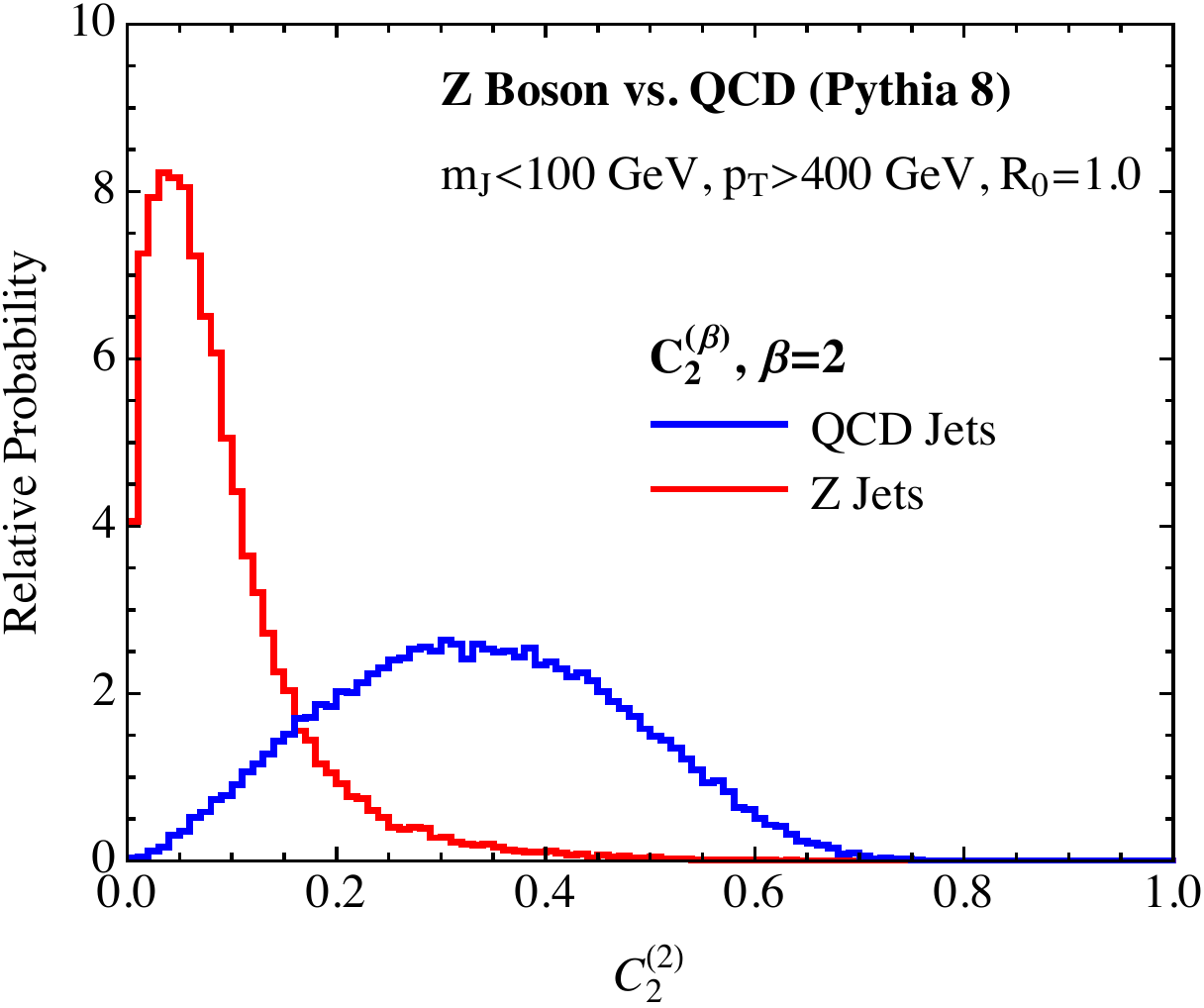}
}\qquad
\subfloat[]{
\includegraphics[width=6.5cm]{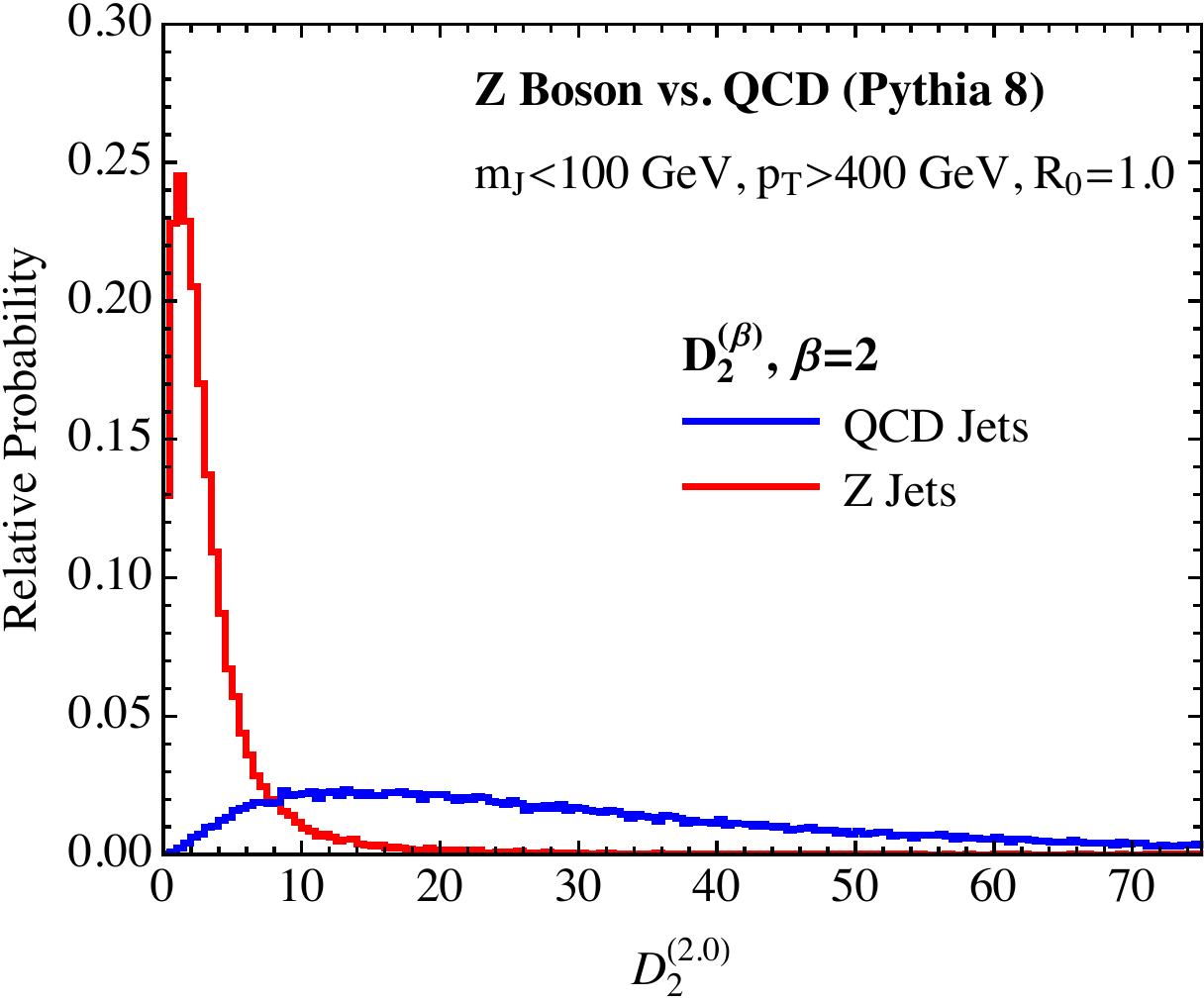}
}
\end{center}
\caption{ Signal and background distributions for the ratio observables $\Cobs{2}{\beta}$ (left) and $\Dobs{2}{\beta}$ (right) for $\beta=0.5,1,2$ from the \madgraph~and \pythia{8} samples. No lower mass cut on the jets is applied but we take $m_J<100$ GeV. 
}
\label{fig:nomasscut}
\end{figure}

\begin{figure}
\begin{center}
\subfloat[]{
\includegraphics[width=6.5cm]{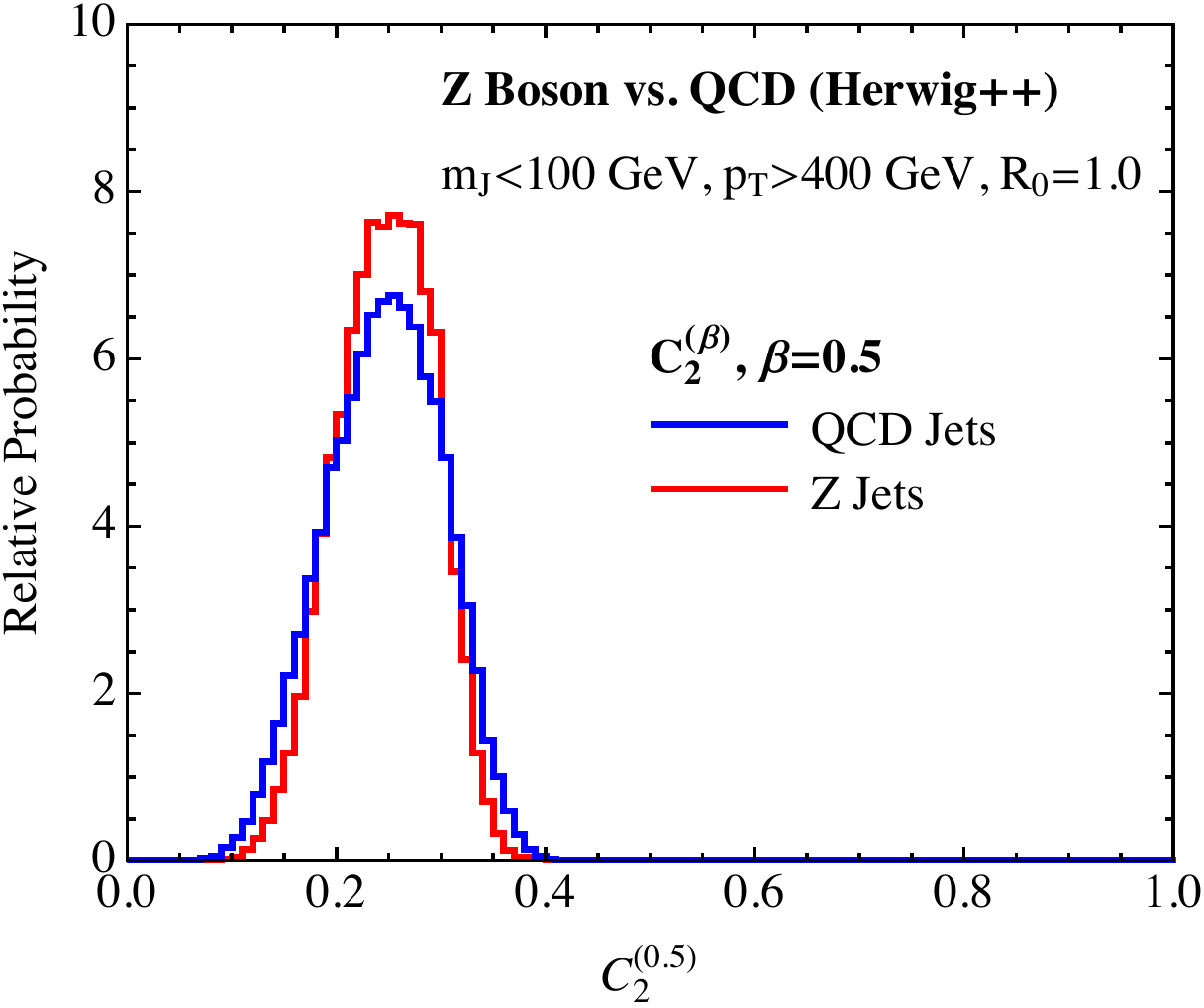}
}\qquad
\subfloat[]{
\includegraphics[width=6.5cm]{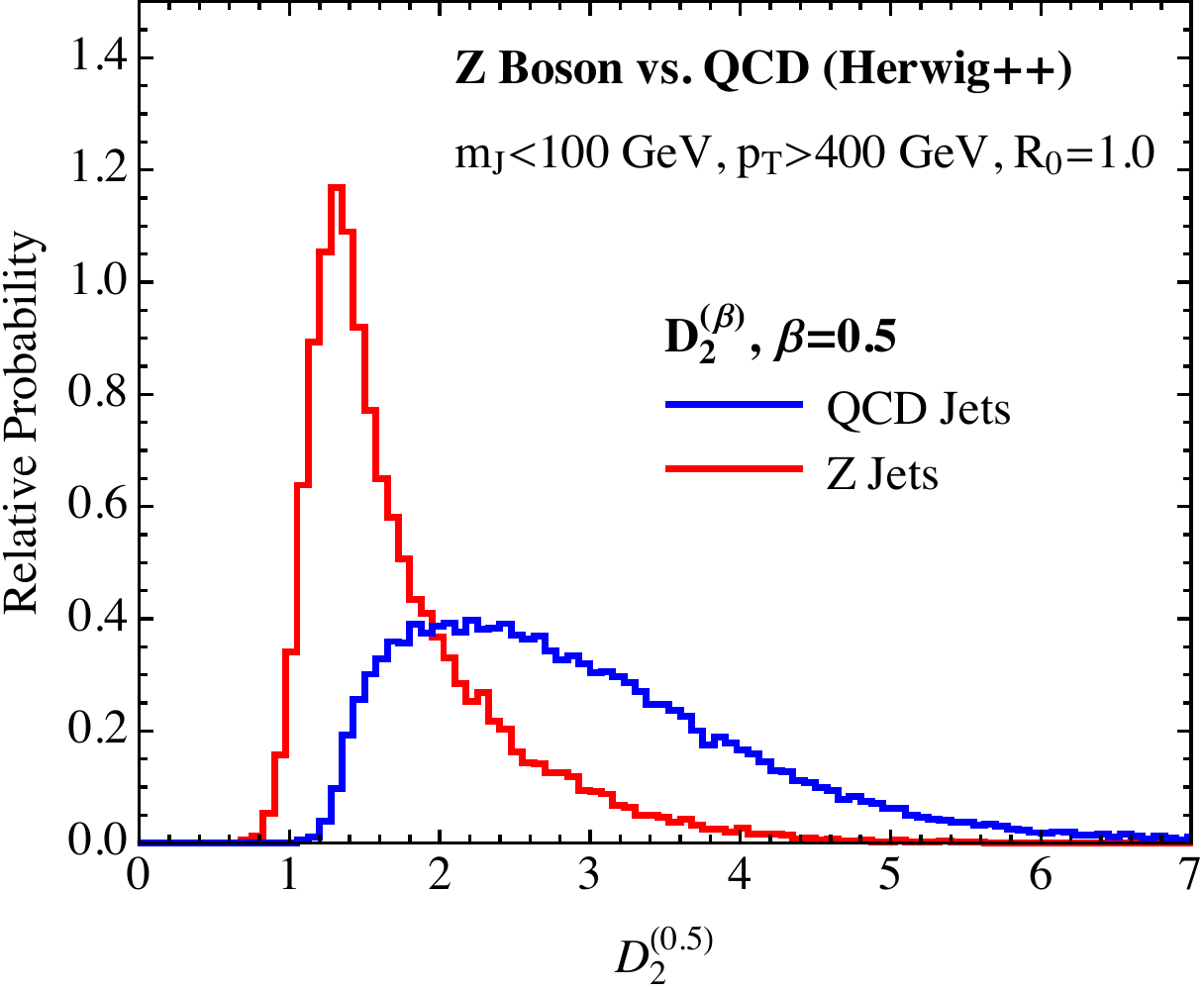}
}\qquad
\subfloat[]{
\includegraphics[width=6.5cm]{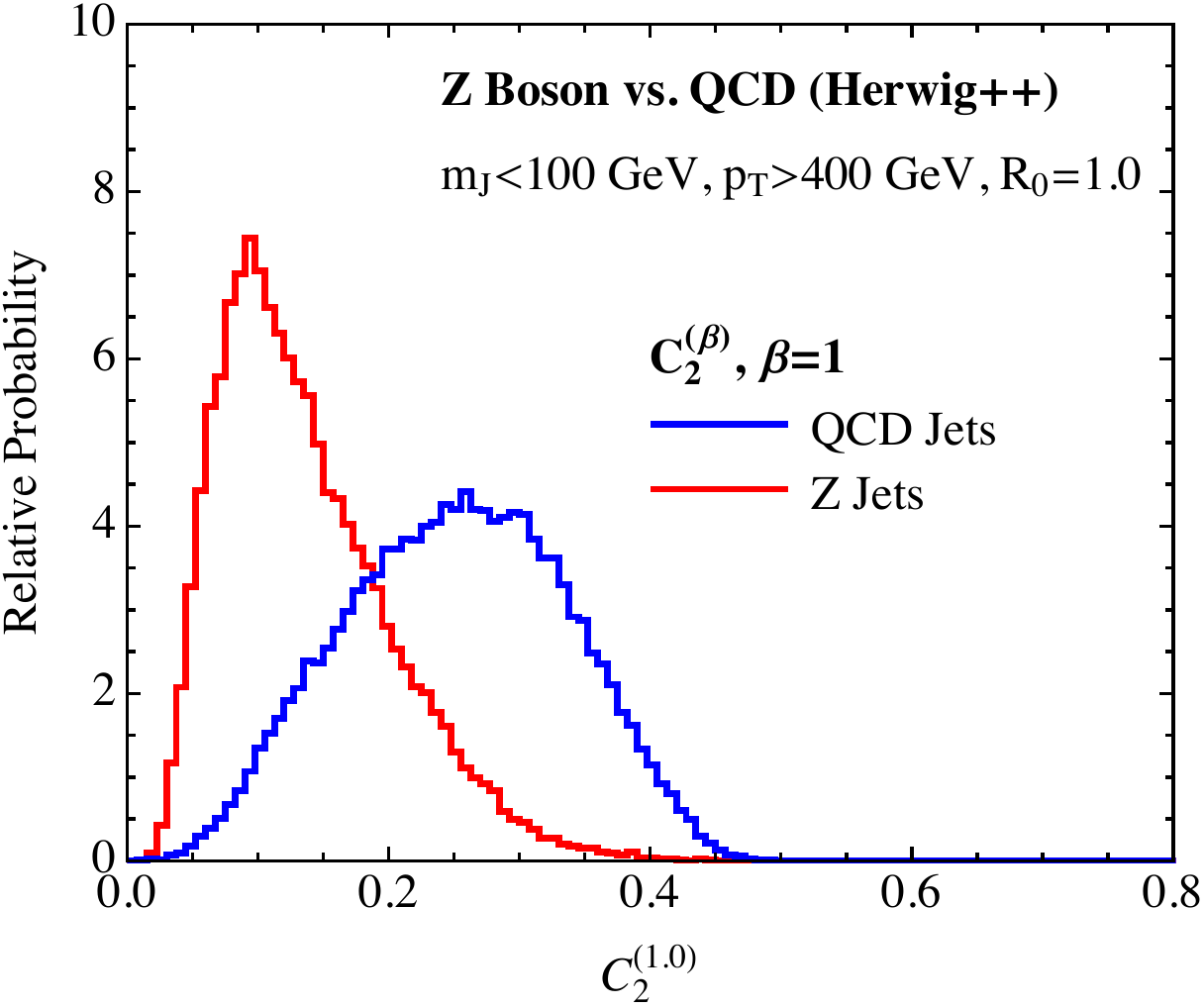}
}\qquad
\subfloat[]{
\includegraphics[width=6.5cm]{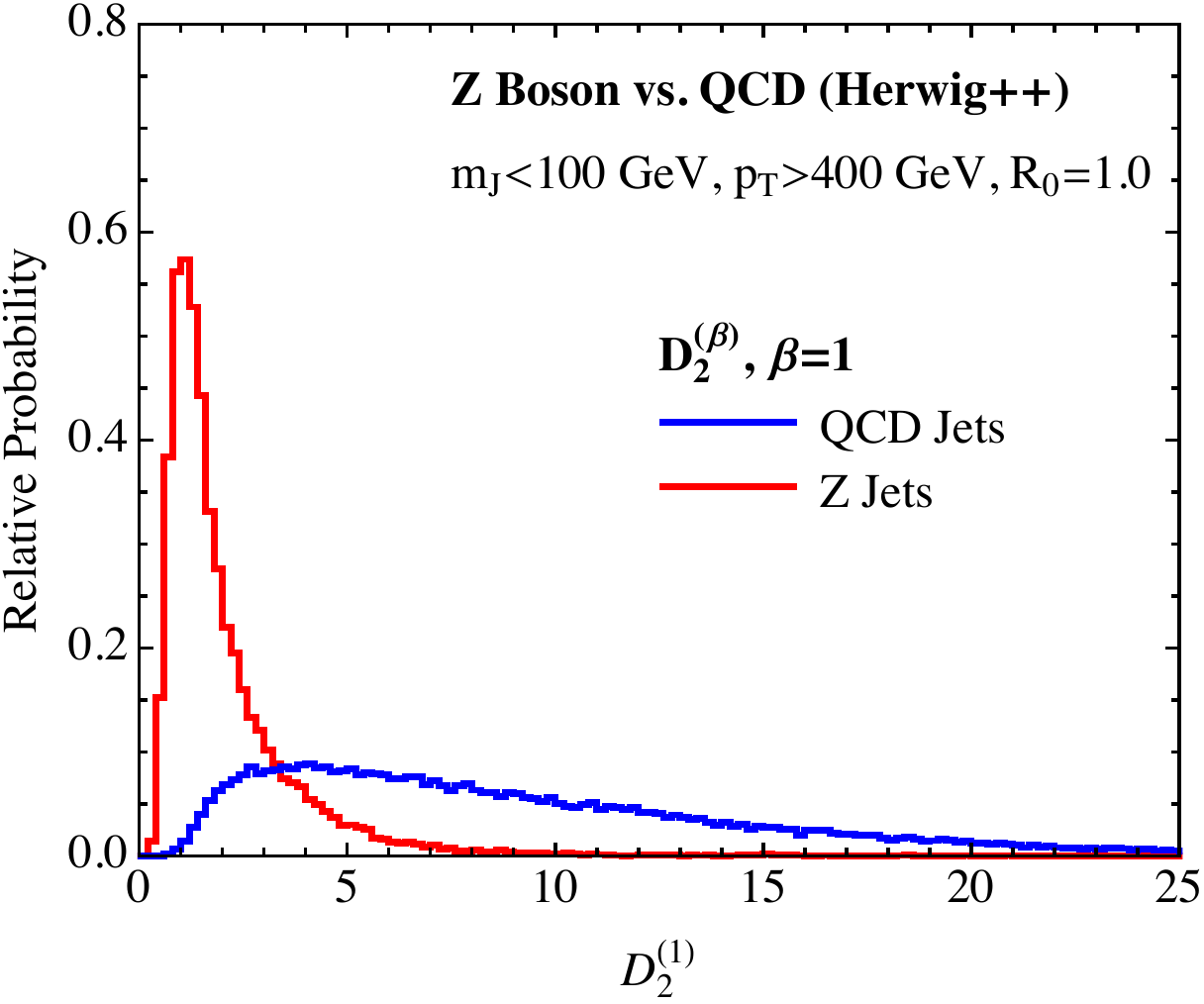}
}\qquad
\subfloat[]{
\includegraphics[width=6.5cm]{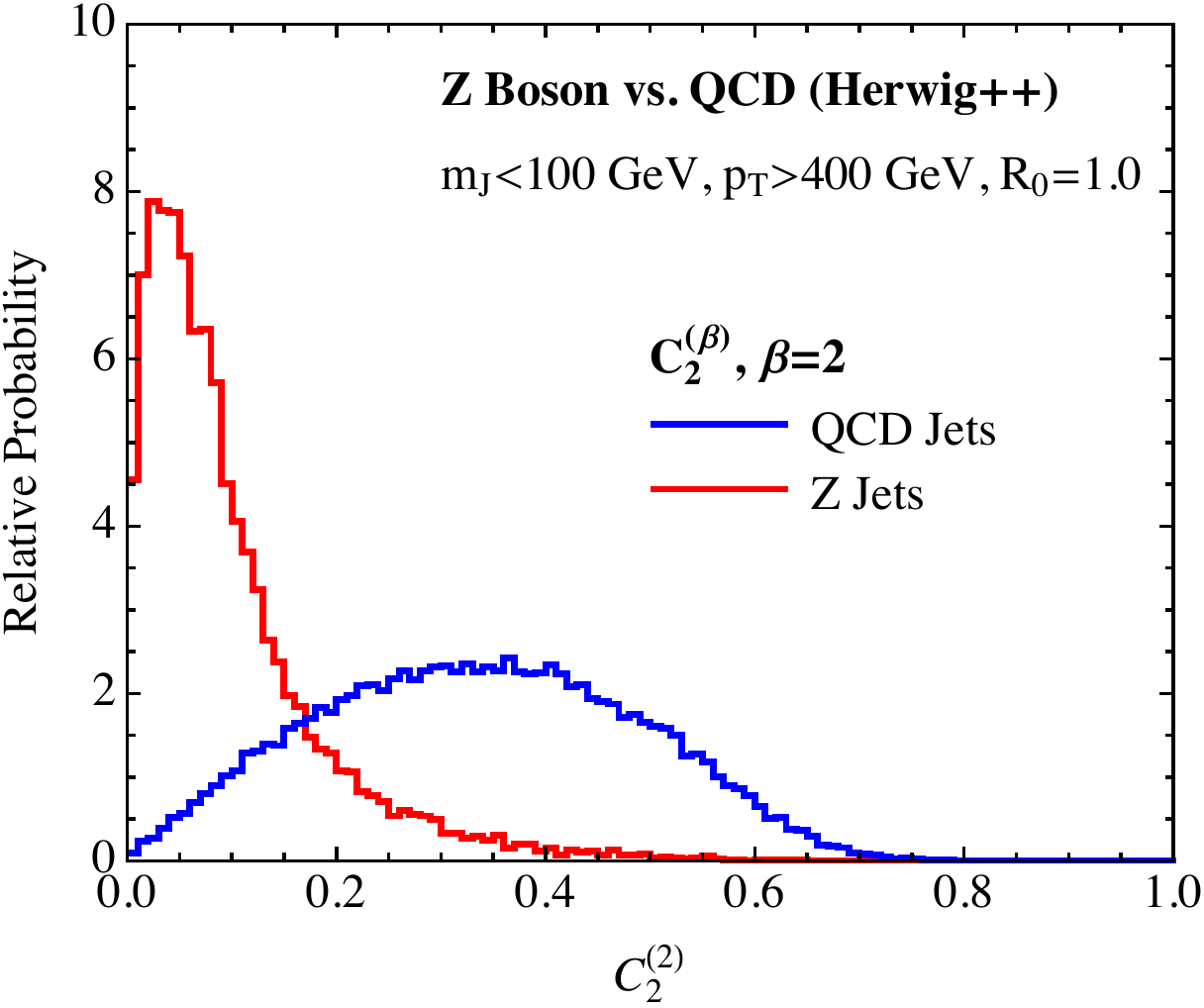}
}\qquad
\subfloat[]{
\includegraphics[width=6.5cm]{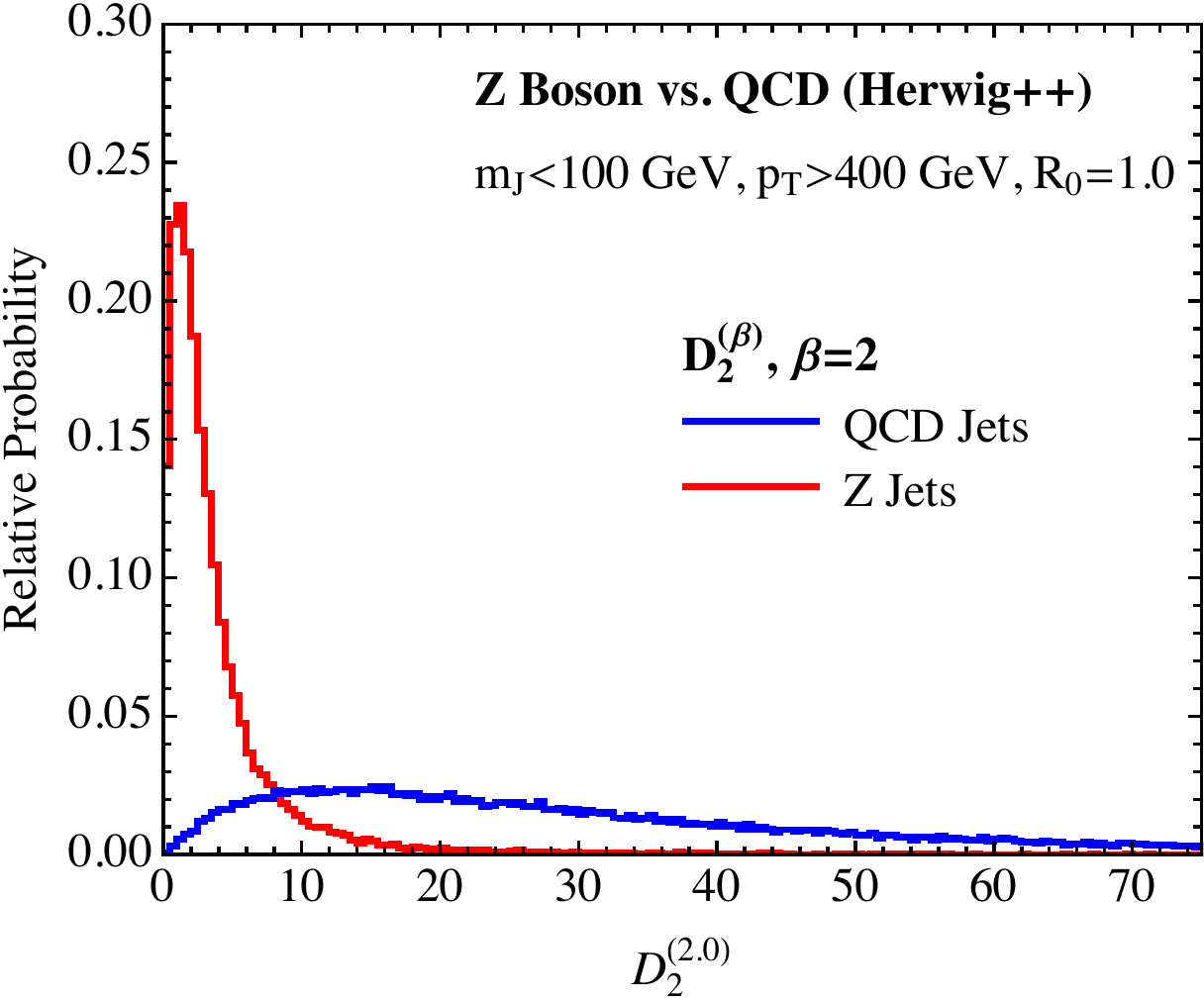}
}
\end{center}
\caption{ 
Same plots as in \Fig{fig:nomasscut}, from the \herwigpp~samples.
}
\label{fig:nomasscut_Herwig}
\end{figure}

\begin{figure}
\begin{center}
\subfloat[]{
\includegraphics[width=6.5cm]{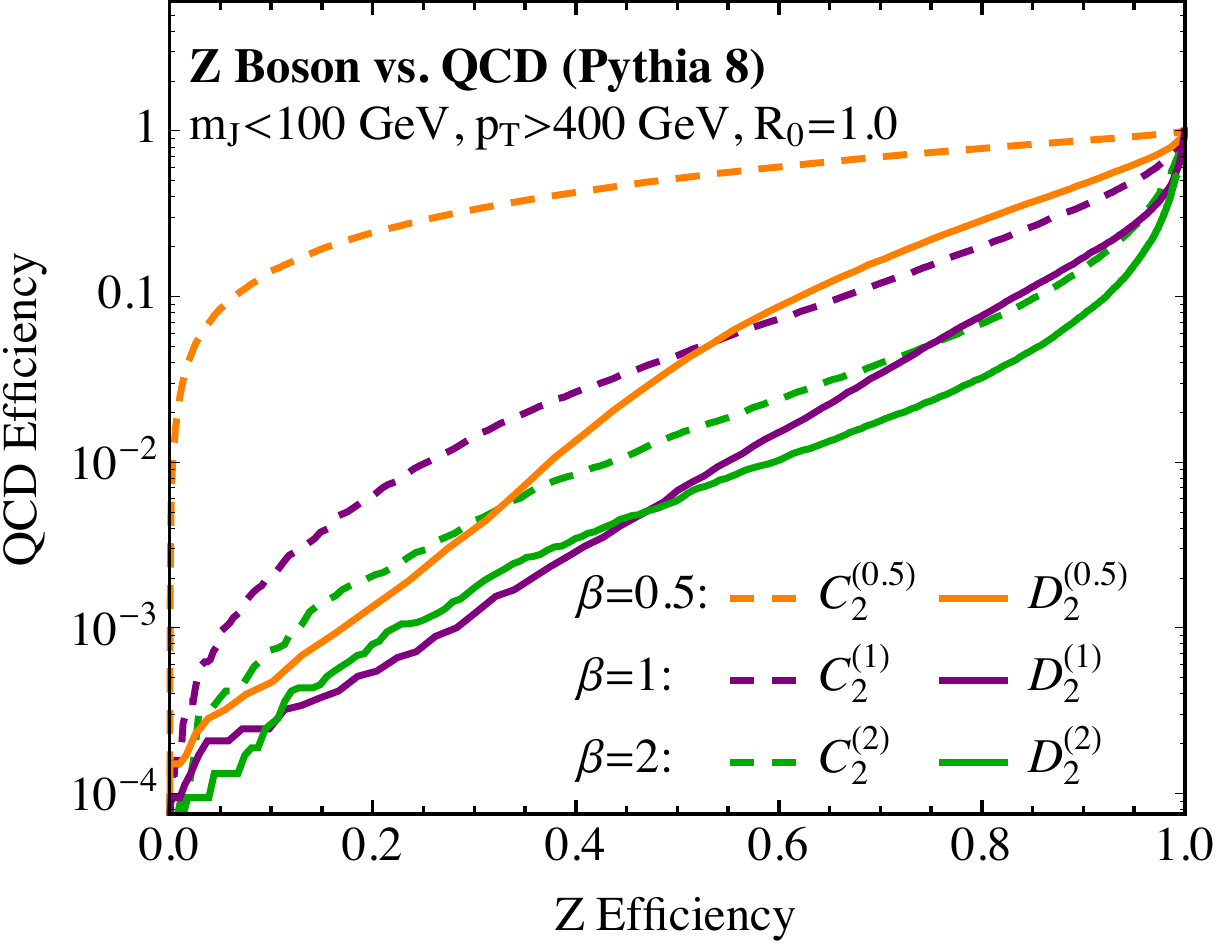}
}\qquad
\subfloat[]{
\includegraphics[width=6.5cm]{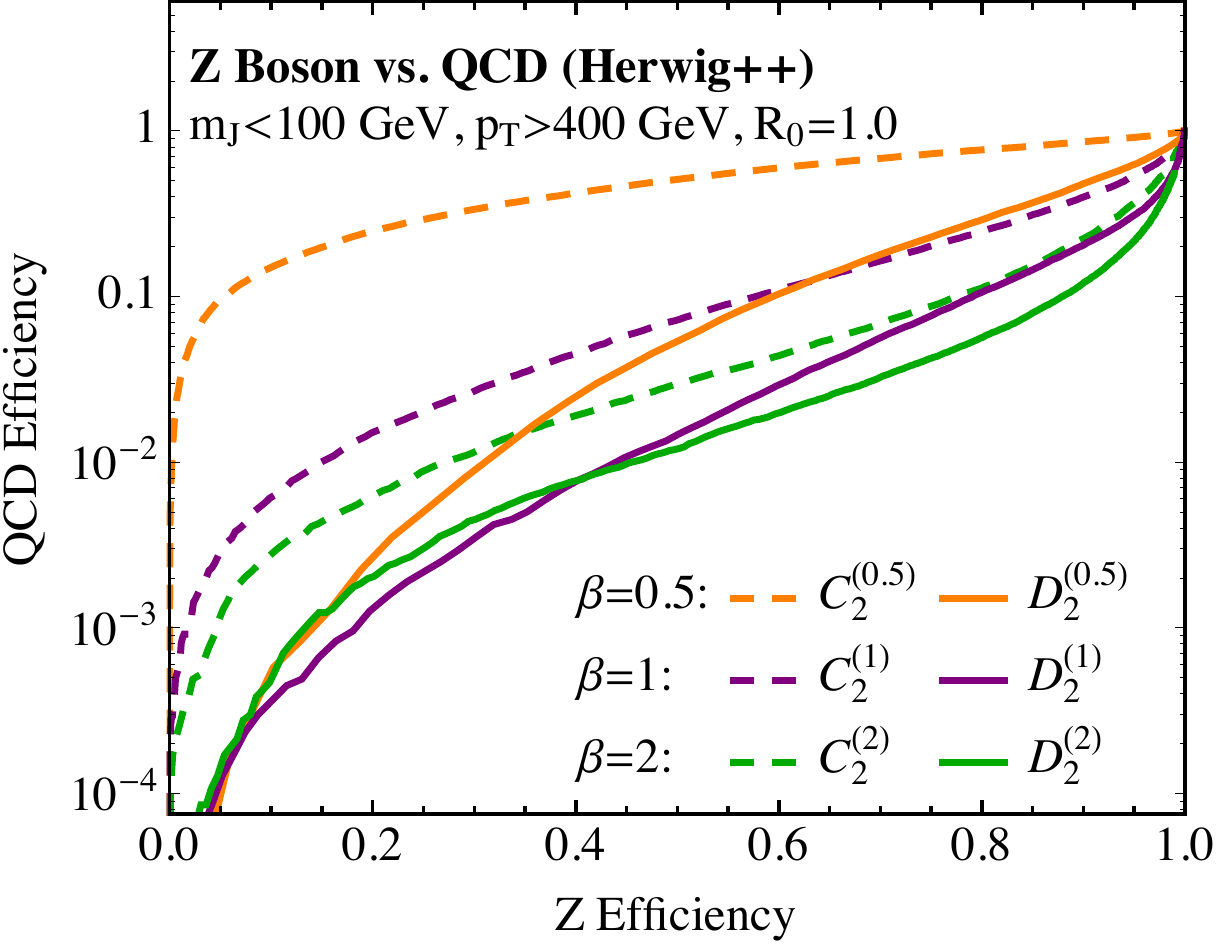}
}
\end{center}
\caption{Signal vs.~background efficiency curves (ROC curves) for $\Cobs{2}{\beta}$ and $\Dobs{2}{\beta}$ for $\beta=0.5,1,2$ for jets with $m_J<100$ GeV, showered with \pythia{8} (left) and \herwigpp~(right). Power counting predictions for the behavior of the ROC curves are robustly reproduced by both Monte Carlo generators. 
}
\label{fig:ROC_C2_nmcut}
\end{figure}

To test these predictions, we will study the different ratio observables formed from $\ecf{2}{\beta}$ and  ${\ecf{3}{\beta}}$ in Monte Carlo simulation. We generated background QCD jets from $pp\to Zj$ events, with the $Z$ decaying leptonically, and boosted $Z$ decays from $pp\to ZZ$ events, with one $Z$ decaying leptonically, and the other to quarks. Events were generated with \madgraph{2.1.2} \cite{Alwall:2014hca} at the $8$ TeV LHC, and showered with either \pythia{8.183} \cite{Sjostrand:2006za,Sjostrand:2007gs} or \herwigpp{2.6.3} \cite{Marchesini:1991ch,Corcella:2000bw,Corcella:2002jc,Bahr:2008pv}, to test the robustness of our predictions to the details of the Monte Carlo generator. Anti-$k_T$ \cite{Cacciari:2008gp} jets with radius $R=1.0$ and $p_T>400$ GeV were clustered in \fastjet{3.0.3} \cite{Cacciari:2011ma} using the Winner Take All (WTA) recombination scheme \cite{Larkoski:2014uqa,Larkoski:2014bia}. The energy correlation functions and $N$-subjettiness ratio observables were calculated using the \texttt{EnergyCorrelator} and \texttt{Nsubjettiness} \fastjet{contrib}s \cite{Cacciari:2011ma,fjcontrib}.

We first compare the discrimination power of $\Cobs{2}{\beta}$ to $\Dobs{2}{\beta}$ with no lower mass cut on the jets for several values of the angular exponent.  We require that $m_J < 100$ GeV which removes a significant fraction of QCD jets that have honest 2-prong structure.  Therefore, we are testing the power of $\Cobs{2}{\beta}$ and $\Dobs{2}{\beta}$ to discriminate between 1-prong and 2-prong jets.  In \Fig{fig:nomasscut}, we show the raw distributions of $\Cobs{2}{\beta}$ and $\Dobs{2}{\beta}$ measured on signal and background for $\beta = 0.5,1,2$.  Especially at small $\beta$, $\Dobs{2}{\beta}$ is much more efficient at separating boosted $Z$s from QCD jets than is $\Cobs{2}{\beta}$.  This is exactly as predicted by the power counting, because $\Cobs{2}{\beta}$ mixes the signal and background regions of phase space, an effect that is magnified at smaller $\beta$.  The discrimination power is quantified in \Fig{fig:ROC_C2_nmcut} where we show the signal vs.~background efficiency curves (ROC curves) for the three choices of $\beta$ for $\Cobs{2}{\beta}$ and $\Dobs{2}{\beta}$.  At low signal efficiency, every $\Dobs{2}{\beta}$ is a better discriminant than any $\Cobs{2}{\beta}$, and the performance of $\Dobs{2}{\beta}$ is much more stable as a function of $\beta$ than $\Cobs{2}{\beta}$.

\begin{figure}
\begin{center}
\subfloat[]{\label{fig:90p_mcut_pythia}
\includegraphics[width=6cm]{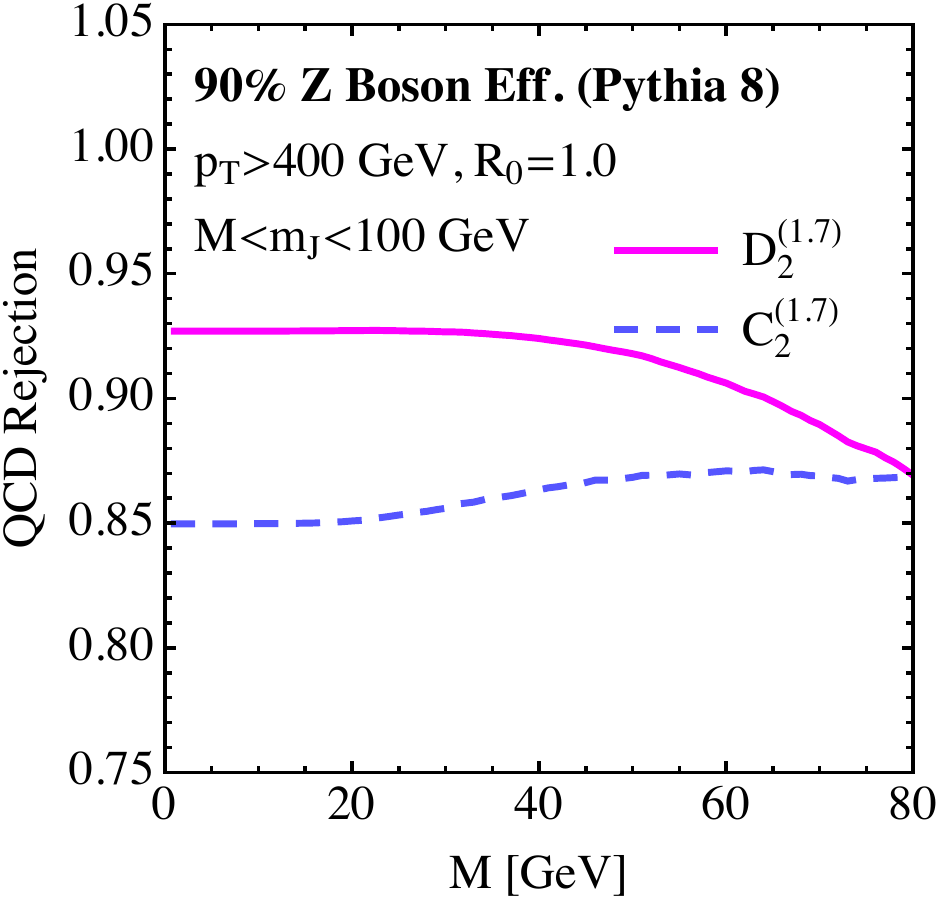}
}\qquad
\subfloat[]{\label{fig:90p_mcut_Herwig}
\includegraphics[width=6cm]{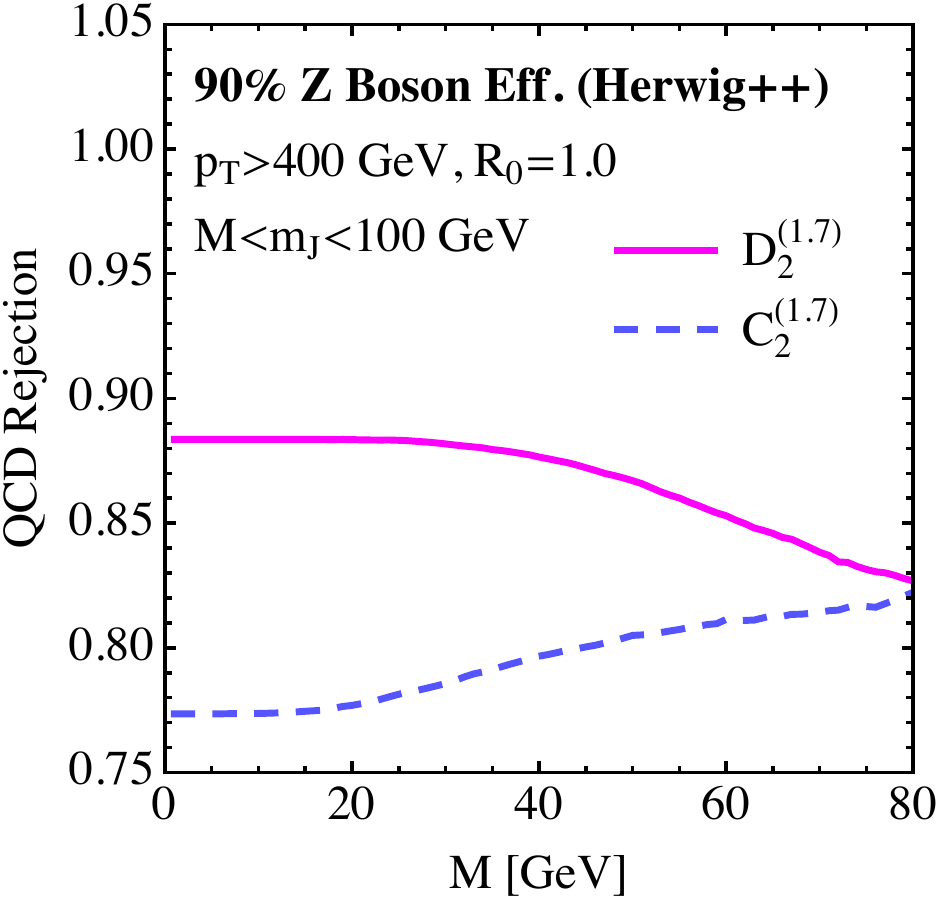}
}
\end{center}
\caption{ QCD rejection efficiency at $90\%$ signal efficiency as a function of the lower mass cut, as predicted by \pythia{8} (left), and \herwigpp (right).  The plots compare the efficiencies of $\Cobs{2}{1.7}$ and $\Dobs{2}{1.7}$.
}
\label{fig:m_dep}
\end{figure}

In the presence of a narrow mass cut window, the power counting analysis of \Sec{sec:masscute2e3} predicted that for $\beta$ near 2, the discrimination power of $\Cobs{2}{\beta}$ and $\Dobs{2}{\beta}$ should be comparable except at high signal efficiency when $\Dobs{2}{\beta}$ should be more discriminating.  To show that this is borne out in Monte Carlo, in \Fig{fig:m_dep} we first plot the rejection efficiency of $\Cobs{2}{1.7}$ and $\Dobs{2}{1.7}$ at 90\% signal efficiency, as a function of the lower mass cut on the jets.\footnote{We use $\beta = 1.7$ as this value was shown in \Ref{Larkoski:2013eya} to be the optimal choice for boosted $Z$ identification.}  When the lower mass cut is near zero, $\Dobs{2}{1.7}$ is significantly more efficient at rejecting QCD background than is $\Cobs{2}{1.7}$, as observed earlier.  As the lower mass cut increases, however, the difference in discrimination power between the two observables decreases in both \pythia{8} and \herwigpp~Monte Carlos. This dependence on the lower mass cut shows that $\Dobs{2}{\beta}$ captures the correct underlying physics of the $(\ecf{2,}{\beta},{\ecf{3}{\beta}})$ phase space, while $\Cobs{2}{\beta}$ does not. The light QCD jets that are added as the mass cut is lowered should be rejected by a variable that partitions the phase space into regions of 1-prong and 2-prong jets, increasing the observed rejection efficiency. This is true for $\Dobs{2}{\beta}$; however, exactly the opposite is true for $\Cobs{2}{\beta}$. 

\begin{figure}
\begin{center}
\subfloat[]{\label{fig:ROC_mcut_pythia_Log}
\includegraphics[width=6cm]{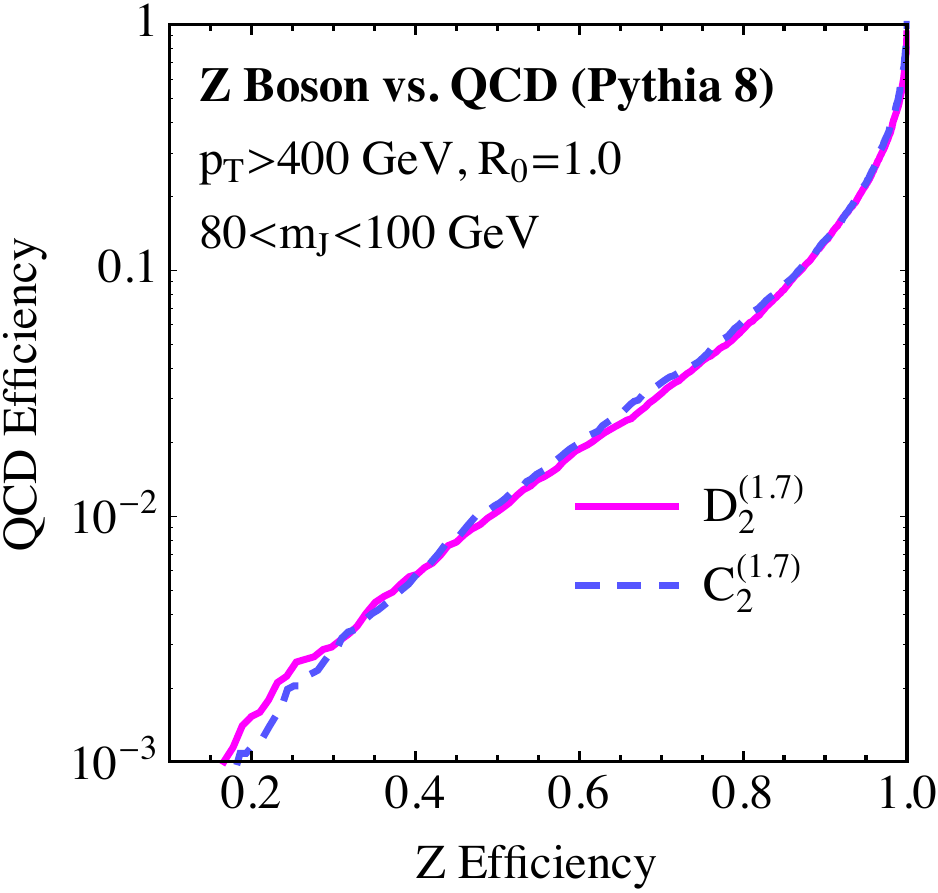}
}\qquad
\subfloat[]{\label{fig:ROC_mcut_Herwig_Log}
\includegraphics[width=6cm]{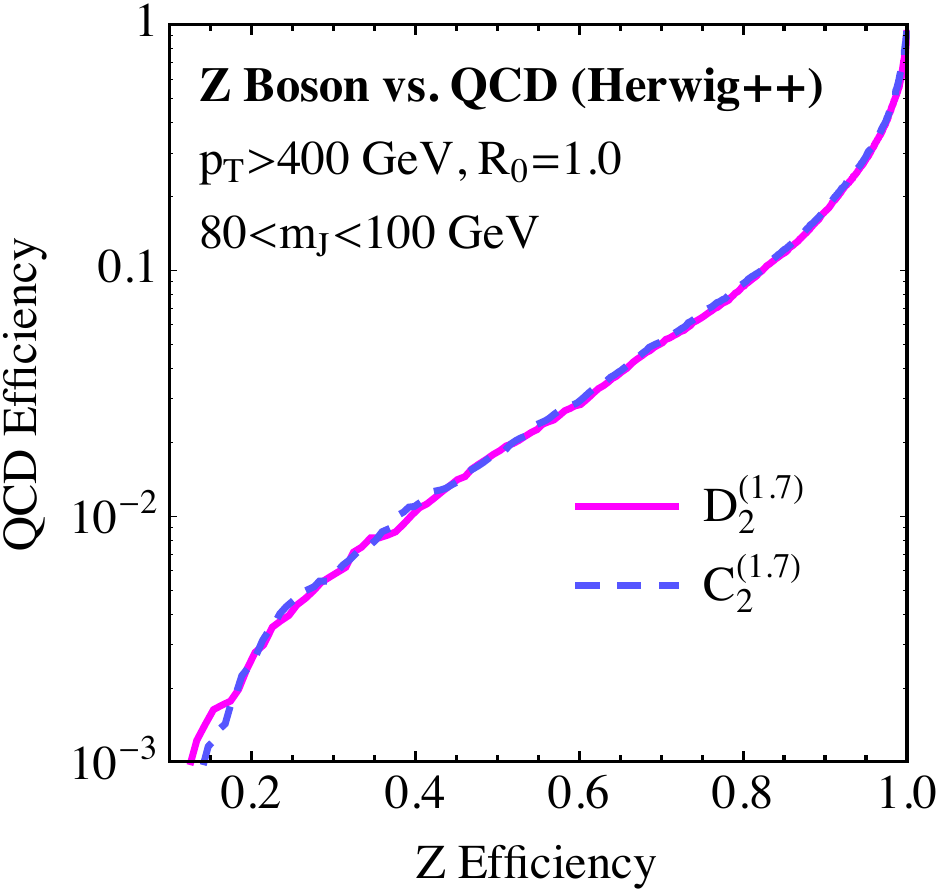}
}\qquad
\subfloat[]{\label{fig:ROC_mcut_pythia}
\includegraphics[width=6cm]{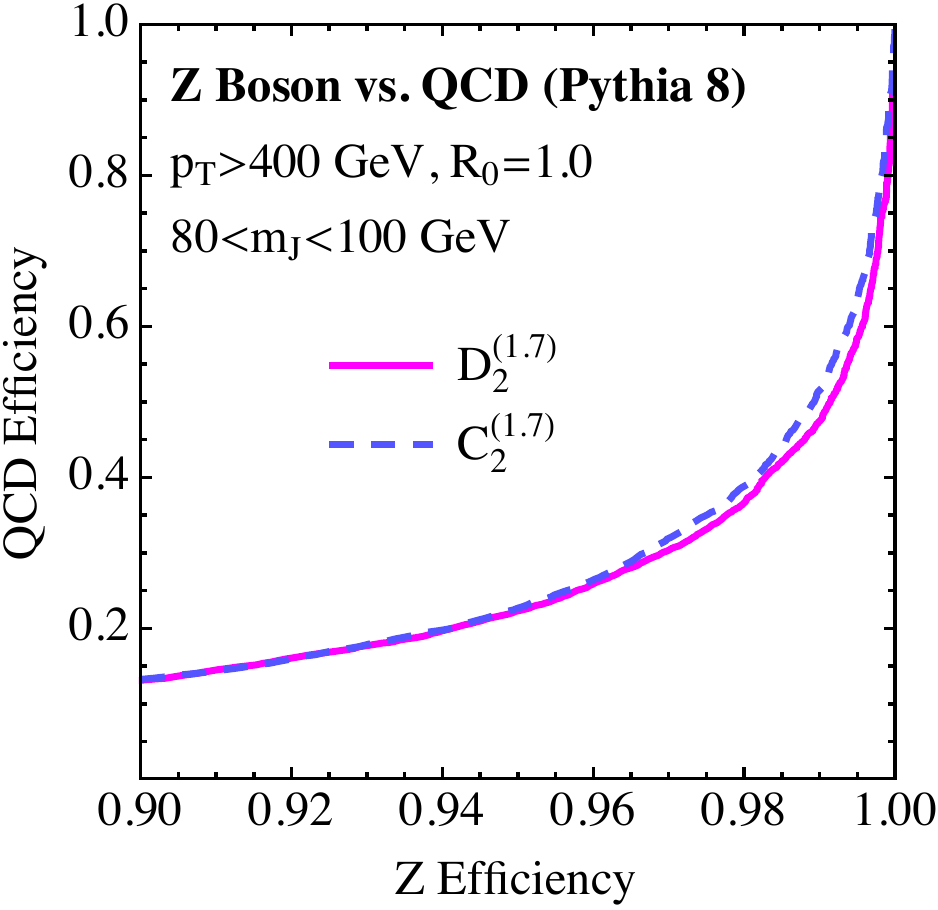}
}\qquad
\subfloat[]{\label{fig:ROC_mcut_Herwig}
\includegraphics[width=6cm]{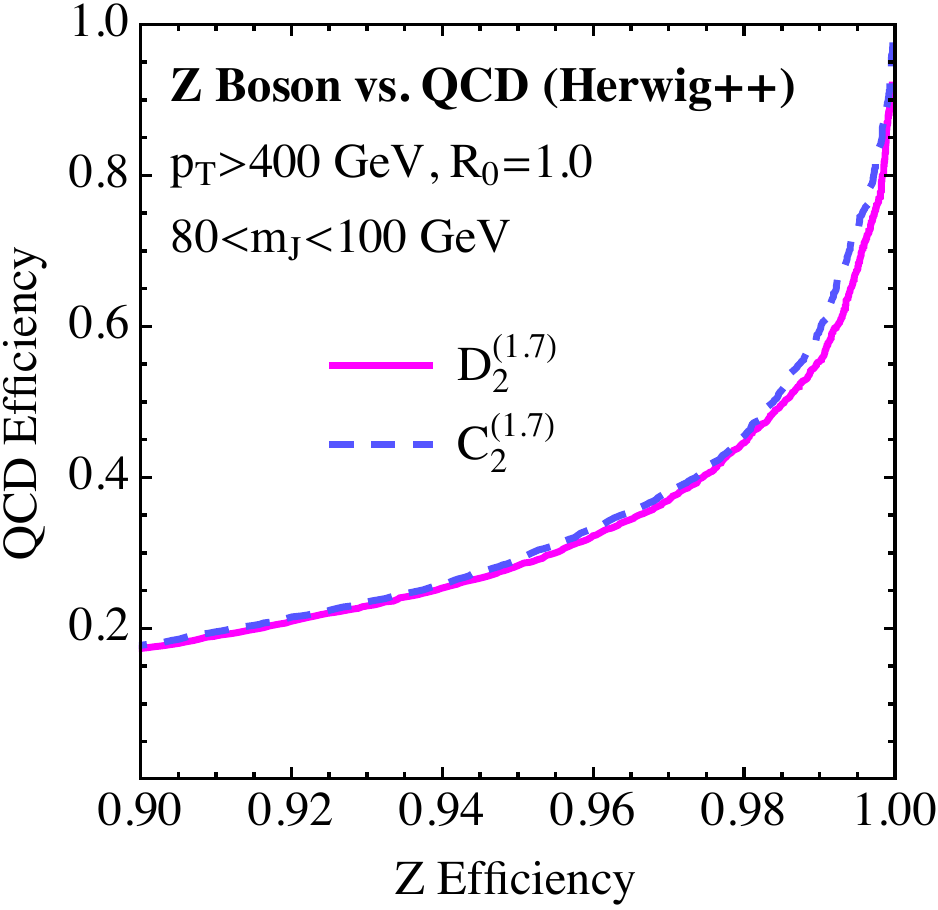}
}\qquad
\subfloat[]{
\includegraphics[width=6cm]{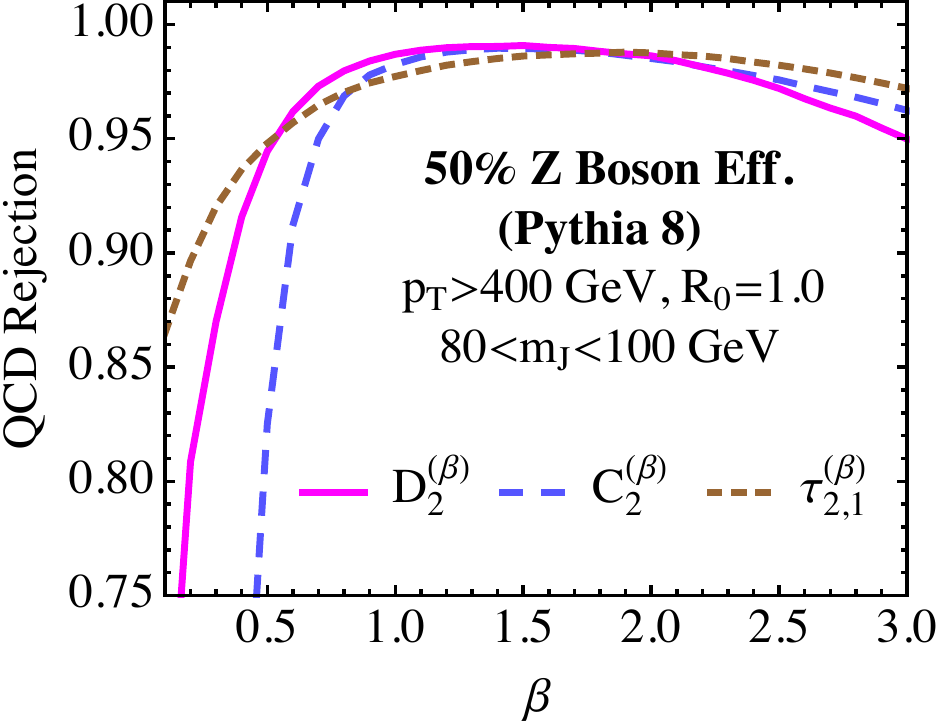}
}\qquad
\subfloat[]{
\includegraphics[width=6cm]{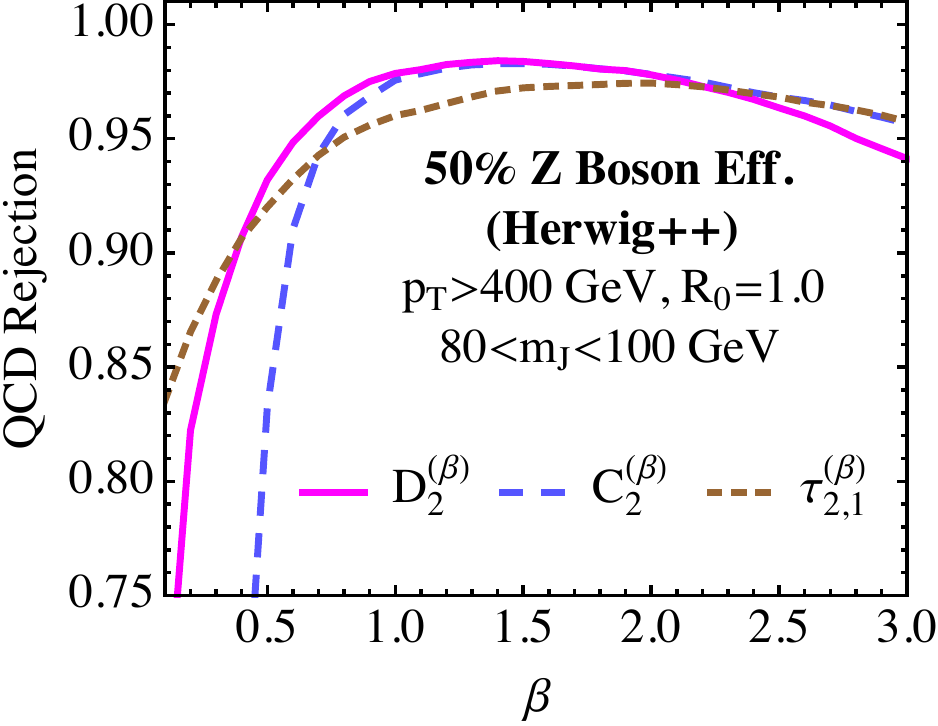}
}
\end{center}
\caption{ Comparison of $\Cobs{2}{\beta}$ and $\Dobs{2}{\beta}$ in the presence of a tight mass cut, $80<m_J<100$ GeV, with the \pythia{8} (left) and \herwigpp (right) samples. ROC curves for $\Cobs{2}{1.7}$ and $\Dobs{2}{1.7}$ demonstrate that with a tight mass cut, both observables perform comparably over a large range of signal efficiencies (top), with $\Dobs{2}{1.7}$ performing slightly better at high signal efficiencies (middle), behavior which is reproduced by both Monte Carlo generators. The QCD rejection rate at $50\%$ signal efficiency as a function of $\beta$ is shown at bottom for $\Cobs{2}{\beta}$, $\Dobs{2}{\beta}$ and $\Nsub{2,1}{\beta}$. 
}
\label{fig:ROC_mcut}
\end{figure}

We now study in more detail the case in which we have constrained the jet mass to lie in the tight mass cut window of $80 < m_J < 100$ GeV.  Over the whole signal efficiency range, $\Cobs{2}{1.7}$ and $\Dobs{2}{1.7}$ have nearly identical ROC curves in both \pythia{8} and \herwigpp, as exhibited in \Fig{fig:ROC_mcut}.  However, focusing in on the high signal efficiency region, we see that indeed $\Dobs{2}{1.7}$ has a slightly better rejection rate than $\Cobs{2}{1.7}$.  This behavior is manifest in both \pythia{8} and \herwigpp, showing that this prediction from the power counting analysis of \Sec{sec:masscute2e3} is robust to the precise details of the parton shower in the Monte Carlo generator. This should be contrasted with the actual numerical value of the QCD rejection, which depends on the generator.  For $\beta\simeq 2$ with a tight mass cut window of $80<m_J<100$ GeV, any discriminating variable of the form $\ecf{3}{\beta}/{\ecf{2}{\beta}}^n$, for $n>0$, provides reasonable discrimination power. The jet mass cut fixes $\ecf{2}{2}$ to a narrow window, and all discrimination power comes from $\ecf{3}{2}$ alone. This demonstrates why  \Ref{Larkoski:2013eya} observed near-optimal discrimination power using $\Cobs{2}{2}$, with $80<m_J<100$ GeV.

The final power counting prediction for the behavior of the observables was that the discrimination power of $\Dobs{2}{\beta}$ should be much more robust than $\Cobs{2}{\beta}$ as $\beta$ decreases from 2, with a mass cut on the jets.  This behavior is reproduced in Monte Carlo, as shown in \Fig{fig:ROC_mcut}, where we have plotted the QCD rejection efficiency at 50\% signal efficiency as a function of $\beta$.  We have also included in these plots the $N$-subjettiness ratio $\Nsub{2,1}{\beta}$ for comparison.  As $\beta\to 0$, the discrimination power of both $\Cobs{2}{\beta}$ and $\Dobs{2}{\beta}$ decrease; however, $\Dobs{2}{\beta}$ maintains high discrimination power to much smaller values of $\beta$ than $\Cobs{2}{\beta}$.  Nevertheless, note that as $\beta\to 0$, $\Nsub{2,1}{\beta}$, while not the optimal discrimination observable, has even more robust discrimination power than $\Dobs{2}{\beta}$.  An understanding of this behavior requires an analyses of ${\cal O}(1)$ numbers, which is beyond what power counting alone can predict.

Thus, we see that all the power counting predictions are realized in both Monte Carlo simulations, demonstrating that parametric scalings are indeed determining the behavior of the substructure observables. We emphasize that the level of agreement between the Monte Carlo generators for the power counting predictions is quite remarkable, given that numerical values for rejection or acceptance efficiencies, for example, do not agree particularly well between the generators. Power counting has allowed us to identify the robust predictions of perturbative QCD.

\subsection{Including Pile-Up}\label{sec:pu}

The power counting analysis of the previous section included only perturbative radiation.  At a high luminosity hadron collider such as the LHC, also important is the effect of multiple proton collisions per bunch crossing, referred to as pile-up.  Pile-up radiation is uncorrelated with the hard scattering event, and as such, has an energy scale that is independent of the hard parton collision energy.  Thus, pile-up can produce a significant amount of contaminating radiation in the event and substantially change jet $p_T$s, masses, or observables from their perturbative values.  An important problem in jet substructure is both to define observables that are less sensitive to the effects of pile-up, as well as to remove or ``groom'', to the greatest extent possible, radiation in a jet or event that most likely is from pile-up.  Several methods for jet grooming and pile-up subtraction have been presented  \cite{Butterworth:2008iy,Ellis:2009su,Ellis:2009me,Krohn:2009th,Soyez:2012hv,Krohn:2013lba,Larkoski:2014wba,Berta:2014eza,Cacciari:2014gra,Bertolini:2014bba}, and are used by the experiments \cite{CMS:2013uea,CMS-PAS-JME-10-013,TheATLAScollaboration:2013ria,TheATLAScollaboration:2013pia,ATLAS-CONF-2012-066}, but we will not consider them here.\footnote{The effects of jet grooming techniques can be understood using power counting techniques, and have been considered in \cite{Walsh:2011fz}.}  Instead, we will demonstrate that power counting can be used to understand the effect of pile-up radiation on the $(\ecf{2}{\beta} ,\ecf{3}{\beta})$ phase space, and therefore on signal and background distributions for observables formed from the energy correlation functions. We envision that similar techniques could be used to develop jet substructure variables with improved resilience to pile-up, but in this paper we will restrict ourselves to an understanding of the behavior of $\Cobs{2}{\beta}$ and $\Dobs{2}{\beta}$.\footnote{While the following analysis is quite general, it is restricted to recoil-free observables defined with a recoil-free jet algorithm, as used in this paper. In the case of a recoil sensitive observable, there is a non-linear response to pile-up due to the displacement of soft and collinear modes with respect to the jet axis. In this case the power counting analysis described here does not apply directly, and a more thorough analysis is required.}

To incorporate pile-up radiation into the power counting analysis, we must make some simplifying assumptions. Because pile-up is independent of the hard scattering event, we will assume that pile-up radiation is uniformly distributed over the jet area.\footnote{This model of pile-up would be removed by area subtraction \cite{Soyez:2012hv}. However, this would also remove perturbative soft radiation depending on the region of phase space. This could be studied in detail using power counting.}  This assumption essentially defines pile-up as another soft mode in the jet, with all angles associated with pile-up scaling as ${\cal O}(1)$. We will denote the $p_T$ fraction of pile-up radiation in the jet as
\begin{equation}
z_{pu} \equiv \frac{p_{T\, pu}}{p_{TJ}} \ .
\end{equation}
No assumption of the relative size of the perturbative soft radiation energy fraction $z_s$ with respect to $z_{pu}$ is made at this point, and indeed the impact of pile-up on the phase space will depend on this relation.

Assuming only that the pile-up $p_T$ fraction $z_{pu} \ll 1$, the two- and three-point correlation functions for 1-prong jets have the scaling
\begin{align}
\ecf{2}{\beta} & \sim  R_{cc}^\beta + z_s + z_{pu} \,, \\
\ecf{3}{\beta} & \sim R_{cc}^{3\beta}+z_s^2 + R_{cc}^\beta z_s +z_{pu}^2 + R_{cc}^\beta z_{pu} \,.
\end{align}
For 2-prong jets, the correlation functions have the scaling 
\begin{align}
\ecf{2}{\beta} & \sim  R_{12}^\beta + z_{pu} \,, \\
\ecf{3}{\beta} & \sim R_{12}^{\beta}z_s+R_{12}^{2\beta} R_{cc}^{\beta}+R_{12}^{3\beta}z_{cs} + R_{12}^\beta z_{pu} + z_{pu}^2\,.
\end{align}
From these scalings, we will be able to understand how pile-up radiation impacts jets in different regions of phase space, and hence the distributions in $\Cobs{2}{\beta}$ and $\Dobs{2}{\beta}$. Note that $z_{pu}$ is a fixed quantity measuring the fraction of pile-up radiation in the jet, and unlike the scalings for the soft, collinear and collinear-soft modes, its scaling is constant throughout the phase space. To understand the impact of the pile-up radiation on different regions of phase space, we will therefore need to understand how the values of $\ecf{2}{\beta}$ and $\ecf{3}{\beta}$ are modified by $z_{pu}$, depending on the different scalings of the contributing modes.

We begin the study of the phase space at small $\ecf{2}{\beta}$.  In the limit when $z_{pu} \gg  z_s$, pile-up dominates the structure of the jet.  In this limit, both 1-prong and 2-prong jets are forced into the region of phase space where $\ecf{3}{\beta} \sim (\ecf{2}{\beta})^2$. Note however that the scaling of the upper boundary of the phase space is robust. We must assume, as we will in what follows, that the value of $z_{pu}$ is such that this region does not extend far into the phase space, or else the energy correlation functions cannot be used to discriminate 1- and 2-prong jets, as their structure is completely dominated by pile-up radiation.

\begin{figure}
\begin{center}
\subfloat[]{
\includegraphics[width=6.5cm]{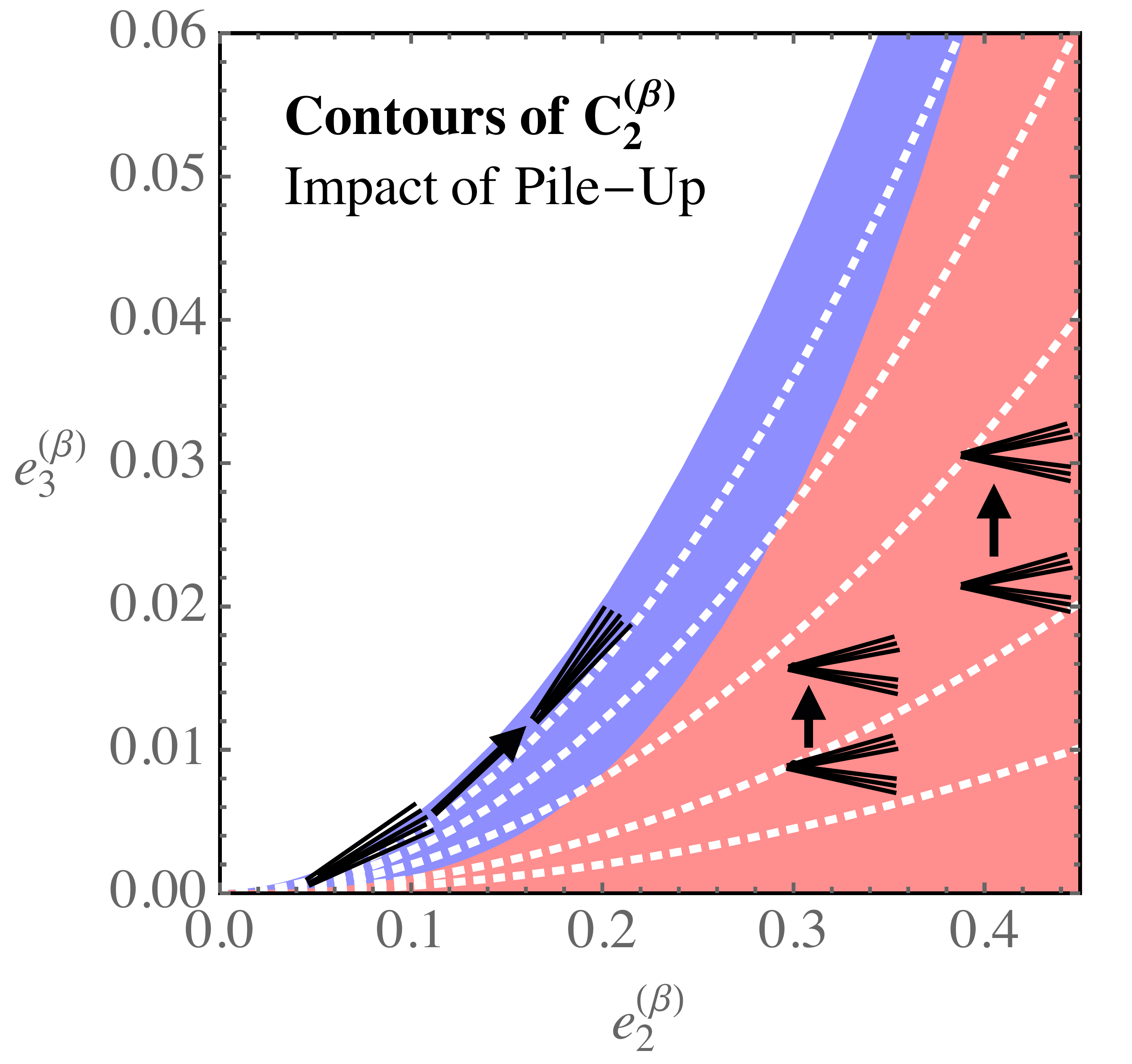}
}\qquad
\subfloat[]{
\includegraphics[width=6.5cm]{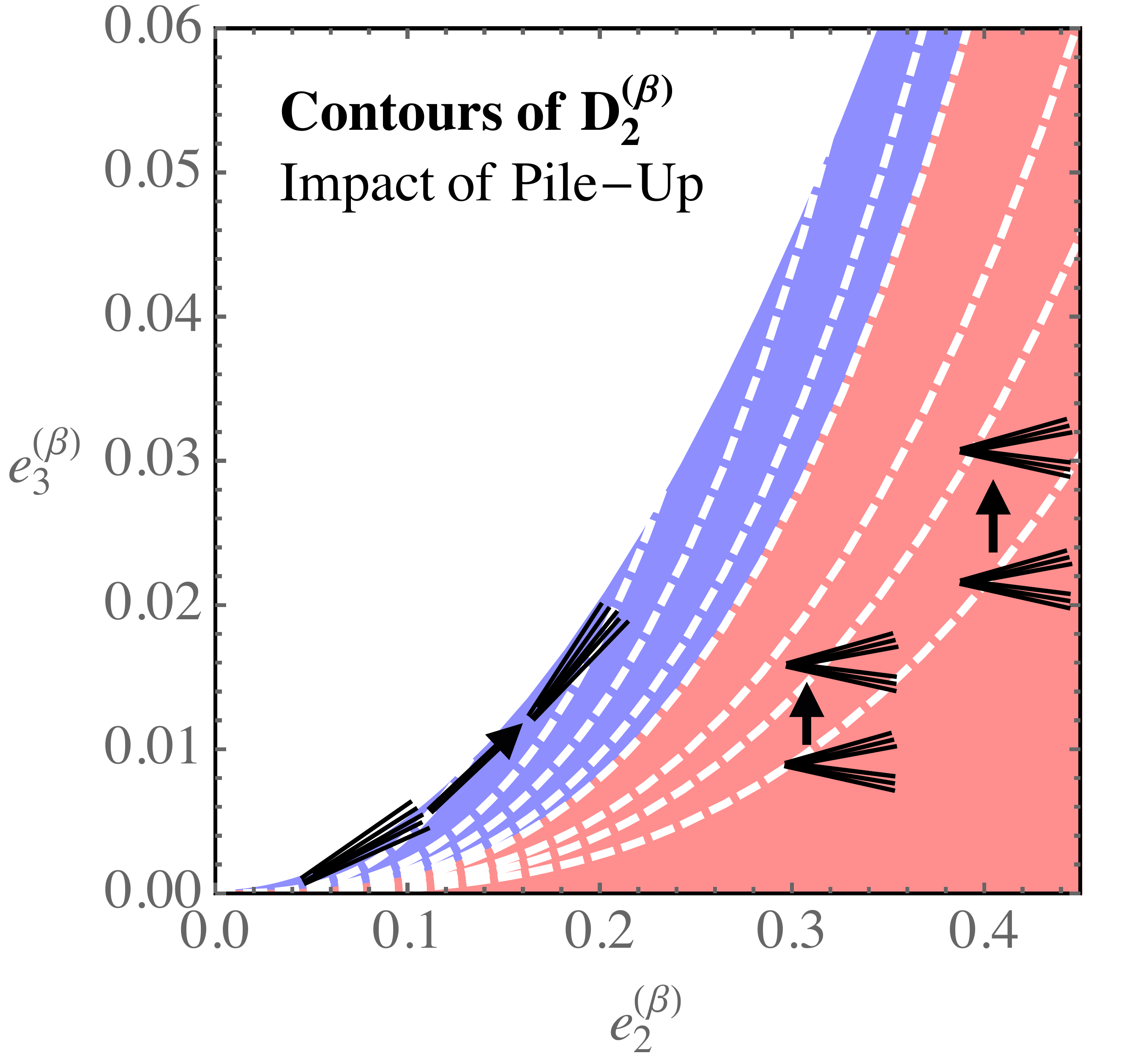}
}
\end{center}
\caption{Illustration of the effect of pile-up on the $\ecfnobeta{2} ,\ecfnobeta{3}$ phase space.   The 1- and 2-prong regions of phase space are denoted by blue or red, respectively, and the arrows show the direction that the jets move in the phase space with the addition of pile-up.  Contours of constant $\Cobsnobeta{2}$ (left) and $\Dobsnobeta{2}$ (right) are shown for reference.
}
\label{fig:ps_pu}
\end{figure}

Moving to slightly larger values of $\ecf{2}{\beta}$, we encounter a region dominated by 1-prong background jets, where $z_{pu} \sim z_s$. Under the addition of pile-up radiation, the two- and three-point correlation functions for 1-prong jets are modified as
\begin{align}
\ecf{2}{\beta} & \to \ecf{2}{\beta} + z_{pu} \,, \\
\ecf{3}{\beta} & \to \ecf{3}{\beta} + \ecf{2}{\beta} z_{pu} +z_{pu}^2 \,.
\end{align}
For $z_{pu}\sim z_s$, the addition of pile-up radiation therefore pushes all 1-prong jets towards the boundary $\ecf{3}{\beta} \sim (\ecf{2}{\beta})^2$. Jets that already satisfy this scaling, maintain it under the addition of pile-up, but move to larger values of $\ecf{2}{\beta}$. This behavior is illustrated in \Fig{fig:ps_pu}, and will imply a very different behavior for the distributions of $\Cobs{2}{\beta}$ and $\Dobs{2}{\beta}$  for background.

At larger values of $\ecf{2}{\beta}$, populated primarily by jets with two hard prongs, pile-up is a power-suppressed contribution to $\ecfnobeta{2}$, but still contributes to $\ecfnobeta{3}$.  That is, pile-up affects the two- and three-point correlation functions measured on 2-prong jets as
\begin{align}
\ecf{2}{\beta} & \to \ecf{2}{\beta} \,, \\
\ecf{3}{\beta} & \to \ecf{3}{\beta} + \ecf{2}{\beta} z_{pu}  \,.
\end{align}
Therefore, pile-up shifts 2-prong jets vertically in the $(\ecfnobeta{2} ,\ecfnobeta{3})$ phase space plane by an amount proportional to the perturbative value of $\ecf{2}{\beta}$.  This behavior is illustrated schematically in \Fig{fig:ps_pu}. At even larger values of $\ecf{2}{\beta}$, we enter a regime where $z_{pu} \ll z_s$. Here the scale of the pile-up radiation is parametrically smaller than the soft perturbative radiation, and so to leading power pile-up can be ignored.

We will now use this understanding of the effect of pile-up radiation on different regions of the $(\ecfnobeta{2} ,\ecfnobeta{3})$ phase space to understand its impact on the distributions for the observables $\Cobs{2}{\beta}$ and $\Dobs{2}{\beta}$ for both the signal and background. We begin by discussing the impact of pile-up radiation on the background distribution of $\Dobs{2}{\beta}$. Recall that at small $\ecf{2}{\beta}$, both $\ecf{2}{\beta}$ and $\ecf{3}{\beta}$ are shifted by the addition of the pile-up radiation, but maintain the parametric scaling $\ecf{3}{\beta} \sim (\ecf{2}{\beta})^2$. This has an interesting effect on the $\Dobs{2}{\beta}$ distribution due to the fact that the functional form of the contours, which are cubic, does not match the quadratic scaling of the upper boundary of the phase space.  The addition of pile-up pushes jets out of the small $\ecf{2}{\beta}$ region of phase space, where $\Dobs{2}{\beta}$ takes large values.  Therefore, an effect of pile-up is to reduce the value the $\Dobs{2}{\beta}$ measured on a jet, compressing the long tail of the perturbative $\Dobs{2}{\beta}$ distribution (exhibited in \Fig{fig:nomasscut}, for example) toward a central value.

We can predict the value of $\Dobs{2}{\beta}$ for background jets in the limit of infinite pile-up.  In this limit, the jet has a single hard core of radiation surrounded by perfectly uniform pile-up radiation. If the energy fraction of each of the $n$ pile-up particles is $z_{pu}$, then the two- and three-point correlation functions take the values
\begin{align}
\ecf{2}{\beta}=n z_{pu} \, , \qquad \text{and} \qquad \ecf{3}{\beta}={n\choose 2}z_{pu}^2\,,
\end{align}
so that, as $n\to \infty$, we have the relation
\begin{align}\label{eq:pure_pu_rel}
\ecf{3}{\beta}=\frac{1}{2}(\ecf{2}{\beta})^2\,. 
\end{align}
Using the definition of $\Dobs{2}{\beta}$, we find that in this limit,
\begin{align}
\Dobs{2}{\beta}=\frac{1}{2\ecf{2}{\beta}}\,.
\end{align}
As the amount of pile-up increases, we expect that the distribution of $\Dobs{2}{\beta}$ accumulates about this value, with a minimal change in the mean, but significant decrease in the width of the distribution. This  behavior is relatively distinct from that of most event shapes under pile-up, which tend to have a shift of the mean as pile-up is increased. The reason that this behavior is pronounced with $\Dobs{2}{\beta}$ is because it is both infrared and collinear unsafe and because the scalings of $\ecf{2}{\beta}$ and $\ecf{3}{\beta}$ in the observable and the upper boundary of the phase space are different. 

For the background distribution of $\Cobs{2}{\beta}$, on the other hand, the parametric scaling of the observable is unaffected by the addition of pile-up, but we expect an ${\cal O}(1)$ shift of the mean of the distribution to larger values. As pile-up increases, from \Eq{eq:pure_pu_rel} we expect the distribution should accumulate about the infinite pile-up limit of $\Cobs{2}{\beta}=1/2$.  We therefore predict that as the pile-up increases, the distribution of $\Cobs{2}{\beta}$ on background jets becomes independent of $\beta$.

We can also understand the behavior of the signal distribution under the addition of soft pile-up radiation. As was discussed, and is shown schematically in \Fig{fig:ps_pu}, signal jets at larger $\ecf{2}{\beta}$ are shifted vertically in the $(\ecf{2}{\beta}$, $\ecf{3}{\beta})$ phase space by pile-up radiation. This predicts that for both $\Cobs{2}{\beta}$ and $\Dobs{2}{\beta}$, the primary effect of the addition of pile-up radiation will be to shift the mean of the distribution to larger values, with a limited modification to its shape. Furthermore, due to the cubic contours for $\Dobs{2}{\beta}$, the shift of the mean of the distribution will be smaller for $\Dobs{2}{\beta}$ than for $\Cobs{2}{\beta}$, implying reduced sensitivity of $\Dobs{2}{\beta}$ to pile-up radiation.

Note that a similar analysis can be straightforwardly applied to the $N$-subjettiness observables $\Nsub{1}{\beta}$ and $\Nsub{2}{\beta}$. Because the analysis proceeds identically, we simply state the result. Under the addition of pile-up radiation, single prong jets at small $\Nsub{1}{\beta}$ experience a shift of both observables, but their parametric scaling remains the same: 
\begin{align}
\Nsub{1}{\beta} & \to  \Nsub{1}{\beta}+z_{pu} \,, \\
\Nsub{2}{\beta} & \to \Nsub{2}{\beta}+z_{pu}\,.
\end{align}
That is, under the addition of pile-up, background jets move along the upper boundary of the phase space, where $\Nsub{2}{\beta}\sim\Nsub{1}{\beta}$.

For jets with two hard subjets, the value of $\Nsub{1}{\beta}$ is not affected, while $\Nsub{2}{\beta}$ shifts as
\begin{align}
\Nsub{1}{\beta} &\to \Nsub{1}{\beta} \,, \\
\Nsub{2}{\beta} &\to \Nsub{2}{\beta}+z_{pu}\, .
\end{align}
This corresponds to a vertical movement in the $\Nsub{1}{\beta},\Nsub{2}{\beta}$ phase space under the addition of pile-up, as was the case for $\ecf{2}{\beta} ,\ecf{3}{\beta}$. We therefore expect a similar behavior for $\Nsub{2,1}{\beta}$ with the addition of pile-up, with a shift of the mean value for the signal distributions and accumulation near $1$ for the background distributions. Unfortunately, power counting alone does not allow us to compare the expected shifts in $\Nsub{2,1}{\beta}$ as compared with those in $\Cobs{2}{\beta}$ and $\Dobs{2}{\beta}$.

\subsubsection{Summary of Power Counting Predictions}
\label{sec:sumpredict_pu}

Here, we summarize the main predictions from our power counting analysis of the impact of pile-up radiation on the $\Cobs{2}{\beta}$ and $\Dobs{2}{\beta}$ distributions. We have:
\begin{itemize}

\item For background $\Dobs{2}{\beta}$ distributions the primary effect of the addition of pile-up radiation is to narrow the distribution; in particular, the long tail of the $\Dobs{2}{\beta}$ distribution is truncated because pile-up moves those jets in the 1-prong region of phase space out of the region of small $e_2$. The peak of the $\Dobs{2}{\beta}$ background distribution should be relatively insensitive to the addition of pile-up, with $\Dobs{2}{\beta}$ accumulating around $\Dobs{2}{\beta} = 1/(2 \ecf{2}{\beta})$ in the limit that uniform pile-up dominates. 

\item For background $\Cobs{2}{\beta}$ distributions the primary effect of the addition of pile-up radiation is a shift of the peak to larger values by an $\mathcal{O}(1)$ amount proportional to the pile-up. The distribution also becomes compressed, and accumulates around $\Cobs{2}{\beta}= 1/2$ in the limit that uniform pile-up dominates.

\item For signal, the primary effect of the addition of pile-up radiation is to translate the mean of the distribution. The displacement of the mean is expected to be smaller for $\Dobs{2}{\beta}$ than for $\Cobs{2}{\beta}$ because of the different scalings for contours of constant $\Cobs{2}{\beta}$ and $\Dobs{2}{\beta}$.

\end{itemize}

\subsubsection{Monte Carlo Analysis}
\label{sec:mc_e2e3_pu}

We now study these predictions in Monte Carlo using the \pythia{8} event samples described in \Sec{sec:mc_e2e3}.\footnote{In this section, we restrict to the \pythia{8} generator, having satisfied ourselves in \Sec{sec:mc_e2e3} that the parametrics of the perturbative phase space are well described by both generators.}  Pile-up was simulated by adding $N_{PV}$ minimum bias events at the 8 TeV LHC, generated with \pythia{8}, to the $pp\to Zj$ and $pp\to ZZ$ samples. To demonstrate the resilience of the distributions to pile-up, we wish to add pile-up radiation to a set of jets with well-defined perturbative properties.  To do this, we cluster jets with the WTA recombination scheme \cite{Larkoski:2014uqa,Larkoski:2014bia} and require that the mass of the jets in the absence of pile-up is $m_J<100$ GeV.  It was shown in \Ref{Larkoski:2014bia} that the jet axis found by the WTA recombination scheme is robust to pile-up and so, when pile-up is included, the perturbative content of the jets will be unaffected. This procedure, although clearly not related to an experimental analysis, provides a measure of the sensitivity of the distributions to soft pile-up radiation. This procedure is similar to that used in \Ref{Soyez:2012hv} to assess the impact of pile-up and pile-up subtraction techniques on a variety of different jet shapes. 

\begin{figure}
\begin{center}
\subfloat[]{
\includegraphics[width=6.5cm]{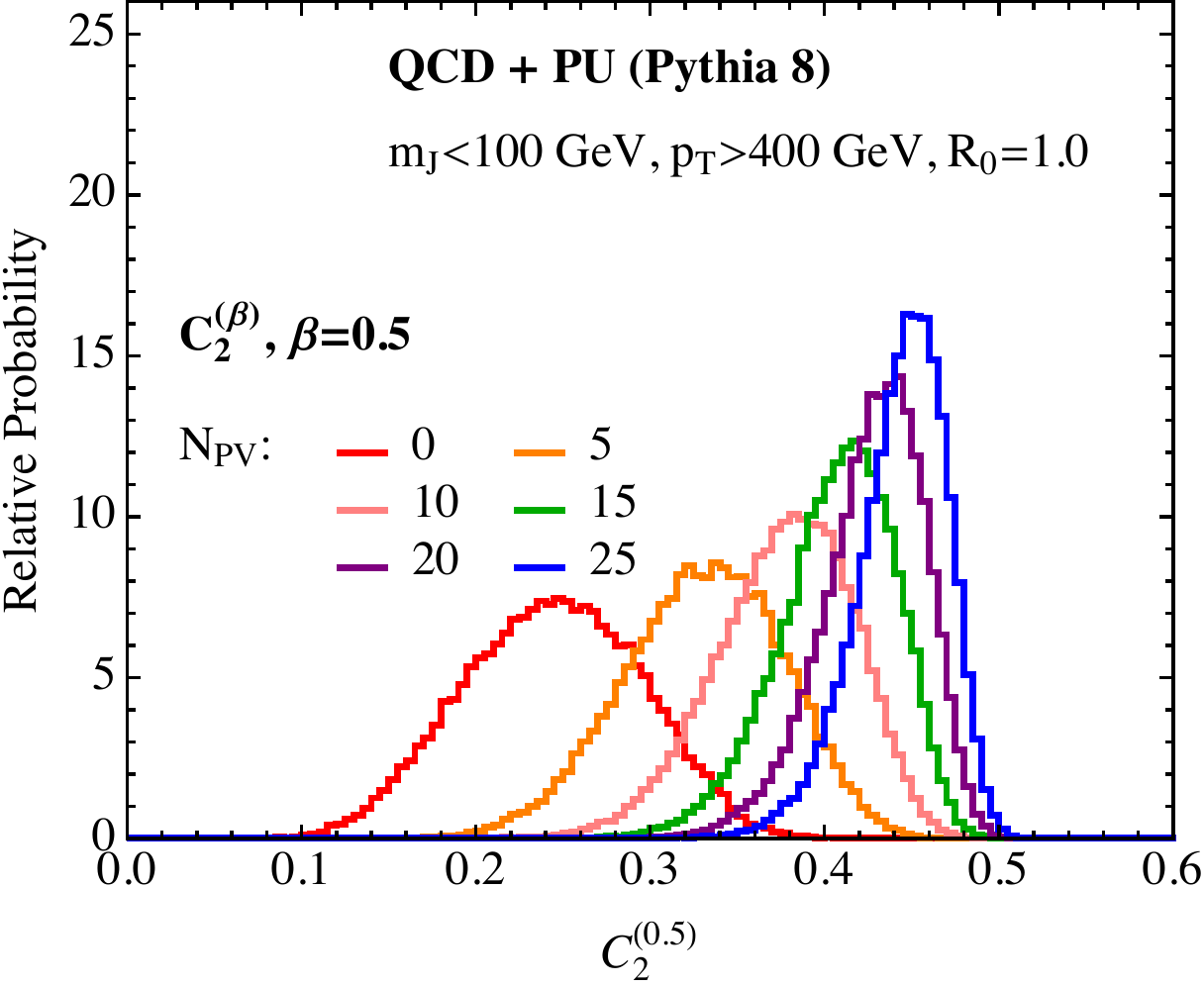}
}\qquad
\subfloat[]{
\includegraphics[width=6.5cm]{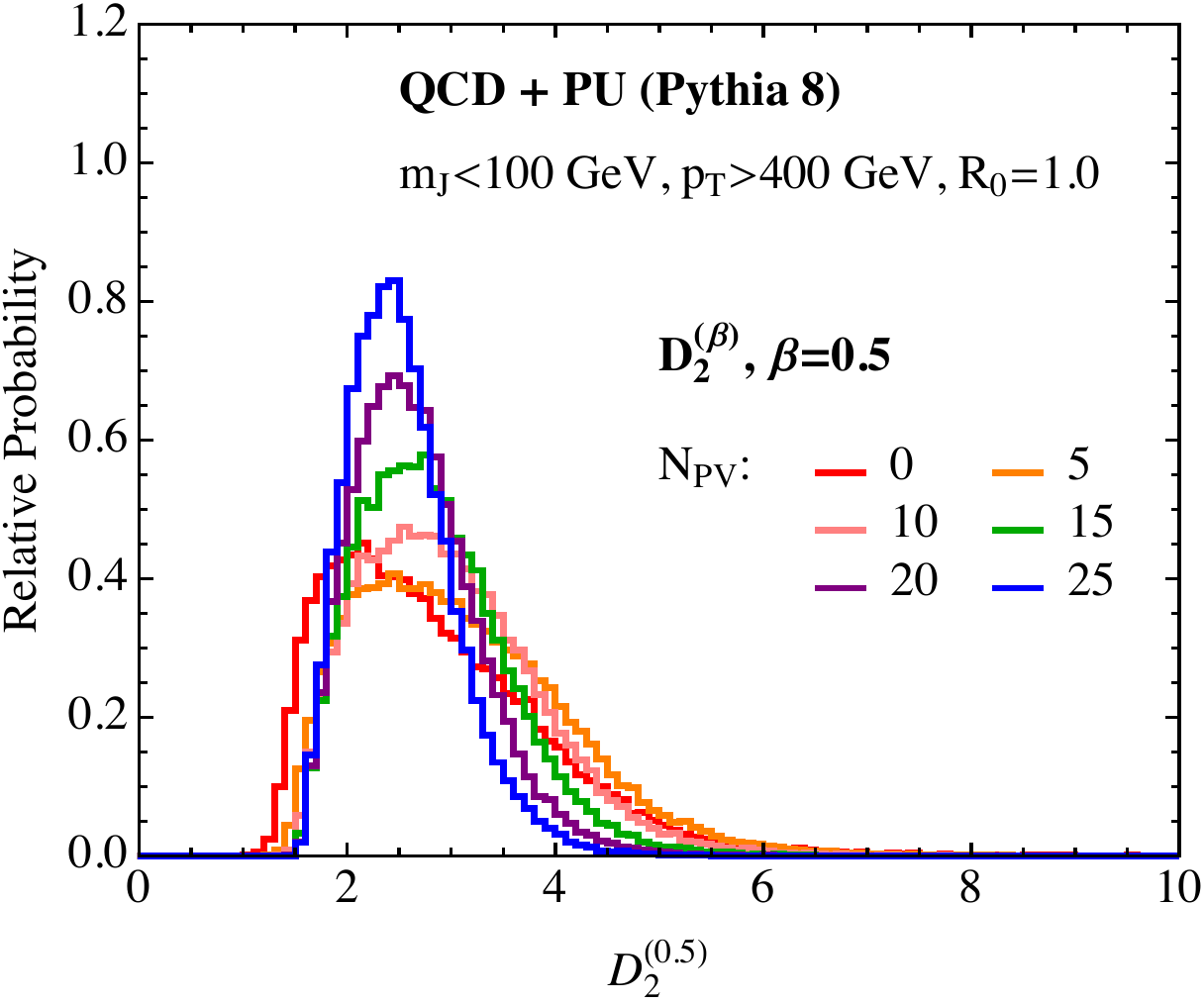}
}\qquad
\subfloat[]{
\includegraphics[width=6.5cm]{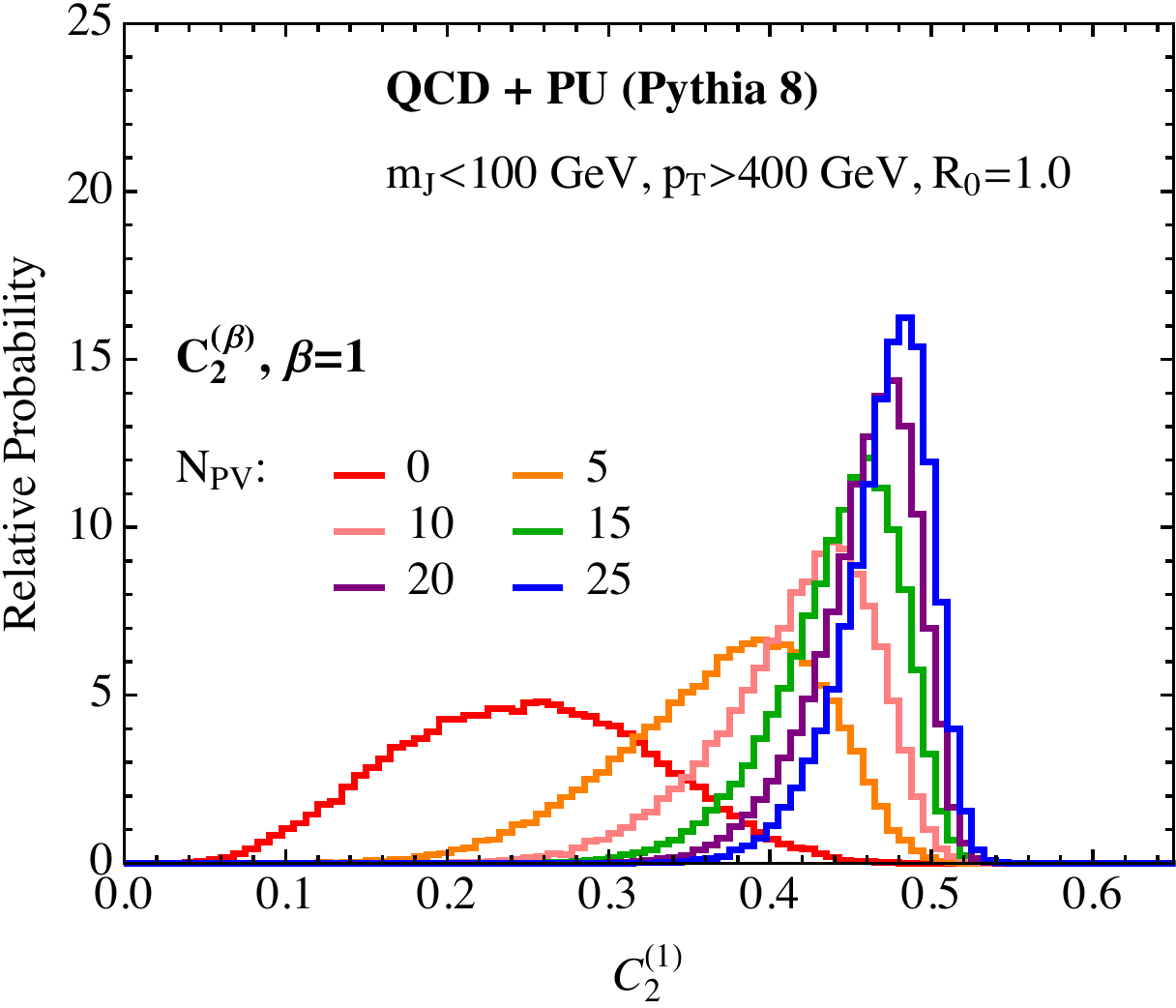}
}\qquad
\subfloat[]{
\includegraphics[width=6.5cm]{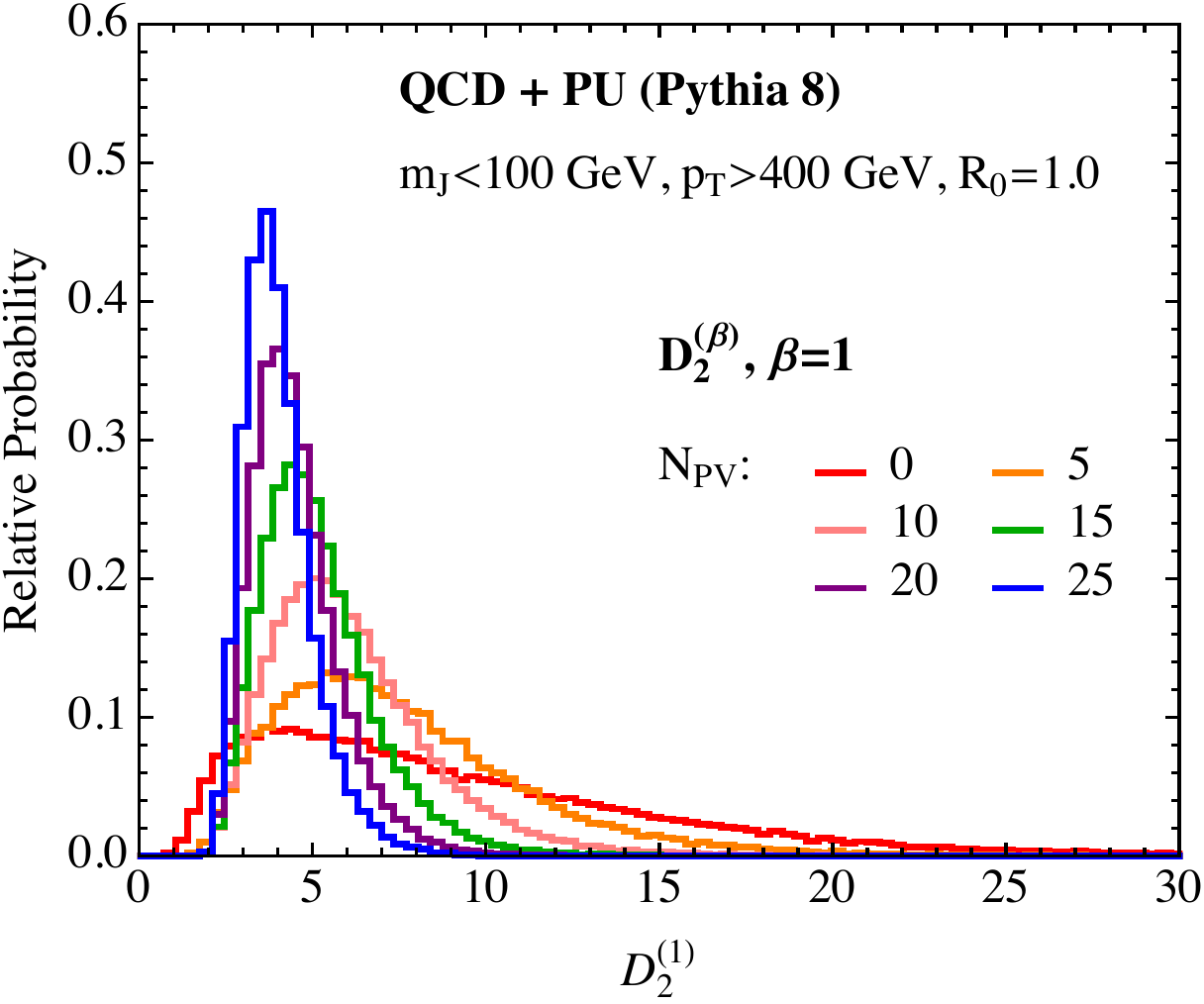}
}\qquad
\subfloat[]{
\includegraphics[width=6.5cm]{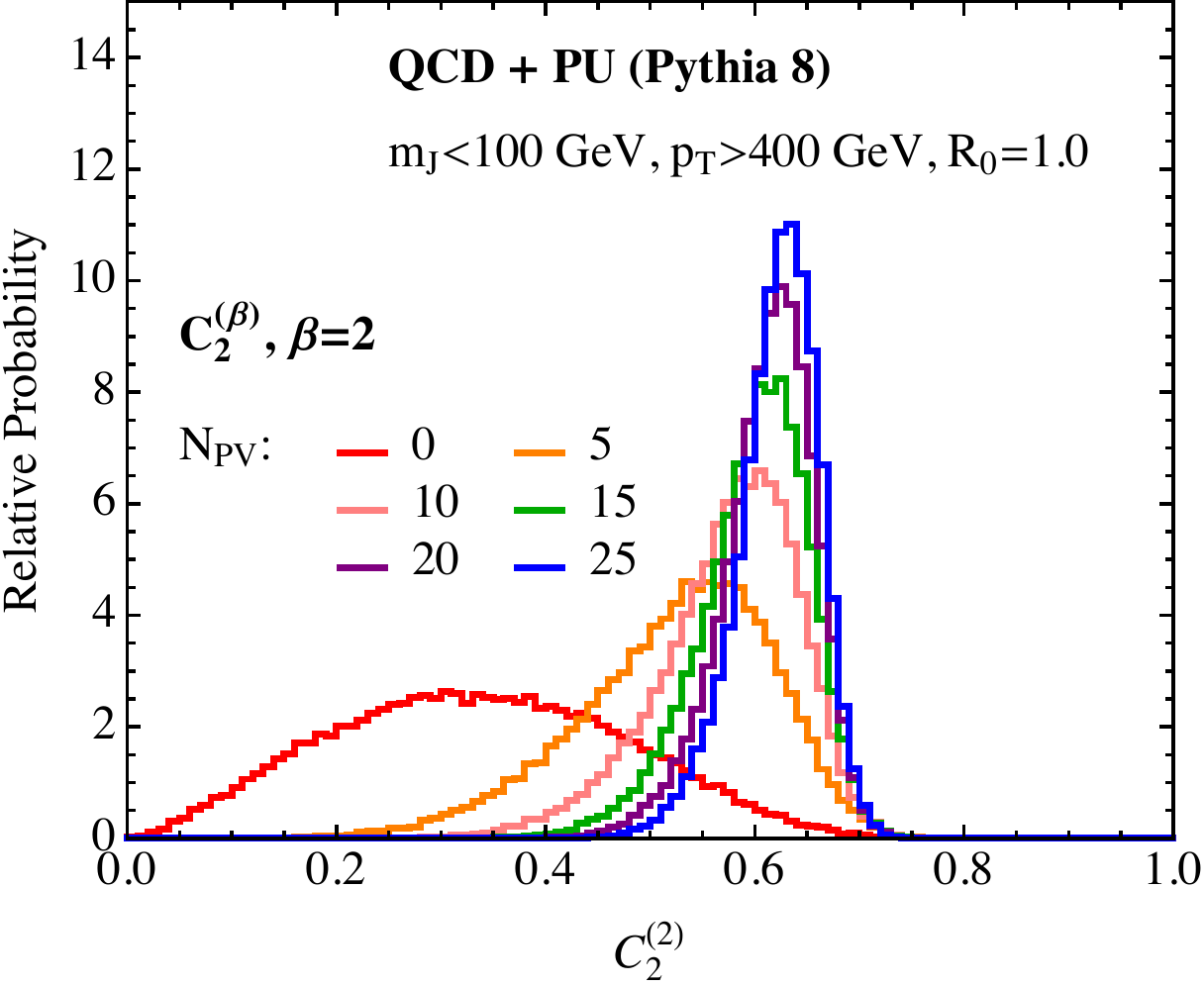}
}\qquad
\subfloat[]{
\includegraphics[width=6.5cm]{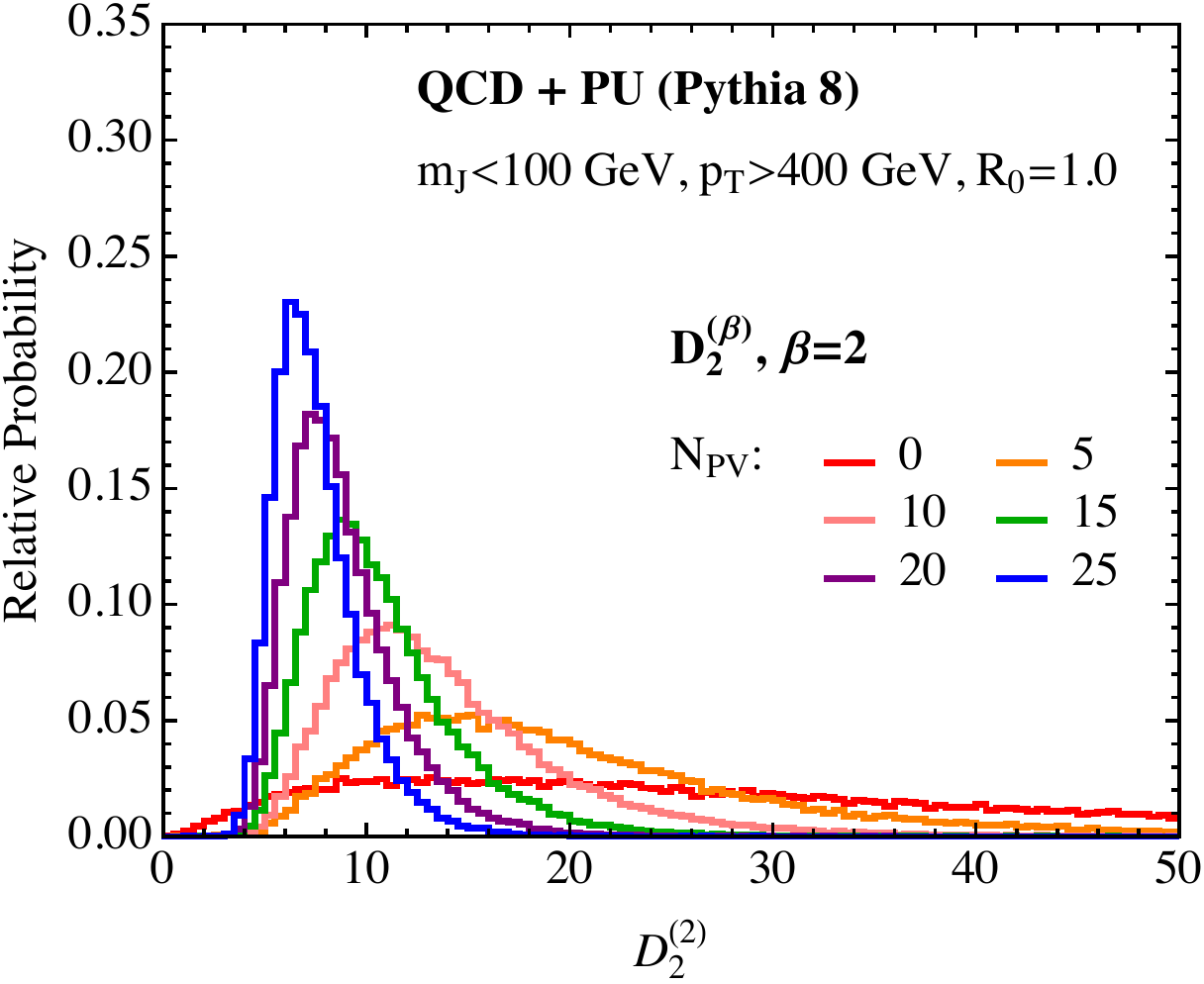}
}
\end{center}
\caption{Effect of pile-up on the distributions of $\Cobs{2}{\beta}$ (left) and $\Dobs{2}{\beta}$ (right) for QCD jets for $\beta = 0.5,1,2$ as measured on the \pythia{8} samples.  The number of pile-up vertices ranges from $N_{PV} = 0 $ (no pile-up) to $ N_{PV}=25$.
}
\label{fig:Bkg_pu}
\end{figure}

\begin{figure}
\begin{center}
\subfloat[]{
\includegraphics[width=6.5cm]{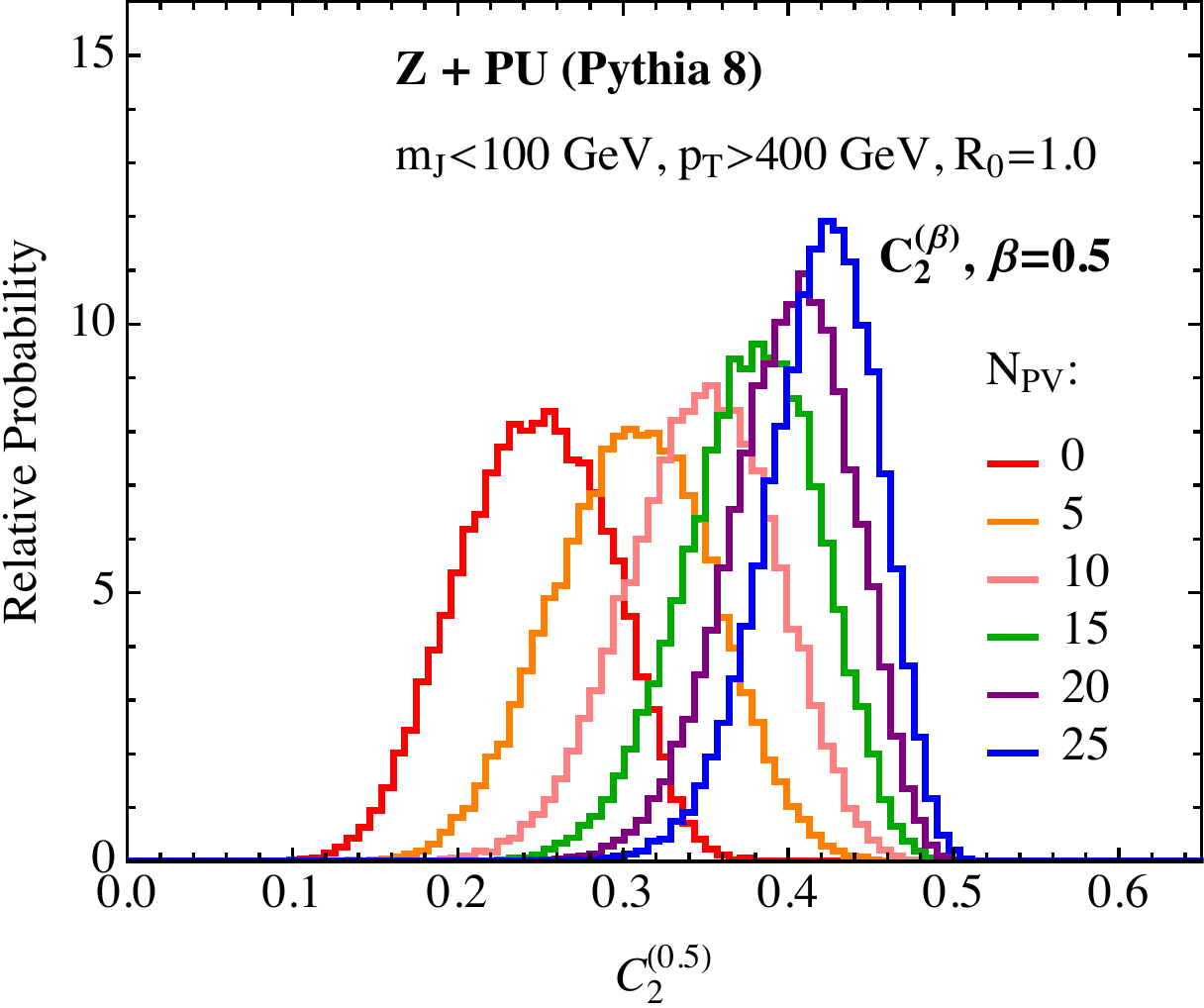}
}\qquad
\subfloat[]{
\includegraphics[width=6.5cm]{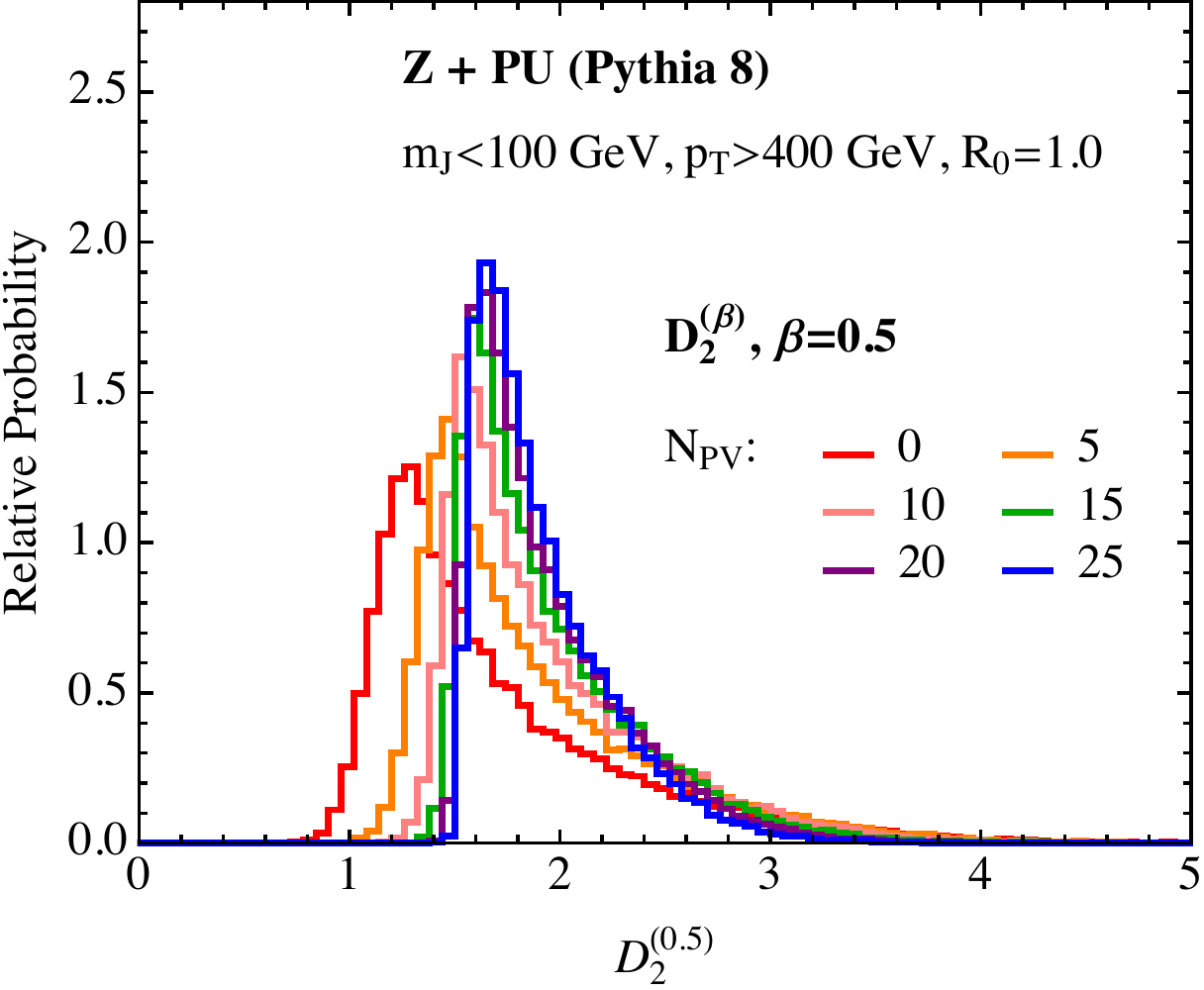}
}\qquad
\subfloat[]{
\includegraphics[width=6.5cm]{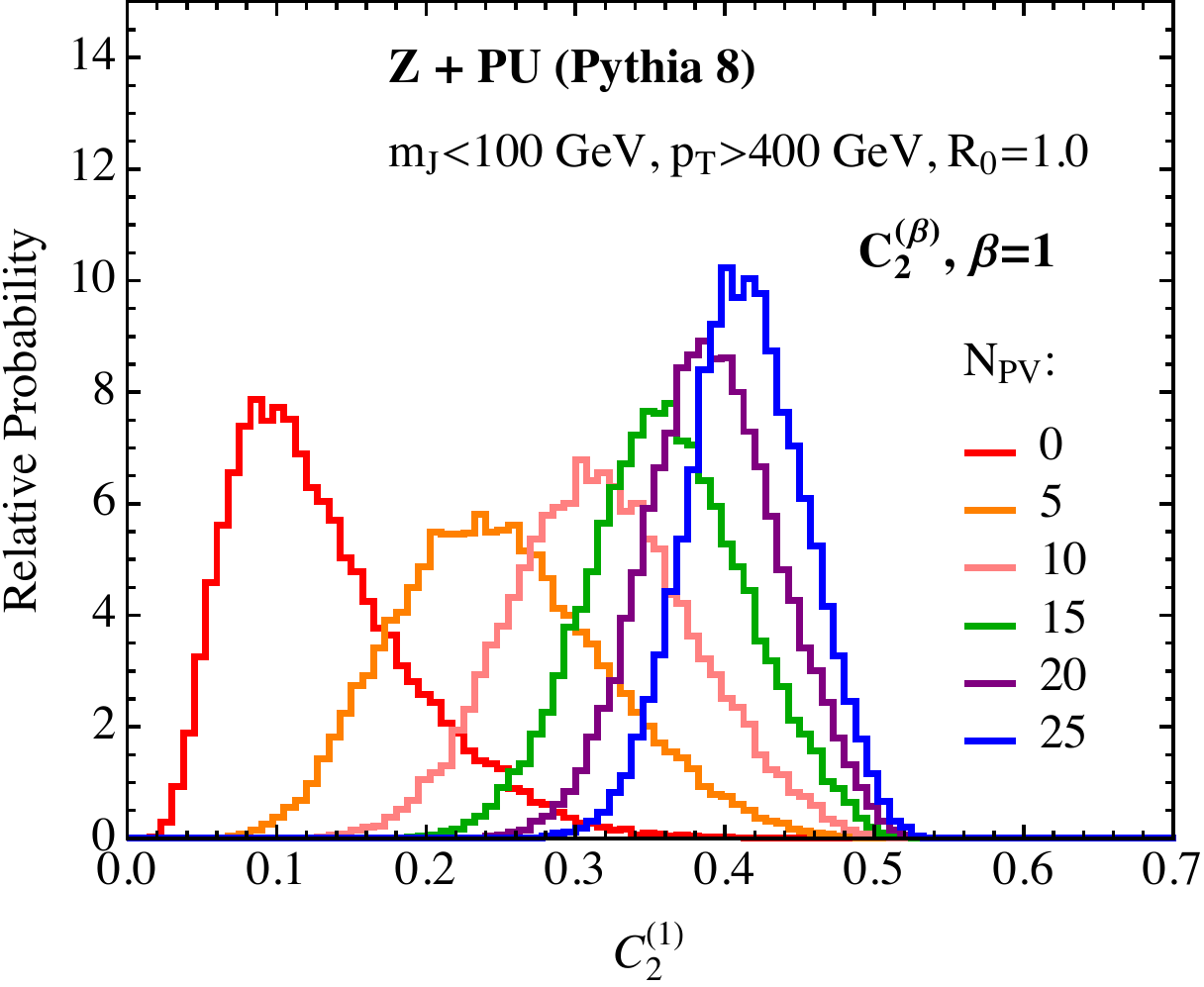}
}\qquad
\subfloat[]{
\includegraphics[width=6.5cm]{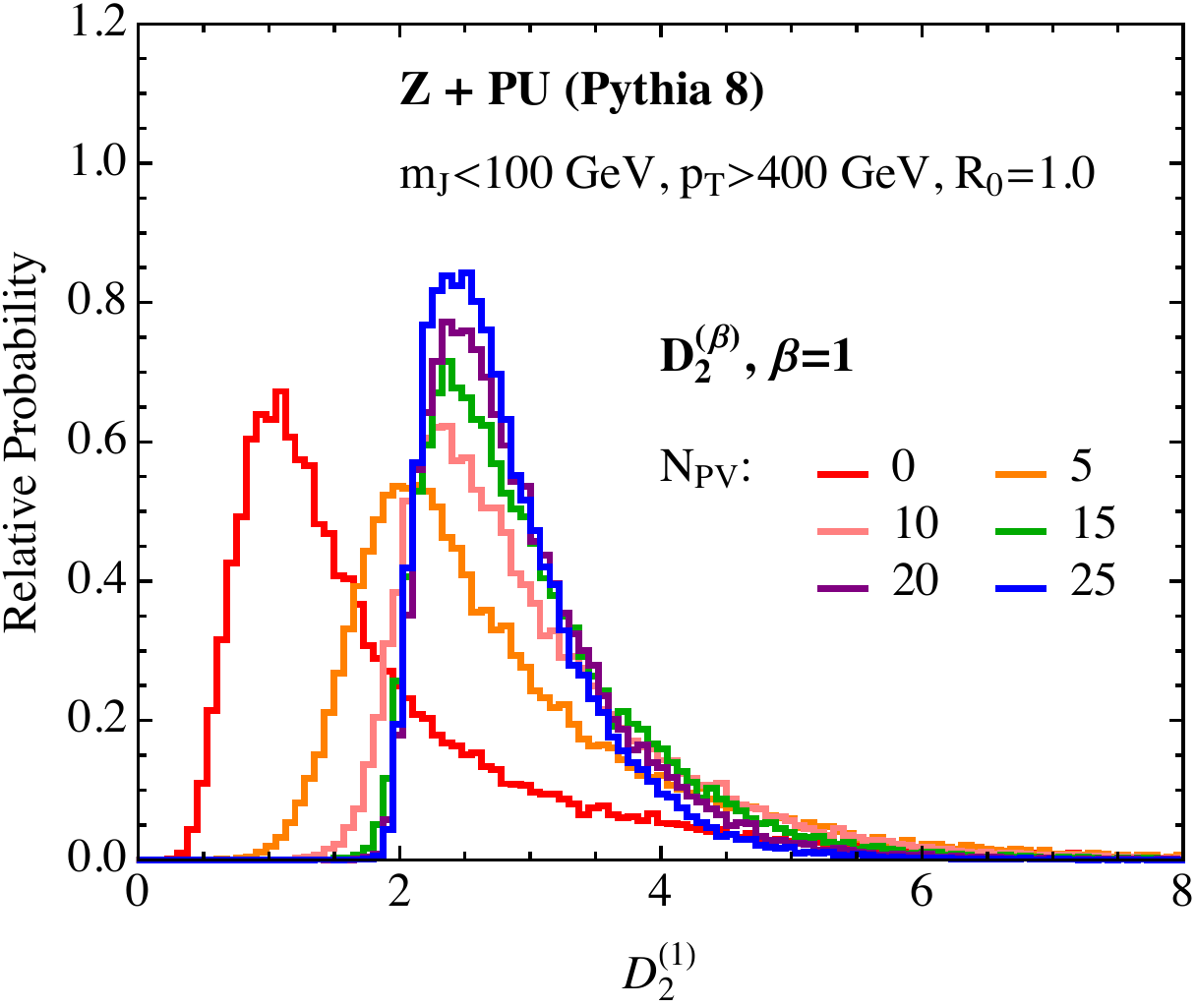}
}\qquad
\subfloat[]{
\includegraphics[width=6.5cm]{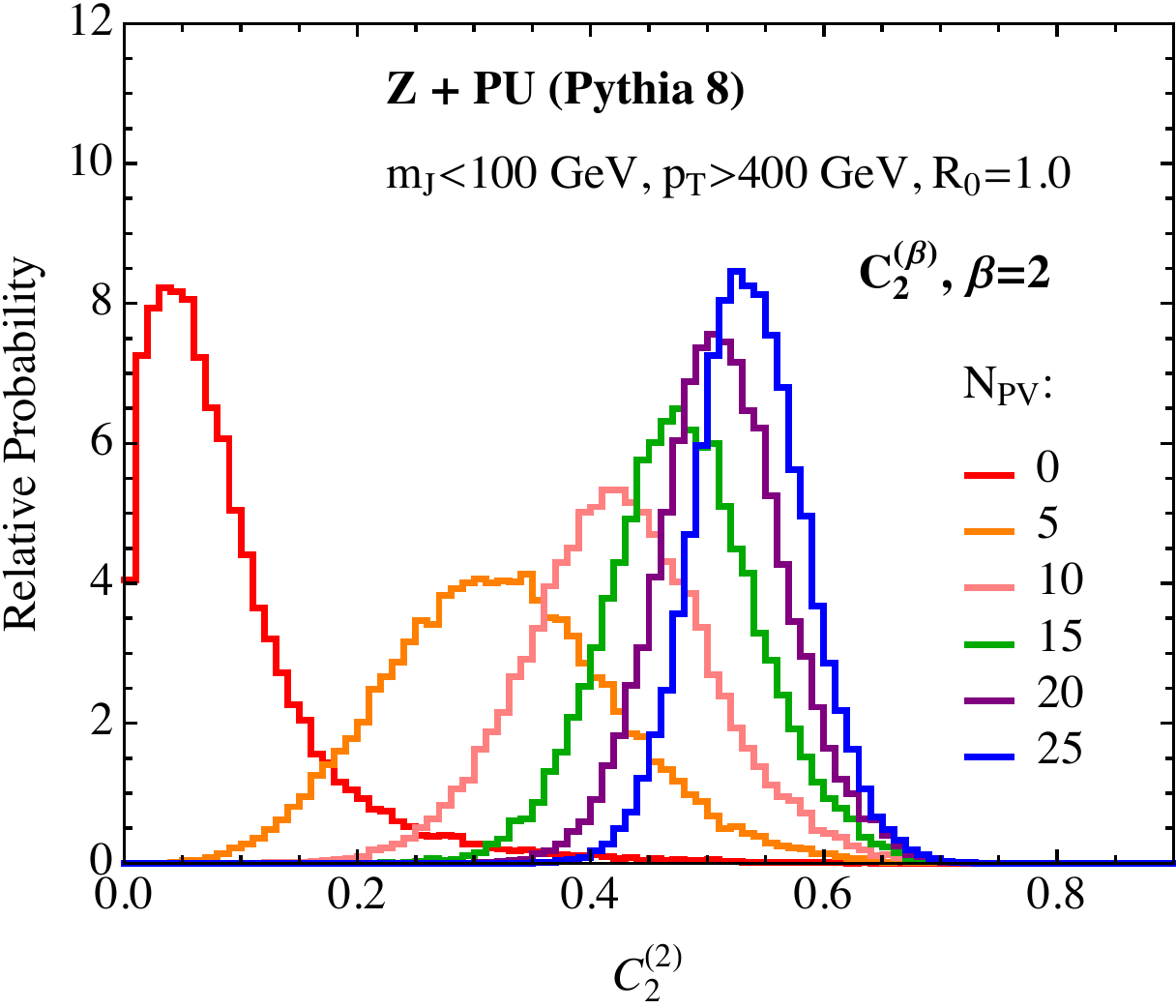}
}\qquad
\subfloat[]{
\includegraphics[width=6.5cm]{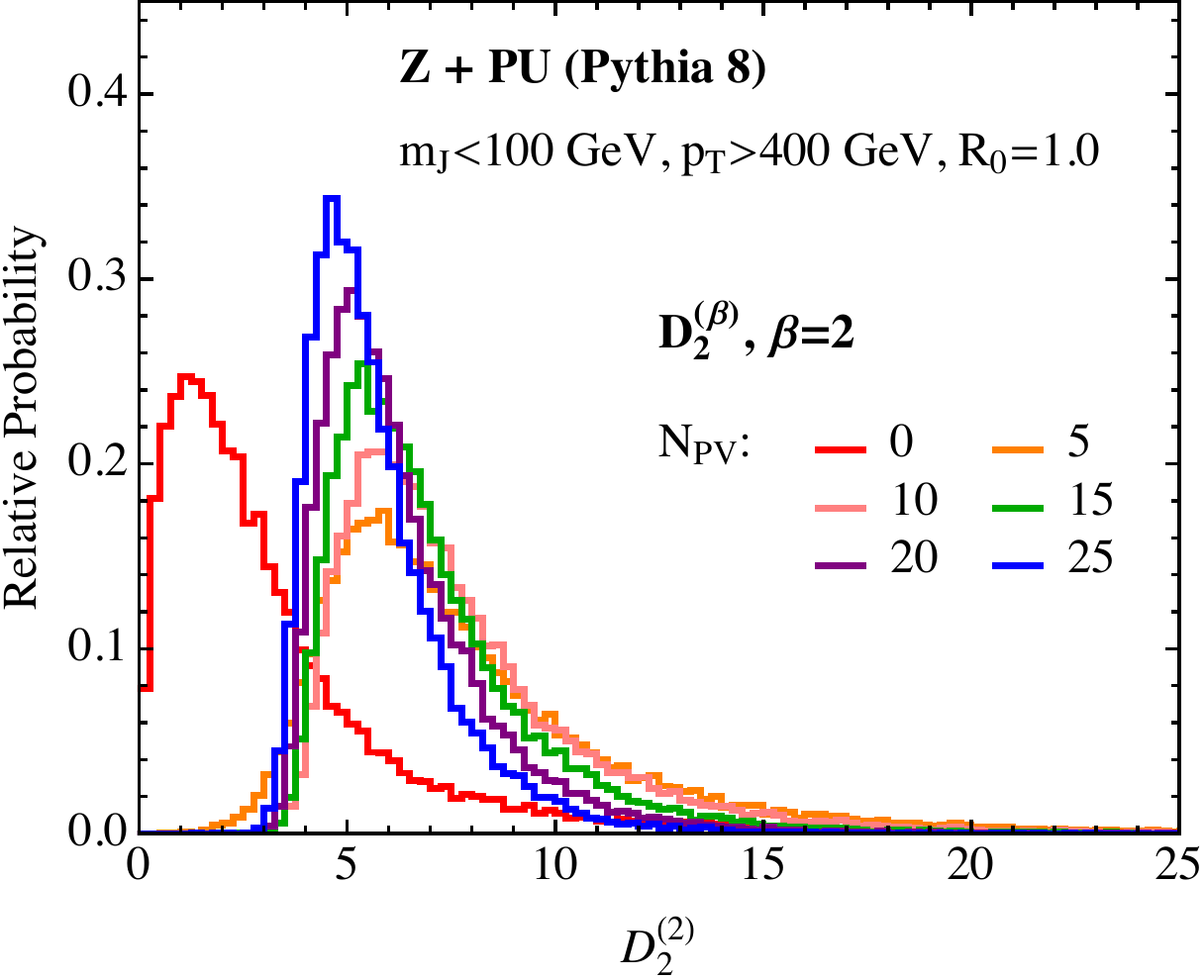}
}
\end{center}
\caption{
The same as \Fig{fig:Bkg_pu}, but measured on boosted $Z$ jets.
}
\label{fig:Sig_pu}
\end{figure}

\begin{figure}
\begin{center}
\subfloat[]{
\includegraphics[width=6.5cm]{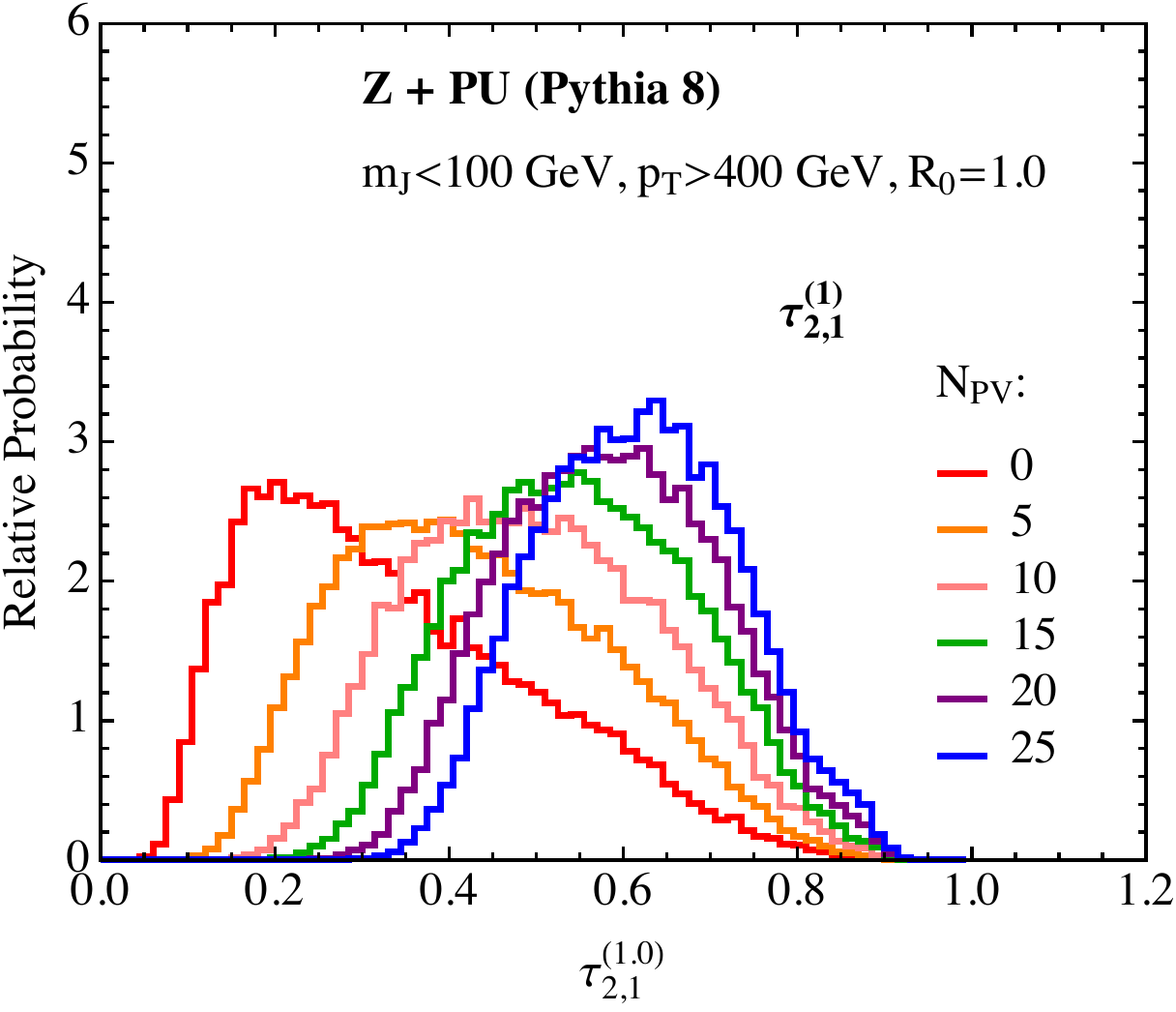}
}\qquad
\subfloat[]{
\includegraphics[width=6.5cm]{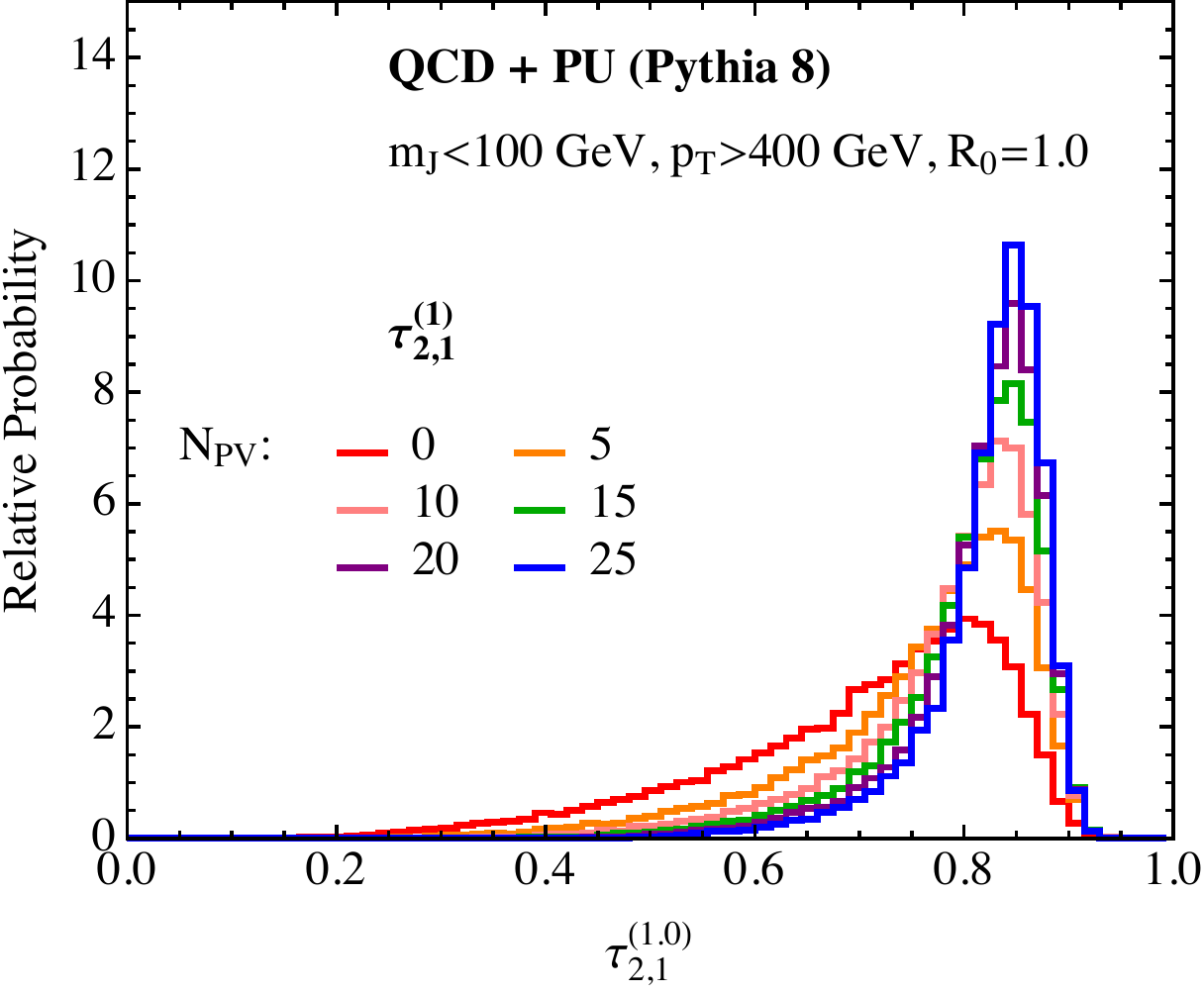}
}
\end{center}
\caption{Effect of pile-up contamination on the measured value of $\Nsub{2,1}{1}$ for signal (left) and background (right) jets from the \pythia{8} samples.  The number of pile-up vertices ranges from $N_{PV}=0$ (no pile-up) to $N_{PV} = 25$. 
}
\label{fig:nsub_pu}
\end{figure}

Using this sample, we can assess the degree to which the power counting predictions of \Sec{sec:pu} are realized in the Monte Carlo simulation. We begin by considering the effect of pile-up on the background distributions. In \Fig{fig:Bkg_pu}, we plot background distributions for $\Cobs{2}{\beta}$ and $\Dobs{2}{\beta}$ with the addition of up to $N_{PV}=25$ pile-up vertices for a few values of $\beta$. For the variable $\Dobs{2}{\beta}$, the power counting analysis of \Sec{sec:pu} predicted that the dominant effect of the addition of pile-up would be a compression of the long tail of the distribution into a peak around $1/2\ecf{2}{\beta}$, with relatively little shift in the mean. This behavior is manifest for all three values of $\beta$ shown. The peak value of the distribution is remarkably stable under the addition of pile-up. On the other hand, for the observable $\Cobs{2}{\beta}$, the mean of the distribution is highly unstable to the addition of pile-up. The dominant effects of pile-up on the distribution of $\Cobs{2}{\beta}$ is a displacement of the mean and the accumulation near the value $\Cobs{2}{\beta}=1/2$. 

In \Fig{fig:Sig_pu} we consider the same set of distributions as for \Fig{fig:Bkg_pu}, but for the signal boosted $Z$ boson sample. In this case, the analysis of the phase space predicted that the dominant effect of the pile-up on the distributions is a shift for both $\Cobs{2}{\beta}$ and $\Dobs{2}{\beta}$, with the shift being smaller for $\Dobs{2}{\beta}$. This behavior is manifest in \Fig{fig:Sig_pu}. The difference in the stability of the mean between $\Cobs{2}{\beta}$ and $\Dobs{2}{\beta}$ is particularly pronounced at small $\beta$. At larger $\beta$, the $\Dobs{2}{\beta}$ distribution exhibits a jump at small amounts of pile-up, and then remains stable as pile-up increases. Unfortunately, we have not been able to understand this behavior completely from power counting. Nevertheless, the improved stability of the distributions of $\Dobs{2}{\beta}$ as compared with $\Cobs{2}{\beta}$ is promising.

For comparison, in \Fig{fig:nsub_pu}, we consider the impact of pile-up on signal and background distributions for $\Nsub{2,1}{\beta}$, for the representative value $\beta=1$. As for $\Cobs{2}{\beta}$ and $\Dobs{2}{\beta}$, we expect that the dominant effect of pile-up on the signal distributions is a shift of the peak value, while for the background distributions, we expect a small shift of the mean and an accumulation of the distribution near $\Nsub{2,1}{\beta}=1$. This is exhibited in the Monte Carlo.

\begin{figure}
\begin{center}
\subfloat[]{
\includegraphics[width=6.5cm]{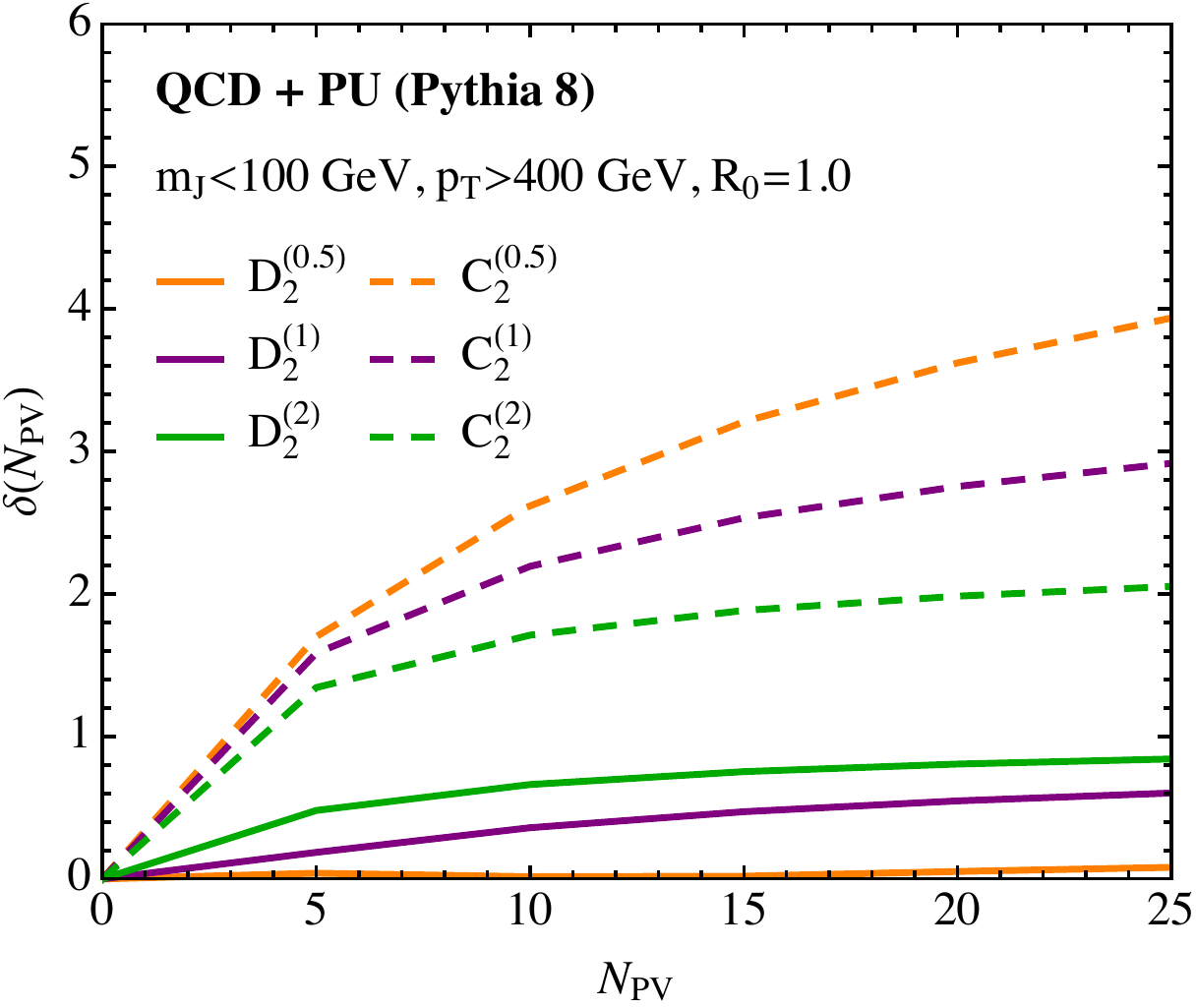}
}\qquad
\subfloat[]{
\includegraphics[width=6.5cm]{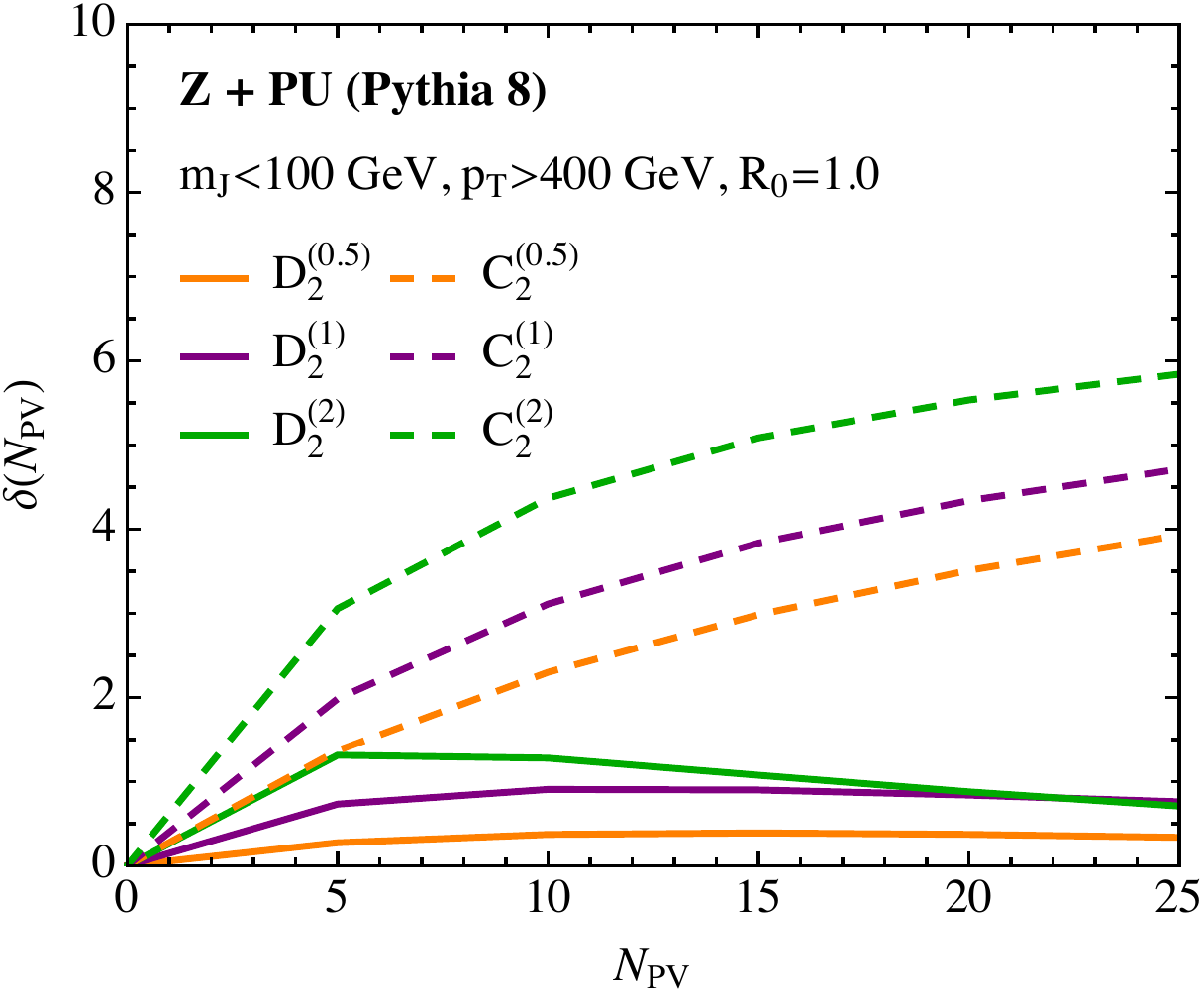}
}
\end{center}
\caption{A comparison of the susceptibility as a function of the number of pile-up vertices $N_{PV}$ of background (left) and signal (right) distributions for $\Cobs{2}{\beta}$ and $\Dobs{2}{\beta}$ to pile-up using the measure $\delta (\text{$N_{PV}$})$ for $\beta = 0.5,1,2$. 
}
\label{fig:C2_pu_mean}
\end{figure}

For a more quantitative study of the stability of the distributions to pile-up, we define
\begin{equation}\label{eq:delta_$N_{PV}$}
\delta_{\Xobs{2}{\beta}}(\text{$N_{PV}$})=\frac{\langle \Xobs{2}{\beta} (\text{$N_{PV}$}) \rangle -\langle \Xobs{2}{\beta} (\text{$N_{PV}$}=0) \rangle}{\sigma_{\Xobs{2}{\beta}}(\text{$N_{PV}$}=0)}\,,
\end{equation}
where $\Xobs{2}{\beta}$ stands for either $\Cobs{2}{\beta}$ or $\Dobs{2}{\beta}$ and $\sigma$ denotes the standard deviation.  This quantity is a measure of how much the mean of the distribution is affected by pile-up, normalized by the width of the distribution, which is important since the observables $\Cobs{2}{\beta}$ and $\Dobs{2}{\beta}$ have support over very different ranges.  While it is clear from \Fig{fig:Bkg_pu} that the dominant effect of pile-up on the background distributions for $\Dobs{2}{\beta}$ is not a shift of the mean, and so the change of the distribution is not accurately captured by the measure of \Eq{eq:delta_$N_{PV}$}, the deviation of the mean is a commonly studied measure of an observable's susceptibility to pile-up. In \Fig{fig:C2_pu_mean} we plot $\delta(\text{$N_{PV}$})$ for the variables $\Cobs{2}{\beta}$ and $\Dobs{2}{\beta}$. As was demonstrated in \Figs{fig:Bkg_pu}{fig:Sig_pu}, the mean of the distributions of $\Dobs{2}{\beta}$ is considerably more stable for both the signal and background distributions.

\section{Power Counting Quark vs. Gluon Discrimination}
\label{sec:qvg}

Unlike the case of boosted $Z$ bosons vs.~massive QCD jets, applying a power counting analysis to quark vs.~gluon jet discrimination demonstrates the limitations of the technique.  Both quark and gluon jets dominantly have only a single hard core, and so the natural discrimination observables are the two-point energy correlation functions, $\ecf{2}{\beta}$.\footnote{To next-to-leading logarithmic accuracy, $\ecf{2}{\beta}$ are identical to the recoil-free angularities \cite{Larkoski:2014uqa}.}  As shown in \Sec{sec:softcollqcd}, power counting the two-point energy correlation functions constrains the soft and collinear radiation as:
\begin{equation}
\ecf{2}{\beta}\sim z_s \sim R_{cc}^\beta \ .
\end{equation}
With power counting alone, this is as far as our analysis can go.  $\ecf{2}{\beta}$ does not parametrically separate quark and gluon jets from one another.

This result is not surprising, however, because there are no qualities of quark and gluon jets that are parametrically different.  Indeed,
\begin{align*}
C_A&\sim C_F \ ,\\
N_C&\sim n_f \ ,\\
\text{spin } 1 &\sim \text{spin }1/2 \ ,
\end{align*}
where $C_A$ and $C_F$ are the color factors for gluons and quarks, $N_C$ is the number of colors, and $n_f$ the number of active fermions.  Predictions of what the best observable for quark vs.~gluon discrimination is requires a detailed analysis of the effects of these order-1 parameters, which has been studied in several papers \cite{Gallicchio:2011xq,Gallicchio:2012ez,Larkoski:2013eya,Larkoski:2014pca}.  However, with the additional input of the form of the splitting functions for quarks and gluons, we can predict that the discrimination power of $\ecf{2}{\beta}$ improves as $\beta$ decreases because smaller $\beta$ emphasizes the collinear region of phase space over soft emissions.  Collinear emissions are sensitive to the spin of the parton in addition to the total color of the jet, and thus are more distinct between quark and gluon jets.  This prediction is borne out by explicit calculation to next-to-leading logarithmic accuracy \cite{Larkoski:2013eya}. An analytic calculation of the improved discrimination power from simultaneous measurement of the recoil-free angularities for two different powers of the angular exponent was calculated in \cite{Larkoski:2014pca}.

Nevertheless, this suggests that power counting does make a definite prediction of quark vs.~gluon discrimination performance.  Because all the physics of quark vs.~gluon jet discrimination is controlled by order-1 numbers, the predicted discrimination should be sensitive to the tuning of order-1 numbers in a Monte Carlo.  It has been observed that \pythia{8} and \herwigpp\  give wildly different predictions for quark vs.~gluon discrimination power \cite{,Larkoski:2013eya,Larkoski:2014pca,CMS:2013kfa,Aad:2014gea}, and presumably the difference is dominated by the tuning of the Monte Carlos.  However, isolating pure samples of quark and gluon jets is challenging experimentally 
\cite{CMS:2013kfa,CMS:2013wea,Aad:2014gea,Gallicchio:2011xc} 
and most of the subtle differences between quarks and gluons only appear at an order formally beyond the accuracy of a Monte Carlo.  Therefore, to solve this issue will require significant effort from experimentalists, Monte Carlo authors, and theorists to properly define quark and gluon jets, to identify the dominant physics, and to isolate pure samples for tuning.

\section{Conclusions}
\label{sec:conc}

In this paper we have demonstrated that power counting techniques can be a powerful guiding principle when constructing observables for jet substructure and for understanding their behavior. Since power counting captures the parametric physics of the underlying theory, its predictions should be robust to Monte Carlo tunings. Using the simple example of discriminating boosted $Z$ bosons from QCD jets with the energy correlation functions, we showed that a power counting analysis identified $\Dobs{2}{\beta}$ as the natural discrimination observable. The scaling of this observable parametrically separates regions of the $(\ecf{2}{\beta},\ecf{3}{\beta})$ phase space dominated by 1- and 2-prong jets.  The distinction between 1- and 2-prong jets is invariant to boosts along the jet direction.

To verify the power counting predictions, we performed a Monte Carlo analysis comparing $\Dobs{2}{\beta}$ with a previously proposed observable, $\Cobs{2}{\beta}$, also formed from the energy correlation functions. We showed that $\Dobs{2}{\beta}$ is a superior observable for discrimination because $\Cobs{2}{\beta}$ inextricably mixes signal-rich and background-rich regions of phase space. All power counting predictions were confirmed by both \herwigpp\ and \pythia{8}, showing that the dominant behavior of the observables is governed by parametric scalings and not by $\mathcal{O}(1)$ numbers. This was contrasted with the case of quark vs.~gluon discrimination for which no parametric differences exist, leading to large discrepancies when simulating quark vs.~gluon discrimination with different Monte Carlo generators.

We also demonstrated that power counting can be used to understand the impact of pile-up on different regions of the phase space, and hence on the distributions of discriminating variables. The distributions for $\Dobs{2}{\beta}$ exhibited improved stability compared with those of $\Cobs{2}{\beta}$, while the background distributions have the interesting feature of being compressed to a central value by the addition of pile-up radiation.

We anticipate many directions to which the power counting approach could be applied. We have restricted ourselves in this paper to a study of observables formed from ratios of energy correlation functions with the same angular exponent. A natural generalization is to ratios of energy correlation functions with different angular exponents, where the optimal observable is given by $\Dobs{2}{\alpha,\beta}=\ecf{3}{\beta}/(\ecf{2}{\alpha})^{3\beta/\alpha}$. Such variables could be useful when considering pile-up in the presence of mass cuts, which are required experimentally. In the presence of a mass cut, an angular exponent of $\ecfnobeta{2}$ near $2$ provides a simple restriction on the phase space, while lowering the angular exponent of $\ecfnobeta{3}$ reduces the effect of soft wide angle radiation. Along these lines, the impact of grooming techniques on the phase space is also simple to understand by power counting \cite{Walsh:2011fz}, and could be used to motivate the design of variables with desirable behavior under grooming.

As another example of considerable interest, the power counting analysis can be extended to the study of top quark discrimination variables by considering the phase space for 1-, 2- and 3-prong jets defined by the two-, three- and four-point energy correlation functions. While a complete analytic calculation for this case is not feasible, a power counting analysis is, and can be used to predict discriminating observables with considerably improved performance compared to those originally proposed in \cite{Larkoski:2013eya}. In the case of a three dimensional phase space, a cut on the jet mass only reduces the phase space to a two dimensional subspace, so that the functional form of the observable remains important. This will be studied further in future work.

Our observation that boost-invariant combinations of the energy correlation functions are the most powerful discriminants can also be exploited for discrimination: we can use boost invariance as a guide for defining the best observables. Together with power counting, this gives a simple but powerful analytic handle to understand and design jet substructure observables.

\begin{acknowledgments}
We thank Jesse Thaler and Iain Stewart, for helpful discussions and Gavin Salam for detailed comments on the manuscript.  This work is supported by the U.S. Department of Energy (DOE) under grant Contract Numbers DE-SC00012567 and DE-SC0011090. D.N. is also supported by an MIT Pappalardo Fellowship.  I.M. is also supported by NSERC of Canada.
\end{acknowledgments}

\bibliography{powercounting}

\end{document}